RWTH Aachen University

Simulation Science

Master's Thesis

# Modeling and simulation of heat source trajectories through phase-change materials

Alexander Gary Zimmerman

Matriculation-no. 351345

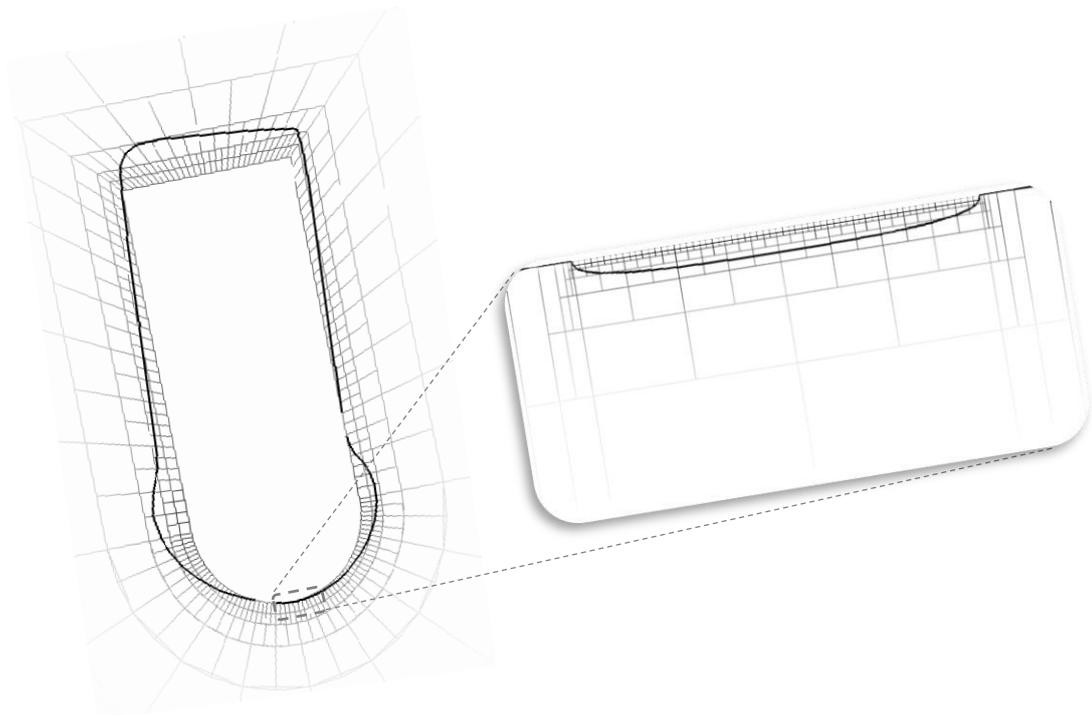

First reviewer: Prof. Marek Behr, Faculty of Mechanical Engineering

Second reviewer: Dr. Julia Kowalski, Junior Research Group Leader, AICES

Aachen, November 2016

# Statutory declaration

I, Alexander Gary Zimmerman, declare that I have independently authored this thesis, that I have only used the declared sources, and that I have explicitly marked all material which has been quoted either literally or by content from these sources.

Aachen, November 2016

<div style="text-align: right;">Alexander Gary Zimmerman</div>

# Abstract

The modeling and simulation of heat source trajectories through phase-change materials is a relevant problem both for space exploration and for terrestrial climate research, among other fields. In space, the DLR and NASA are both interested in exploring beneath the surfaces of icy moons, primarily Enceladus and Europa, where conditions may allow for extraterrestrial life. On Earth, unique subglacial aquatic ecosystems offer potential for geobiological discoveries. Unfortunately, existing ice-drilling technology is dirty and cumbersome. Most importantly, the target environments must not be contaminated; but furthermore, such heavy equipment cannot be reasonably landed for extraterrestrial missions. Melting probes are a clean and compact alternative technology which use heaters to melt through the ice. In recent years, the IceMole melting probe has been developed and deployed. Unique to the IceMole, its trajectory is controlled with differential heating and a small stabilizing drill, allowing the probe to avoid obstacles and navigate to distant targets.

Successful melting probe trajectory control requires advancements not only in the modeling and simulation of the ambient dynamics, e.g. the transport of mass, momentum, and energy within a phase-change material (PCM) domain, but also of the melting probe's coupled rigid body dynamics (RBD). Fundamentally the RBD can be modeled by the equations of motion with force and moment balances; but this approach has already been shown as prohibitively complex for the efficient computation of the coupled RBD and ambient dynamics. In this work, we propose a new approach which views the heat source moving through a series of states that are in equilibrium with the ambient dynamics. Most significantly, the motion of the probe is driven by contact with the evolving phase-change interface (PCI) as the probe melts through the material. From this perspective, the RBD are formulated as an energy minimization problem. For the ambient dynamics, in this work, we do not yet employ a realistic phase-change model, but rather we use a simplified model which yields a qualitatively interesting PCI. This allows us to test the coupled problem before investing in the implementation of more advanced phase-change models.

This thesis formulates the general mathematical problem as two split operators, one for the RBD and one for the ambient dynamics. We couple these operators with feasibility constraints. The constraints ensure that the probe does not penetrate the evolving solid PCM domain. Then we provide a concrete example both for the energy minimization problem, which we formulate and implement as a nonlinear program, and for the unsteady ambient dynamics, which we formulate as a partial differential equation (PDE) and implement as a discrete linear system. In addition to verifying the correctness of the PDE implementation, special care is taken to verify its spatial and temporal orders of accuracy. Finally, we present an algorithm for the temporal coupling of the split operators, which we implement in Python and C++. We demonstrate example trajectories, including the dynamic response of the RB velocity to a rapid change in the heat flux.

# Contents

This page is intentionally left blank.

# List of figures and tables

# Nomenclature

**Symbols**

| Symbol | Description | Space | Units |
|---|---|---|---|
| $a(\cdot, \cdot)$ | Bi-linear diffusion form | $: \mathbb{R}^n \times \mathbb{R}^n \to \mathbb{R}$ | - |
| $\alpha = \dfrac{k}{\rho c_p}$ | Thermal diffusivity | $\mathbb{R}$ | m²/s |
| $\boldsymbol{A}$ | System matrix | $\mathbb{R}^n \times \mathbb{R}^n$ | - |
| $\boldsymbol{b}$ | Body force (e.g. gravity) | $\mathbb{R}^d$ | N |
| $c(\cdot, \cdot)$ | Bi-linear convection form | $: \mathbb{R}^n \times \mathbb{R}^n \to \mathbb{R}$ | - |
| $c_p$ | Specific heat capacity at constant pressure | $\mathbb{R}$ | J / (kg K) |
| $\boldsymbol{C}$ | Convection matrix | $\mathbb{R}^n \times \mathbb{R}^n$ | - |
| $d$ | Number of spatial dimensions | 2 or 3 | |
| $\delta$ | Melt film thickness | $\mathbb{R}$ | m |
| $\boldsymbol{f}$ | Right-hand-side vector | $\mathbb{R}^n$ | - |
| $g$ | Dirichlet boundary function (for boundary value problem) | $: \mathbb{R}^d \to \mathbb{R}$ | - |
| $\boldsymbol{g}$ | Feasibility constraints (for optimization problem) | $: \mathbb{R}^n \to \mathbb{R}^m$ | - |
| $\boldsymbol{\gamma}$ | Discrete set of points on a boundary (for sampling feasibility constraints) | $\mathbb{R}^m$ | |
| $\boldsymbol{\Gamma}$ | Boundary manifold (of a domain manifold $\boldsymbol{\Omega}$) | $\mathbb{R}^d$ | - |
| $h$ | Neumann boundary function | $: \mathbb{R}^d \to \mathbb{R}$ | - |
| | or | | |
| | Enthalpy | $\mathbb{R}$ | J |
| $h_m$ | Latent heat of melting | $\mathbb{R}$ | J / kg |
| $\mathcal{H}: t \times \boldsymbol{\xi} \times \boldsymbol{u} \to \dot{\boldsymbol{\xi}}$ | Operator for rigid body dynamics | | - |
| $i$ | Discrete time index for rigid body dynamics | $\mathbb{N}$ | - |
| $\hat{\boldsymbol{\iota}}$ | First body-fixed axis | $\mathbb{R}^d$ | - |

| Symbol | Description | Space | Units |
|---|---|---|---|
| $j$ | Discrete time index for ambient dynamics | $\mathbb{N}$ | - |
| $\hat{\jmath}$ | Second body-fixed axis | $\mathbb{R}^d$ | - |
| $k$ | Thermal conductivity | $\mathbb{R}$ | W/(m K) |
| $\hat{k}$ | Third body-fixed axis | $\mathbb{R}^3$ | -- |
| $\boldsymbol{K}$ | Laplace/Stiffness matrix | $\mathbb{R}^n \times \mathbb{R}^n$ | - |
| $\mathcal{L}: t \times \boldsymbol{u} \times \boldsymbol{\xi} \times \dot{\boldsymbol{\xi}} \to \boldsymbol{u}_t$ | Operator for ambient field dynamics | | - |
| $m$ | Number of feasibility constraints | $\mathbb{N}$ | - |
| | or | | |
| | Number of time steps for integrating rigid body dynamics | | |
| $\boldsymbol{M}$ | Mass matrix | $\mathbb{R}^n \times \mathbb{R}^n$ | - |
| $n$ | Dimensionality of vector space for finite element discretization | $\mathbb{N}$ | |
| | or | | |
| | Number of optimization design variables | | |
| | or | | |
| | Number of time steps when integrating the ambient dynamics | | |
| $\boldsymbol{\nabla}$ | Spatial gradient operator | $: \mathbb{R} \to \mathbb{R}^d$ | $m^{-1}$ |
| $\boldsymbol{\nabla}^2$ | Laplace operator | $: \mathbb{R} \to \mathbb{R}$ | $m^{-2}$ |
| $\boldsymbol{\omega}$ | Angular velocity | $\mathbb{R}^d$ | rad/s |
| $\boldsymbol{\Omega}$ | Domain manifold | $\mathbb{R}^d$ | - |
| $p$ | Pressure | $\mathbb{R}$ | N/m$^2$ |
| $\phi(t, \boldsymbol{x})$ | Solution basis function | $: \mathbb{R} \times \mathbb{R}^d \to \mathbb{R}$ | - |
| $Pe = \frac{vx_r}{\alpha}$ | Global Peclet number | $\mathbb{R}$ | - |
| $Pe_h = \frac{vh}{2\alpha}$ | Local element Peclet number | $\mathbb{R}$ | - |

| Symbol | Description | Space | Unit |
|---|---|---|---|
| $\Psi$ | Potential energy | $\mathbb{R}$ | J |
| $\boldsymbol{q}$ | Attitude quaternion | $\mathbb{R}^4$ | - |
| $\boldsymbol{r}$ | Shift vector | $\mathbb{R}^d$ | m |
| $\bar{\boldsymbol{r}}$ | Centroid of the rigid body | $\mathbb{R}$ | m |
| $\rho$ | Mass density | $\mathbb{R}$ | kg/m$^3$ |
| $Ste = c_{p,S} \frac{T_h - T_m}{h_m}$ | Stefan number | $\mathbb{R}$ | - |
| $t$ | Time | $\mathbb{R}$ | s |
| $T$ | Temperature | $\mathbb{R}$ | °C |
| $\theta$ | Rotation angle in 2D space | $\mathbb{R}$ | rad |
| $\boldsymbol{u} = [T, \boldsymbol{v}, p]^T$ | Ambient state | $\mathbb{R}^{d+2}$ | |
| $\boldsymbol{v}$ | Velocity | $\mathbb{R}^d$ | m/s |
| $\boldsymbol{x}$ | Coordinates of a point in Cartesian space | $\mathbb{R}^d$ | m |
| | or | | |
| | Optimization design variables | $\mathbb{R}^n$ | |
| $\boldsymbol{\xi} = [\boldsymbol{q}, \boldsymbol{r}]^T$ | Rigid body state | $\mathbb{R}^7$ | |
| $[\ ]_t$ | Partial time derivative | $: \mathbb{R} \to \mathbb{R}$ | $s^{-1}$ |
| $[\ ]_x$ | Partial space derivative | $: \mathbb{R} \to \mathbb{R}$ | $m^{-1}$ |
| $[\ \dot{}\ ]$ | Time derivative (for an ODE) | $: \mathbb{R} \to \mathbb{R}$ | $s^{-1}$ |

**Subscripts, superscripts, and indices**

| | | |
|---|---|---|
| $[\ ]_a$ | Local basis function index | |
| $[\ ]_A$ | Global basis function index | |
| $[\ ]_b$ | Local basis function index | |
| $[\ ]_B$ | Global basis function index | |
| $[\ ]_c$ | Contact boundary | |
| $[\ ]_D$ | Dirichlet | |
| $[\ ]_e$ | Outer boundary | |

| | |
|---|---|
| $[\ ]^h$ | Discrete vector from a continuous function |
| $[\ ]_h$ | Heat source boundary (hull) |
| $[\ ]_L$ | Liquid phase |
| $[\ ]_m$ | Melting interface |
| $[\ ]_N$ | Neumann |
| $[\ ]^i$ | Evaluate the time-dependent function at discrete time $i$ |
| $[\ ]_r$ | Reference values for scaling the dimensionless PDE |
| $[\ ]_S$ | Solid phase |
| $[\ ]_w$ | Wall |
| $[\ ]_0$ | Initial value |
| $[\ ]^-$ | Liquid side of 1D melt interface |
| $[\ ]^+$ | Solid side of 1D melt interface |
| $[\ ]^*$ | Corresponding to the exact Stefan condition |

**Acronyms**

| | |
|---|---|
| BC | Boundary condition (for a PDE) |
| CFD | Computational fluid dynamics |
| CG | Conjugate Gradient (iterative solver) |
| FEM | Finite element method |
| MDAO | Multidisciplinary design, analysis, and optimization |
| MES | Method of exact solution (for verification) |
| MMS | Method of manufactured solution (for verification) |
| ODE | Ordinary differential equation |
| PC | Phase-change |
| PCI | Phase-change interface |
| PCM | Phase-change material |
| PDE | Partial differential equation |
| RB | Rigid body |
| RBD | Rigid body dynamics |
| SPD | Symmetric positive definite (matrix) |
| 2D | Simplified two-dimensional space, corresponding to $d = 2$ |
| 3D | Realistic three-dimensional space, corresponding to $d = 3$ |

This page is intentionally left blank.

# 1 Background and introduction

## 1.1 Melting probes for subglacial exploration

Many potential scientific discoveries wait underneath sheets of ice. Information about our planet's climate history is buried deeply within Antarctic glaciers. Evidence of extraterrestrial life may be swimming in subsurface oceans on Saturn's moon Enceladus (Hsu, Postberg, & al., 2015) or Jupiter's moon Europa, as first evidenced by enormous geysers of water erupting from the moons' icy surfaces. NASA and the DLR are particularly interested in Enceladus. Some future mission concepts plan to sample directly from the geysers. Unfortunately, a sample of these vented molecules could be contaminated by severe radiation around Enceladus, which is poorly protected by Saturn's weak magnetic field. Clean samples exist below kilometers of radiation shielding, i.e. water-ice. This warrants mission concepts that attempt to travel through the ice and sample the subsurface ocean, or rather to sample from a crevasse nearer the surface.

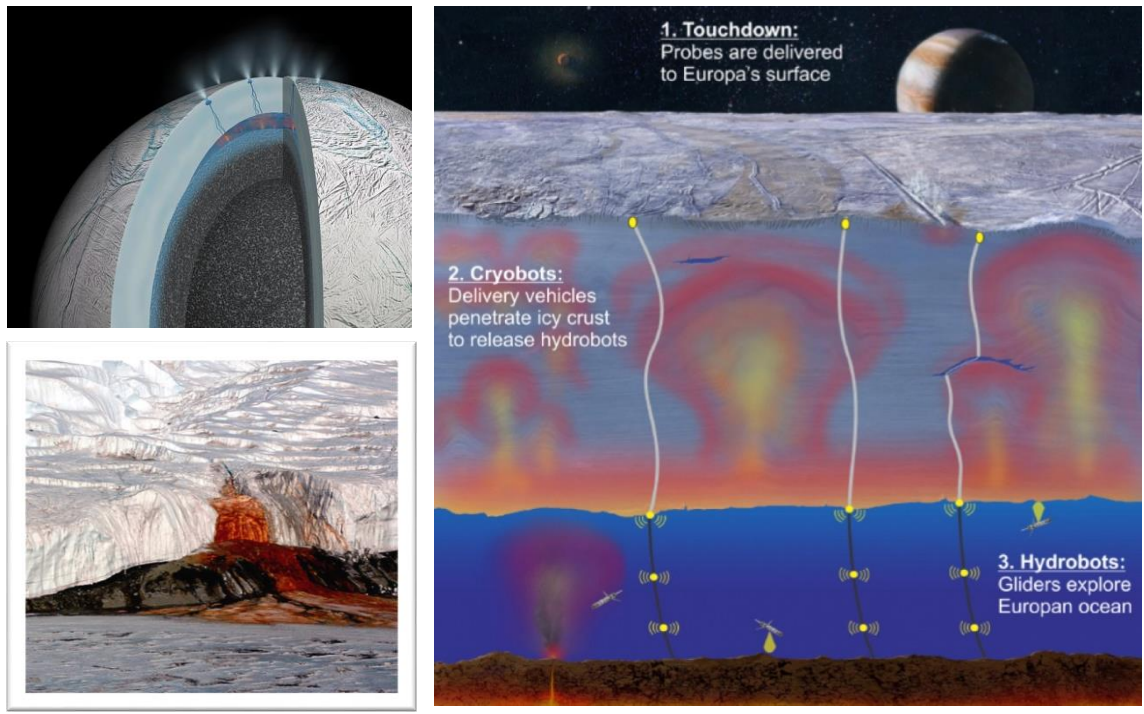

Figure 1.1: Example mission environments.
Top-Left: Rendering of Enceladus geothermal activity (NASA); Bottom-Left: Blood Falls, Antarctica (NSF); Right: Subsurface mission architecture (NASA)

Conventional drilling approaches are too dirty, cumbersome, and inefficient. The exploration of biological environments requires a probe which does not contaminate its surroundings. Furthermore, extraterrestrial missions require a compact payload with low power usage. A popular alternative to drilling is the melting probe, a concept originally attributed to Karl Philberth. Multiple variations of these probes have already been deployed on terrestrial glaciers, and more designs are in development. Two such examples are shown in Figure 1.2. Initial concepts could only be propelled by gravity; but the IceMole (Dachwald, et al., 2014) incorporates a stabilizing drill and employs differential heating to navigate curved trajectories. The drill even allows the IceMole to climb steeply against gravity.

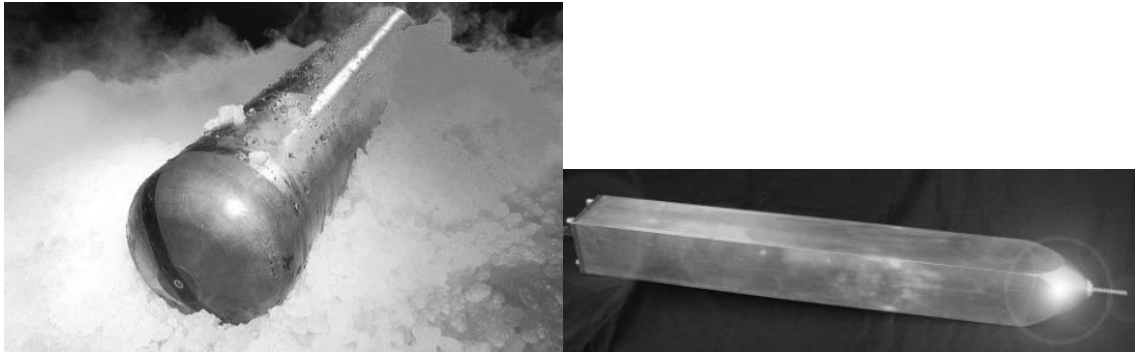

Figure 1.2: Example melting probes
Left: Philbert-probe (NASA); Right: IceMole 2 (Dachwald, et al., 2014)

A key barrier to the development and deployment of melting probes, and most critically to their control, is the lack of modeling and simulation capabilities. Experimental tests are too expensive to sufficiently cover all of the necessary conditions. This practical problem is well understood by the aerospace industry. In the case of the subsurface exploration of icy moons, a true field test will involve landing a probe on a moon of another planet, where the conditions are much different than on Earth. Therefore, to maximize the value of extraterrestrial missions, we need simulations with predictive capability.

## 1.2 The physical forward problem

In general, we consider some rigid body (RB) which is a heat source, moving through a phase-change material (PCM) in time. We refer to the time history of the RB's movement as the trajectory. The physical processes within the PCM domain include melting and sublimation, solid and liquid heat conduction, liquid heat convection, and incompressible fluid flow driven by velocity and pressure gradients. While the methods herein are generalized to any PCM, we primarily consider water-ice, and hence we will refer to the phase diagram in Figure 1.3.

Fundamentally the rigid body dynamics (RBD) are governed by the laws of motion subject to forces and moments from the environment; but we will show (in chapter 4) that the motion can instead be modeled as a series of equilibrium states that minimize an appropriate energy functional. This alternative formulation of the RBD has the potential to greatly reduce the computational complexity of the problem, which would allow it to be applied to larger inverse problems.

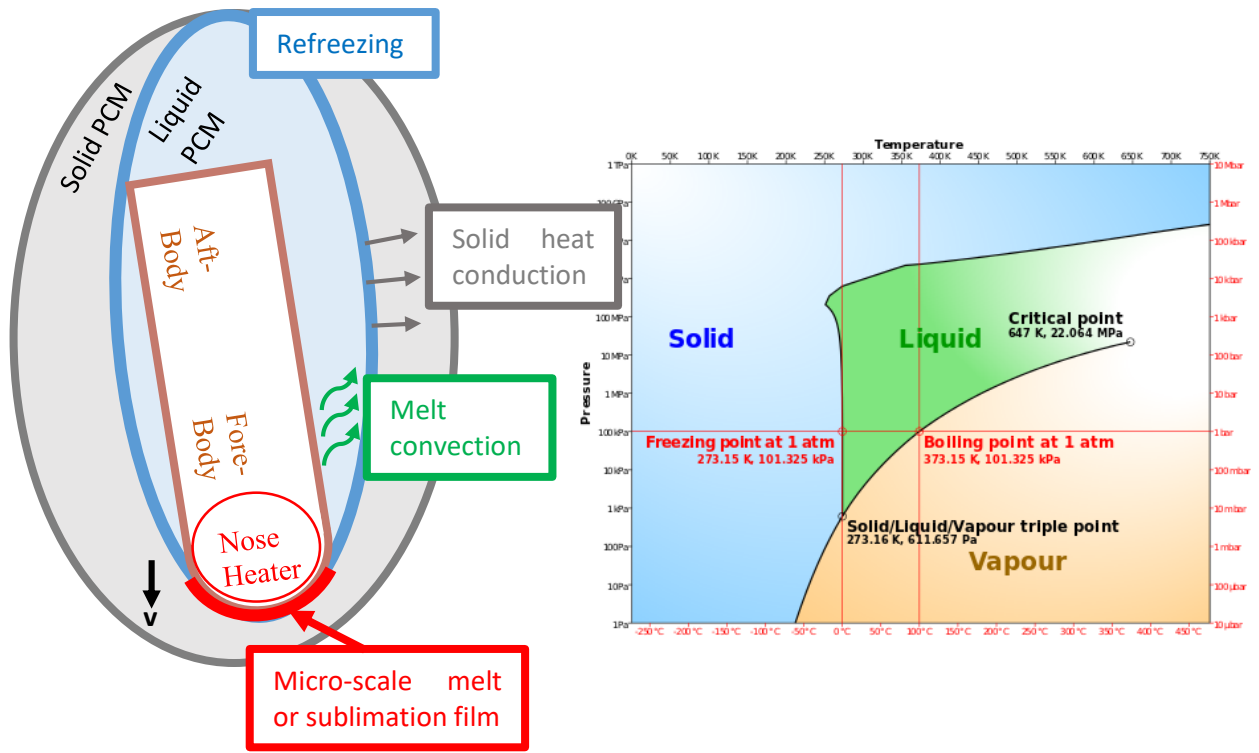

Figure 1.3: The many physical processes of the phase-change problem
Left: Labeled physical processes (own work); Right: Water phase diagram (Cmglee, 2016)

## 1.3 Inverse problems

To motivate the need for forward simulations with low computational complexity, it is important to understand the difference between the forward problem and the many related inverse problems. A solution to the forward problem can be understood as the capability to predict the trajectory of a given melting probe with given control inputs in a given environment. Successfully predicting this trajectory requires accurate simulation of multiple coupled physical processes, some of which are pictured in Figure 1.3, on multiple spatial and temporal scales. Solving the forward problem is already complex; but engineers are also faced with multiple inverse problems that greatly increase the problem size.

Engineers must be able to design these probes to fulfill mission requirements with some degree of confidence. An accurate forward trajectory simulation is just one of many components that are necessary for the design framework. All relevant disciplines provide models for their aspect of the multidisciplinary problem, and typically these models are used together in a multidisciplinary design, analysis, and optimization (MDAO) framework to design a probe that best meets the mission requirements. Every design parameters adds a dimension of complexity to the problem.

Just one of these disciplines, the optimal control of the probe, is itself another inverse problem. The aerospace industry is well acquainted with optimal control problems. The design variables encompass every control input throughout continuous time, which be discretized in some way.

Typically, these solvers become extremely large and sparse nonlinear programs, with many more design variables than other types of design optimization problems.

Furthermore, every discipline is affected by many uncertainties, which introduces a stochastic nature to the MDAO problem. An entire discipline, uncertainty quantification (UQ), involves quantifying how these many uncertainties propagate through the overall model and affect the quantity of interest, which is usually some measure of mission success. If the inverse problem of MDAO was not already large enough, the many dimensions of the probability spaces in uncertainty-based MDAO can quickly create intractable problems.

## 1.4   Objectives and outline

The aforementioned inverse problems all rely on an accurate, robust, and efficient solution to the forward problem. Furthermore, to be successful in applying our models to new mission environments where we have little data, the models must be predictive and therefore based on general physical laws. Such a solution to the forward problem is the subject of this thesis. In the coming chapters, we will present

- Chapter 2: previous work on melting probe trajectory simulation.
- Chapter 3: a general mathematical formulation of the forward problem, decomposing it into two sub-problems, viewed as split operators, and their coupling.
- Chapter 4: the development and implementation of a concrete model of the rigid body dynamics sub-problem.
- Chapter 5: the development and Python implementation of a concrete model of the ambient dynamics sub-problem. Here we greatly simplify the melting model so that we can focus most of our attention on developing the coupled problem.
- Chapter 6: the implementation of the ambient dynamics model in C++, including verification of its correctness, as well as its spatial and temporal orders of accuracy.
- Chapter 7: a brief example of a physically realistic melting simulation, where the phase-change is embedded in a moving melt film boundary.
- Chapter 8: an algorithm for the temporal coupling of the split operators, and the implementation of the coupled problem in Python and C++.

## 2 Previous work and existing methods

Decades of work has already been done toward the modeling and simulation of melting processes, summarized by (Alexiades & Solomon, 1993), and migrating heat sources through melting materials, summarized by (Schüller, 2015). It is helpful to classify the models presented in this chapter into two groups. First we present existing models for the phase-change process itself. Then we briefly present an existing quasi-stationary model for migrating heat sources through phase-change materials.

### 2.1 The melting process

Before we can simulate the trajectory of a heat source through a phase-change material (PCM), we must understand the phase-change process. While general melting probe missions may involve sublimation, we will only present melting models in this thesis.

Melting is the physical process of a substance transitioning[1] from solid phase to liquid phase. The process occurs when the solid's internal energy, measured by the temperature, reaches a melting temperature $T_m$. To increase the internal energy or temperature, heat must be applied. Under constant pressure, the amount of energy required to increase the temperature of a substance is the constant-pressure specific heat capacity $c_p$. Once the substance reaches $T_m$, applied heat continues to increase the internal energy; but instead of increasing the temperature, the heat changes the phase of the substance from solid to liquid. The amount of specific energy required to complete this phase transition is the latent heat of melting $h_m$. The specific heat capacity $c_p$, melting temperature $T_m$, and latent heat of melting $h_m$ are all bulk material properties. These properties are also independent of the direction of the phase-change process, meaning that the same values apply to the freezing process.

#### 2.1.1 The one-dimensional Stefan problem

In one dimension, the melting process is mathematically modeled by the Stefan problem (Alexiades & Solomon, 1993). The Stefan problem models the unsteady conservation of energy near the phase-change interface (PCI). Specifically, mass transfer is ignored, and only isotropic heat transfer by conduction is considered. A sketch of the problem is shown in Figure 2.1. In this figure, $T_h$ is the temperature of the stationary heating surface, $T_S$ is the temperature of the solid PCM before heating, and time-dependent $x(t)$ is the moving position of the phase-change interface (PCI) as the PCM melts.

---

[1] Between the solid and liquid domains in space, there is a sharp transition region; but this size of this region is on the atomic scale and is hence not resolved by any of our continuum models.

To ensure a local energy balance, this model employs the Stefan condition

$$\rho_S h_m \frac{\partial}{\partial t} x_m(t) = q^- - q^+ = -k_L \frac{\partial T(x_m(t)^-, t)}{\partial x} + k_S \frac{\partial T(x_m(t)^+, t)}{\partial x} \quad (2.1)$$

where $h_m$ is the PCM's latent heat of melting, $\rho_S$ is the solid density, $\frac{\partial}{\partial t} x_m(t)$ is the velocity of the PCI, $x_m(t)^-$ and $x_m(t)^+$ respectively denote the left and right hand sides of the PCI, $q^-$ and $q^+$ are the respective heat fluxes into either side of the PCI, and $k_L$ and $k_S$ are the respective thermal conductivities of the liquid and solid states. The key idea is that moving the PCI requires an amount of specific energy equivalent to $h_m$. This is perceived as a step in the heat flux between either side of the PCI. The Stefan number is defined as

$$Ste \equiv c_{p,S} \frac{T_h - T_m}{h_m} \quad (2.2)$$

where $c_{p,S}$ is the solid thermal heat capacity. For a small Stefan number, which is the case for a heat source with a small temperature in water-ice, one may assume a linear change in temperature through the liquid domain of the Stefan problem. This allows us to approximate the gradient as

$$\frac{\partial T(x_m(t)^-, t)}{\partial x} \approx \frac{T_m - T_h}{\delta(t)} \quad (2.3)$$

The Stefan problem neglects mass and momentum conservation laws, and therefore neglects heat and mass transfer by convection. Furthermore, the Stefan problem is not immediately extensible to physical 3D space. This motivates the need for the enthalpy-porosity model.

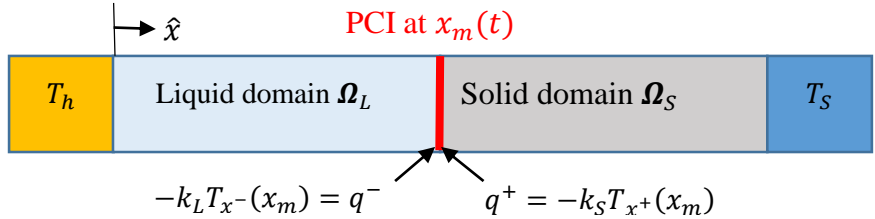

Figure 2.1: The Stefan problem
The Stefan problem requires an energy balance at the time-dependent position of the phase-change interface (PCI) $x_m(t)$ where the temperature equals the melting temperature $T_m$ of the phase-change material (PCM), $T_h > T_m$ is the hot wall temperature, and $T_s < T_m$ is the cold temperature of the solid material.

### 2.1.2 The enthalpy-porosity model

One proposed model for the 3D phase-change process is the fixed-grid enthalpy-porosity model (Belhamadia, Kane, & Fortin, 2012). Whereas only conductive heat transfer was considered in the Stefan problem, the enthalpy-porosity model enforces the incompressible conservation of mass, momentum, and energy, written as

$$\frac{\partial \rho}{\partial t} + \nabla \cdot (\rho \boldsymbol{v}) = 0 \tag{2.4}$$

$$\frac{\partial (\rho \boldsymbol{v})}{\partial t} + \nabla \cdot (\rho \boldsymbol{v} \otimes \boldsymbol{v}) = -\nabla p + \nabla \cdot (\eta \nabla \boldsymbol{v}) + A\boldsymbol{v} + \boldsymbol{S}(T) \tag{2.5}$$

$$\frac{\partial (\rho h)}{\partial t} + \nabla \cdot (\rho \boldsymbol{v} h) = \nabla \cdot (k \nabla T) \tag{2.6}$$

where $\gamma_S$ and $\gamma_L$ are the solid and liquid fractions of the mixture, $\boldsymbol{S}$ is the well-established Boussinesq approximation, and $A$ is the lesser known Kozeny-Carman relation is

$$A = -\frac{C(1-\gamma_L)^2}{\gamma_L^3 + \epsilon} \tag{2.7}$$

and the density $\rho$ and enthalpy $h$ are split by phase

$$\rho = \gamma_L \rho_L + \gamma_S \rho_S \tag{2.8}$$

$$h = f_L h_L + f_S h_S \approx \gamma_L h_L + \gamma_S h_S \tag{2.9}$$

For the pure liquid phase, $\gamma_L = 1$, $A$ vanishes. As the mixture approaches a pure solid, $\gamma_L = 0$, $A$ becomes very large, which vanishes $v$. An arbitrary small number $\epsilon$ prevents division by zero. The factor $C$, in the numerator, of is a constant that accounts for the form of large phase transition regions which may occur during super-cooling, which we shall not need to consider.

With some modifications, a weak form of (2.4)-(2.6) can be derived for numerical solution with the finite element method (FEM). In chapter 5, we will demonstrate FEM for a simpler model which only includes the conservation of energy, and greatly simplifies the phase-change process.

## 2.2 Migrating heat sources through phase-change materials

The majority of the existing work on this topic focuses on analytical models for rectilinear (i.e. usually straight down) melting with simple geometries such as plates and spheres. Recently the rectilinear melting theory has been extended to curvilinear melting (Schüller, Kowalski, & Raback, 2016), i.e. melting along a curved trajectory.

In the regime of close-contact melting, there is a melt film with thickness $\delta$ separating a migrating heat source and the phase-change interface (PCI). In 1D, this looks similar to Figure 2.1, except the hot wall is also moving. For a constant hot wall temperature and constant conditions of the surroundings, the migrating heat source reaches a steady state velocity. At steady state, the hot wall and the PCI move together with the same constant velocity, and hence the melt film thickness $\delta$ is constant. In general, and for time-dependent conditions, the melt film thickness $\delta(t)$ is time dependent. While a typical melting probe is on the scale of a meter in length, $\delta$ is on the micro-meter scale for typical conditions with water-ice.

Accurately predicting a spatially variable melt-film thickness $\delta(x)$, and modeling the dynamics of the migrating heat source, requires integrating the forces acting on the melt film. This requires solving equations for the conservation of mass, momentum, and energy within the domain of the melt-film. This solution is presented by (Schüller, Kowalski, & Raback, 2016) and detailed in the original master's thesis of Schüller (Schüller, 2015).

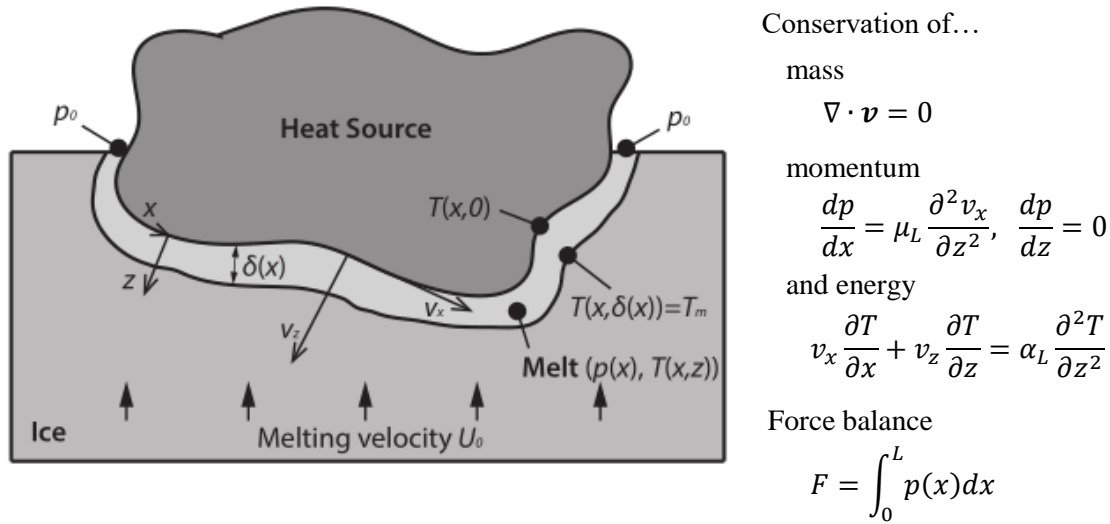

Figure 2.2: Close-contact melting with thin melt film
The diagram is taken from the original master's thesis of Schüller (Schüller, 2015)

According to (Schüller, 2015), almost all publications on close-contact melting model the problem with the conservation laws and force-balance shown on the right side of Figure 2.2. These conservation laws are a simplified form of the Navier-Stokes equations for incompressible flow. The simplifications are from the thin film approximation in the momentum and energy equations, which are well-established in lubrication theory. This assumes that $\delta$ is very thin compared to the length of the heating surface.

The close-contact melting model in (Schüller, Kowalski, & Raback, 2016) was first verified against previous rectilinear melting models, then extended to curvilinear melting, and finally validated against a rectangular plate experiment with some success. As shown in (Schüller, 2015), this theory of curvilinear melting is quasi-stationary and hence can only be applied to the quasi-stationary time intervals of the trajectories in his experiments. The theory does not apply to the unsteady acceleration of the rigid body (RB). As shown in his experiments for a rectangular plate RB, the theory also does not apply for very large changes in attitude.

In fact, all of the models discussed above generally assume that the heat source is either fixed or at a steady velocity. We, in contrast, are generally considering the unsteady movement of the heat source through an unsteady ambient. This is a new area of research, and we have little to existing work to leverage.

# 3 General mathematical formulation of the physical problem

In this chapter, we formulate an abstract mathematical model for integrating the time trajectory of a rigid body (RB) heat source through a phase-change material (PCM) domain. We will refer to the PCM domain also as the ambient domain, since in general the physical processes include not only phase-change processes, but also mass, momentum, and heat transport, as shown in Figure 1.3. We separate the rigid body dynamics (RBD) and the ambient dynamics by splitting them into two operators, which we couple via the shared RB state $\boldsymbol{\xi}$ and ambient state $\boldsymbol{u}$.

The notation both for the RBD and for the ambient dynamics are shown in Figure 3.1. The notation is consistently chosen for the 2D (i.e. $d = 2$) case and the realistic 3D (i.e. $d = 3$) case. The left half of Figure 3.1 focuses on the rigid body (RB). The RB surface is equivalent to the heating surface $\Gamma_h$. The state of the moving RB is named $\boldsymbol{\xi}$. The state is decomposed into the orientation $\boldsymbol{q}$ and the position of the translated body-fixed origin $\boldsymbol{r}$. The orientation is represented by a scalar rotation angle in 2D. In 3D, it can either be represented by Euler angles or a quaternion. Given that the surface is a rigid body, we neglect any structural deformations. Therefore, the surface $\Gamma_h$ is dependent only on the reference surface $\Gamma_{h,0}$, the orientation $\boldsymbol{q}$, and the translation $\boldsymbol{r}$. The phase-change interface $\Gamma_m$ is shown to highlight the interaction with the ambient. In chapter 4 we will show how $\Gamma_h$ and $\Gamma_m$ interact in the form of feasibility constraints on the RB position.

The right half of Figure 3.1 focuses on the ambient. The entire ambient domain is named $\boldsymbol{\Omega}$. For the purposes of discussion, we further decompose this into solid and liquid subsets of the spatial domain, respectively $\boldsymbol{\Omega}_S$ and $\boldsymbol{\Omega}_L$, such that $\boldsymbol{\Omega} = \boldsymbol{\Omega}_S \cup \boldsymbol{\Omega}_L$. The spatial coordinates of points within a domain are named $\boldsymbol{x} \in \boldsymbol{\Omega} \subset \mathbb{R}^d$. The state of the ambient is named $\boldsymbol{u}(t, \boldsymbol{x})$. The boundaries are separated into the environment boundary $\Gamma_e$, the heating surface $\Gamma_h$, the phase-change interface (PCI) $\Gamma_m$, and the contact surface $\Gamma_c$. The contact surface is defined such that the distance between the probe surface and the melt interface is within the micro-scale melt film thickness, i.e. $\Gamma_c = \left\{ \boldsymbol{x} : \boldsymbol{x} \in \Gamma_h \land \min_{\boldsymbol{x}_m \in \Gamma_m} \|\boldsymbol{x} - \boldsymbol{x}_m\| \leq \delta_r \right\}$, where $\delta_r$ is a characteristic thickness on the order of the melt film thickness. This distinction is important when one wishes to consider models for close-contact melting, as discussed in section 2.2.

The rigid body and ambient states $\boldsymbol{\xi}(t)$ and $\boldsymbol{u}(t)$ are both time-dependent. This implies that the surfaces $\Gamma_h(t)$ and $\Gamma_m(t)$ are also time dependent. The interaction of these two time dependent surfaces is the main subject of this thesis. In chapter 8 we will show how $\Gamma_h(t)$ and $\Gamma_m(t)$ interact in the form of feasibility constraints on the rigid body dynamics (RBD).

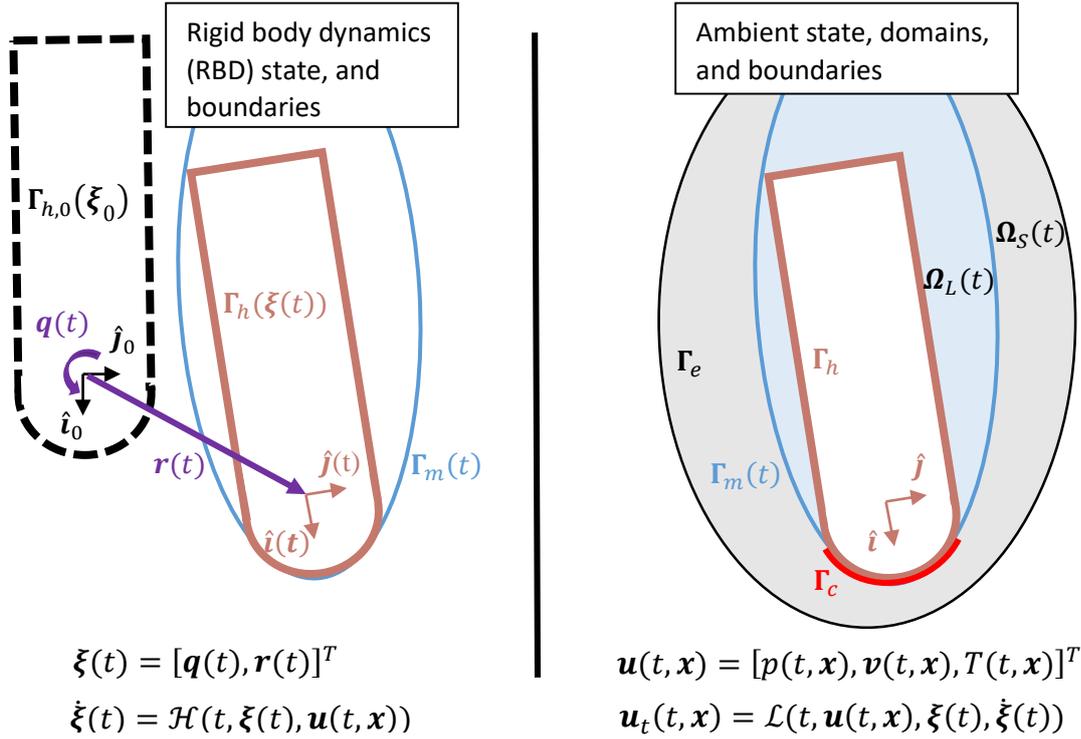

Figure 3.1: Domains, boundaries, and state variables of the split operator problem
Left: RBD diagram. Right: Ambient diagram Domains that are volumes in 3D and areas in 2D are named $\boldsymbol{\Omega}$, while boundaries that are surfaces in 3D and curves in 2D are named $\boldsymbol{\Gamma}$. The translation vector is named $\boldsymbol{r} \in \mathbb{R}^d$, while The zero subscript $[\ ]_0$ denotes the initial state. The body-fixed axes are named $\hat{\boldsymbol{\imath}}, \hat{\boldsymbol{\jmath}}$, and $\hat{\boldsymbol{k}}$.

This two-way coupling is generally not easy to solve. To simplify, we decompose the problem into the RBD which govern the time evolution of $\boldsymbol{\xi}(t)$ and hence $\boldsymbol{\Gamma}_h(t)$, and the ambient dynamics which govern the time evolution of $\boldsymbol{u}(t)$ and hence $\boldsymbol{\Gamma}_m(t)$. We abstractly refer to these processes with the operators $\mathcal{H}$ and $\mathcal{L}$, respectively. Given that the driving forces for the RBD come from the ambient field $\boldsymbol{u}$, in general we write

$$\dot{\boldsymbol{\xi}}(t) = \mathcal{H}(t, \boldsymbol{\xi}(t), \boldsymbol{u}(t, \boldsymbol{x})) \qquad (3.1)$$

where $\dot{\boldsymbol{\xi}}$ is the time derivative of the RB state. Similarly, both $\boldsymbol{\xi}$ and $\dot{\boldsymbol{\xi}}$ influence the ambient dynamics $\boldsymbol{u}$, and so in general we write

$$\boldsymbol{u}_t(t, \boldsymbol{x}) = \mathcal{L}(t, \boldsymbol{u}(t, \boldsymbol{x}), \boldsymbol{\xi}(t), \dot{\boldsymbol{\xi}}(t)) \qquad (3.2)$$

where $\boldsymbol{u}_t$ is the partial time-derivative of the ambient state. In chapters 4 and 5, we specify $\mathcal{H}$ and $\mathcal{L}$ concretely, depending on which aspects of the problems we wish to model. Typically, $\mathcal{L}$ will be a PDE (partial differential equation). Methods for coupling the PDE problem to the RBD can be viewed in two categories:

**Invasive methods:** "Invasive" refers to the task of having to substantially modify the PDE solver. This is generally avoided, since it is expensive to develop PDE solvers. For many models we might choose for $\mathcal{L}$, PDE solvers may already exist which do not consider the moving RB, i.e. $\boldsymbol{\xi}$ must be considered as a constant to apply the existing method. Modifying the solver to model $\boldsymbol{\xi}(t)$ would be invasive. That being said, the nature of the moving heat source problem may warrant a unique invasive approach.

**Non-invasive methods:** Alternatively, "non-invasive" refers to leaving an existing PDE solver mostly untouched, and handling the coupling problem with some external driver. View this as a split operator approach, meaning that $\mathcal{H}$ and $\mathcal{L}$ must be separately solved in series, and coupled with some auxiliary conditions. This generally leads to low-order temporal accuracy, and therefore larger problem sizes when time accuracy is needed. Often even more critical, transferring data between $\mathcal{H}$ and $\mathcal{L}$ will always require expensive interpolation and extrapolation operations.

In this work, we employ a non-invasive split operator approach. Thus we will first develop models for the operators $\mathcal{H}$ and $\mathcal{L}$ separately in chapters 4 and 5, and then we loosely couple them with a driver in chapter 8. From this general setting, we can study the ambient independently from the probe's velocity by considering $\dot{\boldsymbol{\xi}} = 0$, and we can study equilibrium melting velocities by considering $\dot{\boldsymbol{\xi}} = \boldsymbol{v}$. In the latter case, as we will see in chapter 7, it is convenient to change our reference frame when solving $\mathcal{L}$. We will formulate $\mathcal{L}$ such that the velocity of the RB is interpreted as convection of the ambient field $\boldsymbol{u}$.

## 3.1  $\mathcal{H}$: Rigid-body dynamics

In 2D, $\boldsymbol{q}$ is a scalar rotation angle. In 3D, $\boldsymbol{q}$ can either be the three Euler angles or an attitude quaternion. The Euler angles admit a geometric singularity, but the quaternion representation admits multiple vectors that represent the exact same geometric state (i.e. the quaternion does not uniquely represent the state). Three parts of the four-part quaternion specify the 3D rotation axis, while the remaining fourth part specifies the rotation angle about that axis. In 2D the rotation axis is fixed, which makes it no different than if one were using the single 2D Euler angle.

In this thesis, only 2D test cases are presented, so we will skip the detailed discussion; but, since the end goal is to use quaternions, and since the quaternion has a simple interpretation in 2D, we will use the attitude quaternion to write down the general 3D formulation in this section.

Therefore, we refer to the position and attitude of the RB at time *t* as the state vector and its time derivative as

$$\boldsymbol{\xi}(t) := \begin{bmatrix} \boldsymbol{q}(t) \\ \boldsymbol{r}(t) \end{bmatrix} \tag{3.3}$$

$$\dot{\boldsymbol{\xi}}(t) = \begin{bmatrix} \dot{\boldsymbol{q}}(t) \\ \dot{\boldsymbol{r}}(t) \end{bmatrix} \tag{3.4}$$

The time-dependent position of each rigid body (RB) point $\boldsymbol{x}$ is the conjugation of the point by $\boldsymbol{q}$, then shifted by $\boldsymbol{r}$

$$\boldsymbol{x}(t) = \boldsymbol{q}(t)\boldsymbol{x}_0\boldsymbol{q}(t)^{-1} + \boldsymbol{r}(t) \tag{3.5}$$

where the quaternion

$$\boldsymbol{q} = \cos\left(\frac{\theta}{2}\right) + \sin\left(\frac{\theta}{2}\right)\left(a_x\hat{\boldsymbol{\imath}} + a_y\hat{\boldsymbol{\jmath}} + a_z\hat{\boldsymbol{k}}\right) \tag{3.6}$$

represents a rotation of angle $\theta$ about an axis $\hat{\boldsymbol{a}} = a_x\hat{\boldsymbol{\imath}} + a_y\hat{\boldsymbol{\jmath}} + a_z\hat{\boldsymbol{k}}$. $\theta$ is measured clockwise from a line of sight in the direction of the axis of rotation.

Generally rigid body dynamics problems in the literature are solved by integrating the equations of motion with force and moment balances, for example in the work of (Schüller, Kowalski, & Raback, 2016). This can be prohibitively expensive when solving large inverse problems, as we discussed in section 1.3. In chapter 4 of this work, we develop an alternative which does not require resolving the forces and moments, which has potential to greatly reduce the computational complexity of the problem.

## 3.2 $\mathcal{L}$: Ambient dynamics

For the ambient state within the domain $\boldsymbol{\Omega} = \boldsymbol{\Omega}_L \cup \boldsymbol{\Omega}_S$, we consider a continuum field of a phase-change material (PCM) whose state is defined by scalar temperature, vector velocity, and scalar pressure

$$\boldsymbol{u}(t, \boldsymbol{x}) := \begin{bmatrix} T(t, \boldsymbol{x}) \\ \boldsymbol{v}(t, \boldsymbol{x}) \\ p(t, \boldsymbol{x}) \end{bmatrix} \tag{3.7}$$

The phase state will depend on the temperature and pressure per Figure 1.3. Different physical processes dominate the different domains shown in Figure 3.1. In principle, one must model heat conduction in all $\boldsymbol{\Omega}$, but convection only in $\boldsymbol{\Omega}_L$. At $\boldsymbol{\Gamma}_c$, the phase-change will either be solid-liquid (melting) or solid-gas (sublimation) depending on the pressure. A liquid-solid (refreezing) phase-change occurs at $\boldsymbol{\Gamma}_m$.

The time evolution of this field is fundamentally governed by the conservation laws of mass, momentum, and energy. These laws can generally be formulated as transport equations and written as PDE's (partial differential equations). In section 2.1.2, we briefly described the enthalpy-porosity model (2.4)-(2.6), which is a concrete example of $\mathcal{L}$ that accounts for all three conservation laws, and the phase-change.

In this thesis, beginning in chapter 5, we only consider the energy balance including heat transfer by conduction and convection. We will neglect pressure and prescribe the velocity field, therefore omitting the mass and momentum balances. Furthermore, we will not include the latent heat of melting $h_m$ in the energy balance. We instead simply assume that the material is solid when below the freezing temperature. While such a model will not be useful for quantitative predictions, it is qualitatively interesting and will serve as a suitable way to test the coupling of the rigid body dynamics $\mathcal{H}$ to some ambient field evolving according to $\mathcal{L}$. A critical next step for this research will be to realistically model the contribution of $h_m$ within the domain $\mathbf{\Omega}$.

# 4  $\mathcal{H}$: Equilibrium motion modeled as energy minimization

Solving the equations of motions is prohibitively expensive, particularly because the total force and moment which drive the rigid body dynamics (RBD) must be integrated from the ambient solution, which itself depends on the RBD (Schüller, Kowalski, & Raback, 2016). The high computational complexity of solving that coupled problem motivates the search for a new paradigm. We begin the search with this chapter. To inspire a simple model, consider the early melting probes which were only propelled by gravity. The movement is constrained entirely by the evolving solid wall of ice. Figure 4.1 shows a simple example.

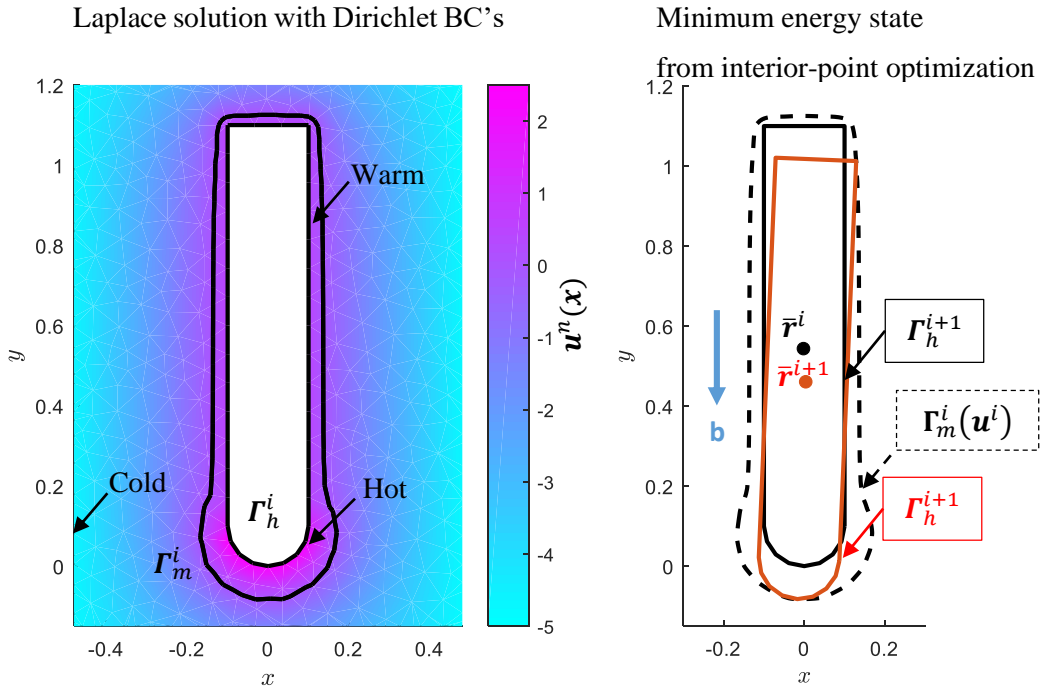

Figure 4.1: Example result for the minimum energy rigid body dynamics (RBD) model. On the left of Figure 4.1, the hull of the rigid body (RB) at some discrete time step $i$ is $\boldsymbol{\Gamma}_h^i$. The phase-change interface (PCI) is $\boldsymbol{\Gamma}_m^i$, and is equivalent to the temperature contour $T = T_m$.

For this particular visualization, the ambient is at steady-state. In fact, this is the solution to the Laplace problem using MATLAB's PDE toolbox (described in more detail in section 4.3.3). Beginning in chapter 5, we will instead solve the convection-diffusion equation, which includes the heat equation as a special case, to model the unsteady ambient.

On the right of Figure 4.1, $\boldsymbol{\Gamma}_h^{i+1}$ is the hull corresponding to the minimum energy RB state $\boldsymbol{\xi}^{i+1}$ when the ambient state $\boldsymbol{u}^i$ (and hence the PCI $\boldsymbol{\Gamma}_m^i$) is held constant. The centroid of the RB is $\bar{\boldsymbol{r}}$. The state $\boldsymbol{\xi}^{i+1}$ minimizes the RB's gravitational potential within the constant gravitational field $\boldsymbol{b}$, subject to constraints such that the RB cannot penetrate the solid PCM. As physical intuition would suggest, the body simply falls down into contact with the solid ice, and tips to one side. This behavior is consistent with the laws of motion if the probe takes infinitesimal steps through equilibrium states, introducing a modeling error $\mathcal{O}(\Delta t)$ since the deviation from equilibrium increased with $\Delta t$. With Figure 4.1 in mind, we can more easily explain the concrete formulation.

## 4.1 The general minimization form

To reduce our general RBD form (3.1) to a minimization problem, we must discretize the time derivative. In this work, we choose a first-order forward Euler approximation for integrating in time, i.e.

$$\frac{\xi^{i+1} - \xi^i}{\Delta t} = \mathcal{H}(t, \xi, u^i) + \mathcal{O}(\Delta t) \tag{4.1}$$

This allows us to model a time step in the RBD as

$$\boldsymbol{\xi}^{i+1} = \underset{\xi}{\text{argmin}} \int_{\Gamma_h(\xi)} f(\boldsymbol{u}^i) \, d\Gamma_h \tag{4.2}$$

such that

$$g_L \leq \int_{\Gamma_h(\xi)} g(\boldsymbol{u}^i) \, d\Gamma_h \leq g_U \tag{4.3}$$

where $f$ is some energy functional and $g$ is an inequality constraint functional. The subscripts $L$ and $U$ respectively denote lower and upper bounds. We could also discuss equality constraints; but they are not needed for the work in this thesis, and so we will not include the extra equation.

## 4.2 Minimize gravitational potential subject to feasibility constraints

In this work, we consider a model with an energy function that is independent of the ambient state, but with inequality constraints that are dependent on the ambient state. Specifically, the constraints depend on the phase state of the ambient. We constrain the rigid body (RB) such that it cannot penetrate the solid PCM.

### 4.2.1 Feasibility constraints $g$: No solid penetration

First consider the inequality constraint $g$. In general, the phase state of the ambient will depend on both the pressure and the temperature. In this thesis, as we will discuss in more detail in chapter 5, we are only modeling the temperature variable $T(\boldsymbol{x}, t)$ in the ambient dynamics. This means that the phase-state of the ambient $u^i(\boldsymbol{x})$ at a discrete time is entirely defined by the temperature field $T^i(\boldsymbol{x})$ at that time.

One could weakly impose the no penetration constraint by integrating (4.3) with an appropriate $g$. In this work, we instead impose a discrete set of strong inequality constraints $\boldsymbol{g}$. We write the general discrete constraints as

$$g_L \leq \boldsymbol{g}(\boldsymbol{\xi}, T^i) \leq g_U \tag{4.4}$$

Hence we evaluate the discrete feasibility constraints by sampling the temperature field $T^i(\boldsymbol{x})$ at discrete points[2] $\boldsymbol{x}$ on the continuous surface $\boldsymbol{\Gamma}_h$. We name this discrete set $\boldsymbol{\gamma}$, where $\boldsymbol{\gamma} \subset \boldsymbol{\Gamma}_h$.

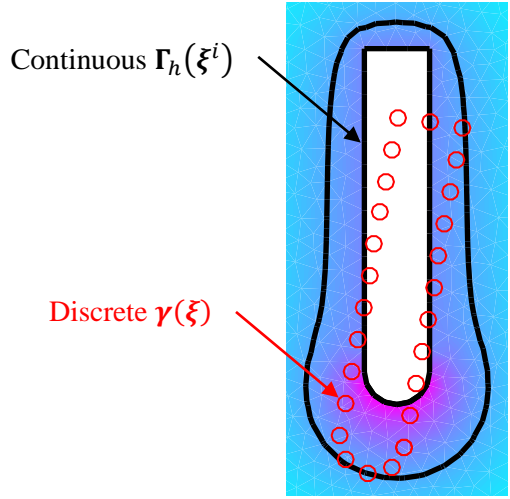

Figure 4.2: Sampling discrete feasibility constraints
$\boldsymbol{\Gamma}_h(\boldsymbol{\xi}^i)$ is the heating surface at initial state $\boldsymbol{\xi}^i$. $\boldsymbol{\gamma}(\boldsymbol{\xi})$ is a discrete subset of the candidate heating surface $\boldsymbol{\Gamma}_h(\boldsymbol{\xi})$. To obtain a discrete set of feasibility constraints, the temperature field is sampled onto the points $\boldsymbol{x} \in \boldsymbol{\gamma} \subset \boldsymbol{\Gamma}_h(\boldsymbol{\xi})$.

We choose the lower bound to be the melting temperature and no upper bound. Altogether this gives us the discrete feasibility constraints

$$\boldsymbol{g}(\boldsymbol{\xi}, T^i) = T^i(\boldsymbol{x}) \ \forall \boldsymbol{x} \in \boldsymbol{\gamma} \subset \boldsymbol{\Gamma}_h(\boldsymbol{\xi}), \qquad g_L = T_m, \qquad g_U = \infty \tag{4.5}$$

### 4.2.2 Energy function $f$: gravitational potential

Now consider the energy functional $f$. We wish to minimize gravitational potential $\Psi$ given a gravitational field $\boldsymbol{b}$. We can assume $\boldsymbol{b}$ is uniform in space at the scale of a melting probe. Given a rigid body, the energy functional only depends on the position of the body's center of mass $\bar{\boldsymbol{r}}$.

$$\int_{\boldsymbol{\Gamma}_h(\boldsymbol{\xi})} f \, d\boldsymbol{\Gamma}_h = \Psi(\boldsymbol{b}, \boldsymbol{\Gamma}_h(\boldsymbol{\xi})) = -\boldsymbol{b} \cdot \bar{\boldsymbol{r}}(\boldsymbol{\xi}) \tag{4.6}$$

Assuming a homogeneous material density, the center of mass is equivalent to the geometric centroid, which we obtain with the volume integral

$$\bar{\boldsymbol{r}} = \frac{\int_{\boldsymbol{\Omega}_h} \boldsymbol{r} \, d\boldsymbol{\Omega}_h}{\int_{\boldsymbol{\Omega}_h} d\boldsymbol{\Omega}_h} \tag{4.7}$$

---

[2] For the purposes of this thesis, this discrete set is chosen somewhat arbitrarily. For more complicated topologies of $\boldsymbol{\Gamma}_h$ and $\boldsymbol{\Gamma}_m$ a more rigorous method may be needed to maintain a reasonable problem size.

where $\mathbf{\Omega}_h$ is the RB domain contained by $\mathbf{\Gamma}_h$. Via Stokes's Theorem, $\oint_\Gamma \omega = \int_\Omega d\omega$, we could generally obtain the centroid of the manifold $\Omega \in \mathbb{R}^d$ by integrating over its oriented surface manifold $\Gamma$ in $\mathbb{R}^{d-1}$. For now, consider only a 2D example. Parameterizing the surface as a 1D manifold in 2D space, and applying Green's theorem, yields

$$A = \int_0^1 r_0(t)\left(\frac{d}{dt}r_1(t)\right) dt, \quad \bar{r} = \frac{1}{2A} \begin{Bmatrix} \int_0^1 r_1^2 \left(\frac{d}{dt}r_0(t)\right) dt \\ \int_0^1 r_0^2 \left(\frac{d}{dt}r_1(t)\right) dt \end{Bmatrix} \quad (4.8)$$

Given that the surface is a rigid body, we only need to integrate this function once to obtain a reference centroid $\bar{r}_0$, and the time-dependent center of mass $\bar{r}(t)$ is defined by the state $\boldsymbol{\xi}(t)$. The solution to the minimization problem, shown in Figure 4.1, matches our physical intuition so long as the probe's density is heavier than the surrounding melt. Note how the body tips to one side, just as a physical rigid body should behave.

In 2D space, the state is $\boldsymbol{\xi} = [\theta, r_0, r_1]$. This already makes visualizing $f: \mathbb{R}^3 \to \mathbb{R}$ difficult. $\mathbb{R}^1$ slices are shown in Figure 4.3. Only $r_0 = 0$ is shown, since in this specific case, $r_0$ does not influence $f$.

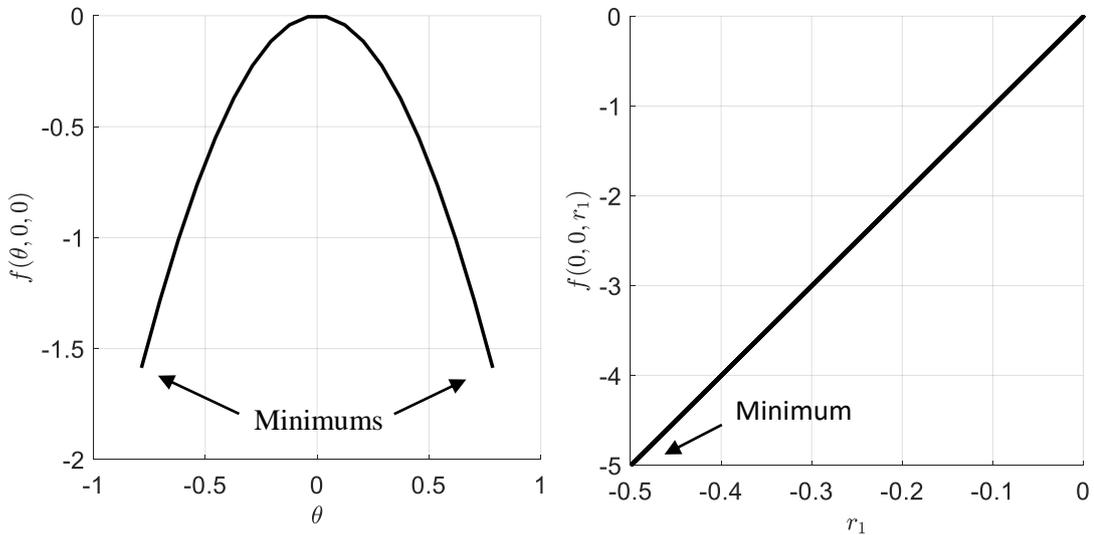

Figure 4.3: $\mathbb{R}^1$ slices of the energy function $f: \mathbb{R}^3 \to \mathbb{R}$
Left: Sliced at $(r_0, r_1) = (0,0)$. Right: Sliced at $(\theta, r_0) = (0,0)$. Here $f$ is the minimum gravitational potential energy function (4.6). This visualization ignores feasibility constraints, and hence the feasible minimums of the constrained problem are different than shown.

Depending on the symmetry of $\boldsymbol{u}$, the constrained minimum may not be unique. For the case of Figure 4.3, there are two minimums in the $(\theta, r_0, r_1)$ design space. As an edge case, with a specific gravity vector and symmetry of $\boldsymbol{u}$, 3D physical space may admit an infinite set of minimums. Mathematically this non-uniqueness may pose a problem. In reality, there are stochastic aspects of the system which cause a single outcome.

Furthermore, Figure 4.3 shows an obvious saddle point. In reality, small perturbations cause the physical system to move away from such a saddle point. In this thesis, the existence of this point has not been an issue; but if it were to become an issue, one could perturb $\boldsymbol{\xi}$ or $\boldsymbol{u}$, perhaps stochastically.

Altogether we write the time-discrete rigid body dynamics, driven by minimizing gravitational potential subject to no solid penetration constraints, as

$$\dot{\boldsymbol{\xi}}^{i+1} = \frac{\boldsymbol{\xi}^{i+1} - \boldsymbol{\xi}^i}{t_{i+1} - t_i} \tag{4.9}$$

$$\boldsymbol{\xi}^{i+1} = \underset{\boldsymbol{\xi}}{\operatorname{argmin}} -\boldsymbol{b} \cdot \bar{\boldsymbol{r}}(\boldsymbol{\xi}) \tag{4.10}$$

$$T_m \leq T^i(\boldsymbol{x}) < \infty \quad \forall \boldsymbol{x} \in \boldsymbol{\gamma} \subset \boldsymbol{\Gamma}_h(\boldsymbol{\xi}) \tag{4.11}$$

$$\boldsymbol{\xi}_L \leq \boldsymbol{\xi} \leq \boldsymbol{\xi}_U \tag{4.12}$$

with the initial guess $\boldsymbol{\xi} = \boldsymbol{\xi}^i$.

## 4.3 $\mathcal{H}$ as a constrained nonlinear program

In the remainder of this chapter, we will rewrite (4.10)-(4.12) in the standard form of a constrained nonlinear program, and then implement this program in Python. The optimization algorithms we wish to employ generally find local solutions to problems of the form

$$\boldsymbol{x}^* = \underset{\boldsymbol{x}}{\operatorname{argmin}} f(\boldsymbol{x}), \quad \boldsymbol{x} \in \mathbb{R}^n, \quad f \colon \mathbb{R}^n \to \mathbb{R} \tag{4.13}$$

subject to the constraints

$$\boldsymbol{g}_L \leq \boldsymbol{g}(\boldsymbol{x}) \leq \boldsymbol{g}_U, \quad \boldsymbol{g} \colon \mathbb{R}^n \to \mathbb{R}^m \tag{4.14}$$

and bounds on the design space

$$\boldsymbol{x}_L \leq \boldsymbol{x} \leq \boldsymbol{x}_U \tag{4.15}$$

where $\boldsymbol{x}$ are named the design variables, $f$ is the objective function, and $\boldsymbol{g}$ are the inequality constraints. The constraints restrict a feasible region

$$\boldsymbol{\mathcal{F}} = \{\boldsymbol{x} | \boldsymbol{g}_L \leq \boldsymbol{g}(\boldsymbol{x}) \leq \boldsymbol{g}_U\} \tag{4.16}$$

For this work, the design variables will be the members of the subset of $\boldsymbol{\xi}$ which are degrees of freedom of the rigid body (RB) trajectory. For example, in 2D, the rotation axis is fixed, so there is only a single rotational degree of freedom (the angle). In this case, the RB state has a total of three degrees of freedom, i.e.

$$x_{2D} = \xi_{2D} = \begin{bmatrix} \theta \\ r_0 \\ r_1 \end{bmatrix} \in \mathbb{R}^3 \qquad (4.17)$$

The full 3D problem is easiest to understand in the context of the quaternion. In 3D, the rotation axis itself has three degrees of freedom, so the state has a total of seven, i.e.

$$x_{3D} = \xi = \begin{bmatrix} q \\ r \end{bmatrix} \in \mathbb{R}^7 \qquad (4.18)$$

This is a good opportunity to review the right side of Figure 4.1. In this case, the constraints are dependent on all three design variables from (4.17), while the objective function, which was chosen as the gravitational potential, is only affected by $\theta$ and $r_1$.

Note that in the grand scope of optimization problems, seven is a very small number of design variables. The complexity of our problem shows up in the constraints. The sampled temperature field for evaluating the constraints $g$ (4.5) will generally come from the solution to a partial-differential equation (PDE), and therefore $g$ which will be expensive to evaluate.

### 4.3.1 Local vs. global optimization, and gradient-based vs. gradient-free

As described by (Biegler, 2010), the minimum $x^*$ is global if $f(x^*) \leq f(x) \ \forall x \in \mathcal{F}$, and local if $f(x^*) \leq f(x) \ \forall x \in \mathcal{N}(x^*) \cap \mathcal{F}$, where $\mathcal{N}(x^*) = \|x - x^*\| < \epsilon, \ \epsilon > 0$. It is simple to understand this practically from the perspective of a gradient-based optimizer starting from a single point in the design space. The steepest descent direction depends entirely on the gradient (or rather the Jacobian for multi-dimensional $x$) of the objective function at this point. If any minimum exists, then eventually one descends to a point where the gradient is positive in all directions. Now travelling in any direction would increase the function value, and so a local minimum has been found. Unfortunately, from this perspective, there is no way to know if a lower point exists somewhere in the global domain.

Generally, gradient-based optimization algorithms can only guarantee finding local minimums Gradient-free algorithms (e.g. genetic algorithms or particle swarm optimization) can sample the global space much more thoroughly, but they do not guarantee strict optimality in any way, and their convergence rates are slow. Lucky for us, the problem in this thesis is well suited for gradient-based methods that find local minima. This is particularly true in this work because of the quasi-equilibrium motion assumption that is fundamental to the framework.

When using gradient-based methods, we can benefit when the temperature field is smooth enough to admit continuous first derivatives if we wish to use its exact Jacobian, and second derivatives if we wish to use its exact Hessian. First order continuity is also necessary for numerically solving the partial-differential equation (PDE) which models the temperature field with the finite element method (FEM) as discussed in chapter 5, so we can reasonably assume that the exact Jacobian will be available, but not necessarily the exact Hessian. Indeed, the gradients of the temperature field are readily available from MATLAB's `PDETool` when we use this tool for the example in section 4.3.3. In practice it is somewhat rare that exact Hessians are needed by users of nonlinear program solvers, so these solvers are usually equipped to numerically approximate the Hessian.

Here we briefly discuss the two methods employed in this thesis, SLSQP (Sequential Least-Squares Quadratic Programming) and interior point (a.k.a. barrier) optimization. Sequential quadratic programming (SQP) and interior point methods are presented by (Biegler, 2010); but literature on "SLSQP" is more elusive.

SQP is a popular algorithm, primarily due to its fast convergence properties which are inherited from Newton methods. SQP solves the problem as a series of Newton iterations, with each iteration solving a quadratic program, i.e. a quadratic model of the objective function, subject to a linearization of the constraints. According to the SciPy documentation (The Scipy community, 2016), SLSQP is an extension of SQP based on the Fortran subroutine originally implemented by Dieter Kraft (Kraft, 1988).

Interior point (a.k.a. barrier) optimization relaxes the constraints and solves the prime-dual equations, which amounts to solving a sequence of relaxed problem, also with a Newton-based method. This algorithm is better suited for optimization problems with a large number of constraints. In this thesis, the topologies of the heating surface $\mathbf{\Gamma}_h$ and the phase-change interface (PCI) $\mathbf{\Gamma}_m$ have been well behaved, and we have not yet extended the test cases to 3D, so there have not been an excessive number of constraints. That being said, full 3D simulations with more realistic melting models and more complicated probe geometries may necessitate the use of interior point optimization. MATLAB's `fmincon` defaults to an interior point method. In this work, we also successfully employed ipopt (Wächter & Biegler, 2006) which is openly available from the COIN-OR Initiative (COIN-OR, 2016).

### 4.3.2 Python implementation

With the goals of using free software and a flexible language that requires few lines of code, the rigid body dynamics operator $\mathcal{H}$ was implemented in Python. We will also use this Python program to drive the entire coupled problem, calling the implementation of $\mathcal{L}$ as a sub-process; but we will save that discussion for chapter 8. The SLSQP minimization method from SciPy (The Scipy community, 2016) is used, and only with approximate (finite differenced) gradients.

The main components of the code, and their dependencies, are shown in Figure 4.4. The code is object-oriented. The user writes a Python script that imports the `trajectory` module, instantiates the `Trajectory` class, modifies parameters that are members of the `Trajectory` instance, and then either calls `Trajectory.run_steps` for some number of steps, or manually iterates with `Trajectory.run_step`. The entire code is shared publicly at (Zimmerman, dimice-python-cpp, 2016).

In chapter 8 we will discuss in detail how the code from Figure 4.4 iterates the coupled problem through time. In the current chapter we focus only on the preparation and solution of (4.9)-(4.12) for a single discrete time step. This procedure is contained in the method `Trajectory.run_step`, the most relevant lines of which are shown in Code 4.1. To better understand the behavior of the constrained optimization approach, two examples of searches for feasible and optimal states are shown in Figure 4.5.

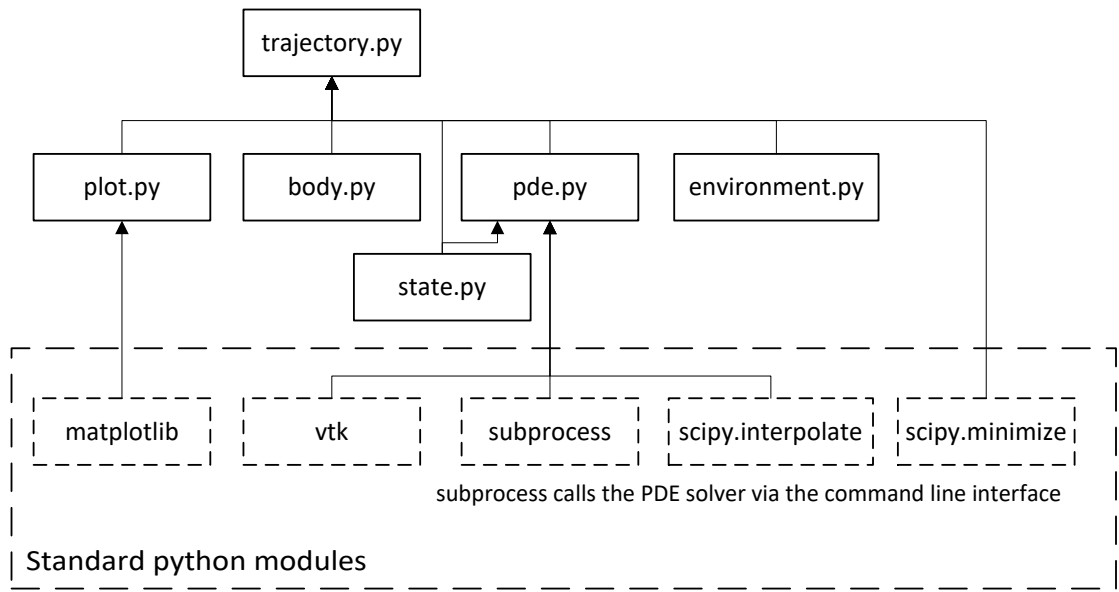

Figure 4.4: Python implementation overview
The source code is shared on GitHub (Zimmerman, dimice-python-cpp, 2016)

As shown, the PDE solver is called for the current state of the rigid body. Then `scipy.interpolate` is used to construct an interpolant from the PDE solution, so that this interpolant can be sampled for the constraints $g$. The objective function $f$ is defined in this version simply as the vertical coordinate of the center of gravity, which of course assumes there is a uniform gravitational field pointing straight downward, as in Figure 4.1. This calls the method `body.get_center_of_gravity` from an instance of the `Body` class. Next the constraint function $g$ is defined, which also calls a method of the `Body` class so that the temperature interpolant can be sampled at points on the body's hull in the candidate state.

```
data = pde.solve(state)

T = scipy.interpolate.LinearNDInterpolator(data.x0, data.x1, data.x2,
        data.u, fill_value=environment.temperature)

def f(x):

    gravity_aligned_axis = 1
    return body.get_center_of_gravity(x)[gravity_aligned_axis]

def g(x):

    return T(body.get_hull_points(x))

out = scipy.minimize(fun=f, x0=state, constraints={'type': 'ineq', 'fun': g})

state = out.x
```

Code 4.1 : Abridged Python script for solving $\mathcal{H}$

These are the relevant lines from `Trajectory.run_step` in (Zimmerman, dimice-python-cpp, 2016).

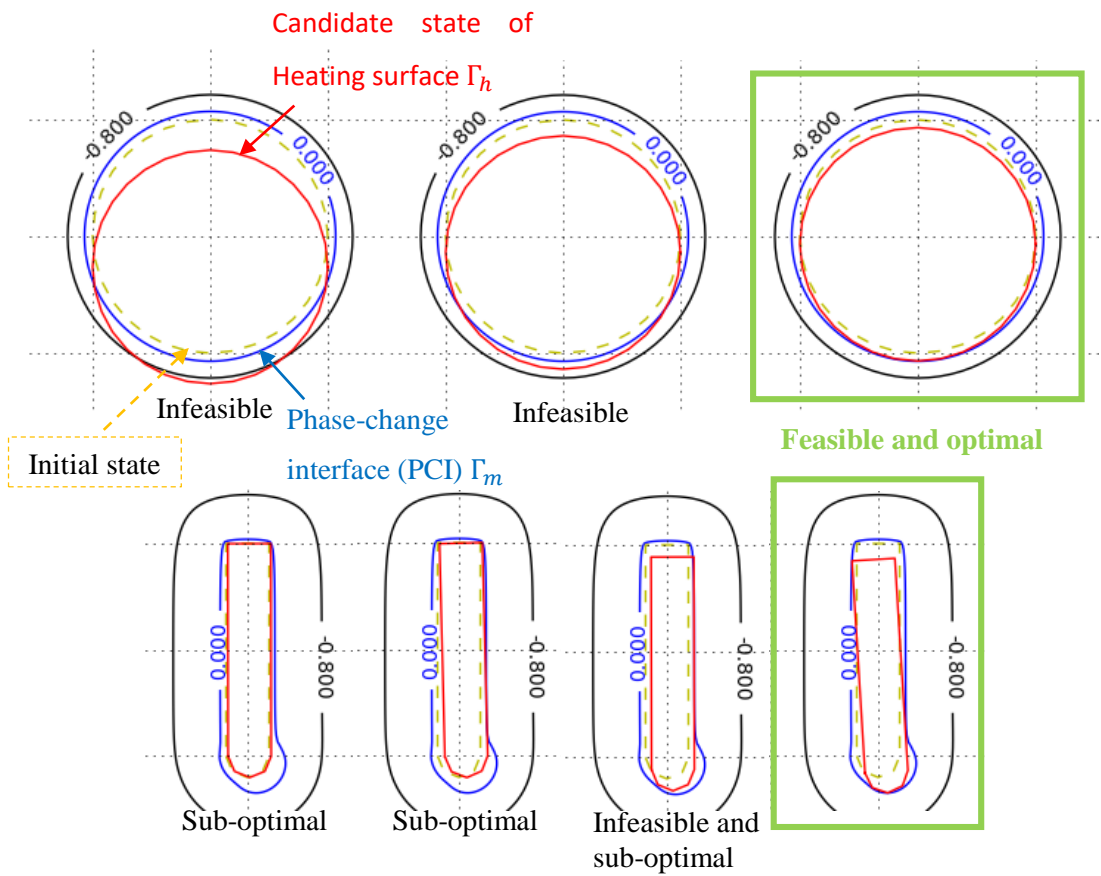

Figure 4.5 : Visualization of the minimizer searching for the minimum feasible state
Top: Circular geometry. Bottom: 2D sphere-cylinder (i.e. circle-rectangle) geometry.
To obtain these, the candidate state was plotted whenever the objective function was called. The candidate state is feasible when the heating surface does not intersect the PCI. It is optimal when there is no other feasible state with a lower vertical position.

### 4.3.3 Tracking a moving phase-change interface

To demonstrate the capability of the rigid body dynamics (RBD) to track a moving phase-change interface (PCI), a prototype was written in MATLAB. To test the capability of the minimal gravitational potential model, only some qualitatively useful temperature field was needed. This was easily obtained with MATLAB's built-in `PDETool`.

The flow of the MATLAB prototype is shown in Figure 4.6. With `PDETool` it was simple to sketch the 2D sphere-cylinder geometry in Figure 4.1, set Dirichlet boundary conditions on different parts of the rigid body (RB) surface and on the outside edges $\mathbf{\Gamma}_e$, and to solve the Laplace equation, producing a smooth temperature field. MATLAB's built-in `ScatteredInterpolant` was then used to create an interpolant instance that could easily be sample for the feasibility constraints.

After solving a single step as shown in Figure 4.5, the temperature interpolant was transformed with the same operations as the rigid body. This transformed interpolant was then used as the ambient state for the next pseudo-time step. In this way, the pseudo-trajectory could be marched through multiple steps, demonstrating the capability for the RBD model to track a moving PCI.

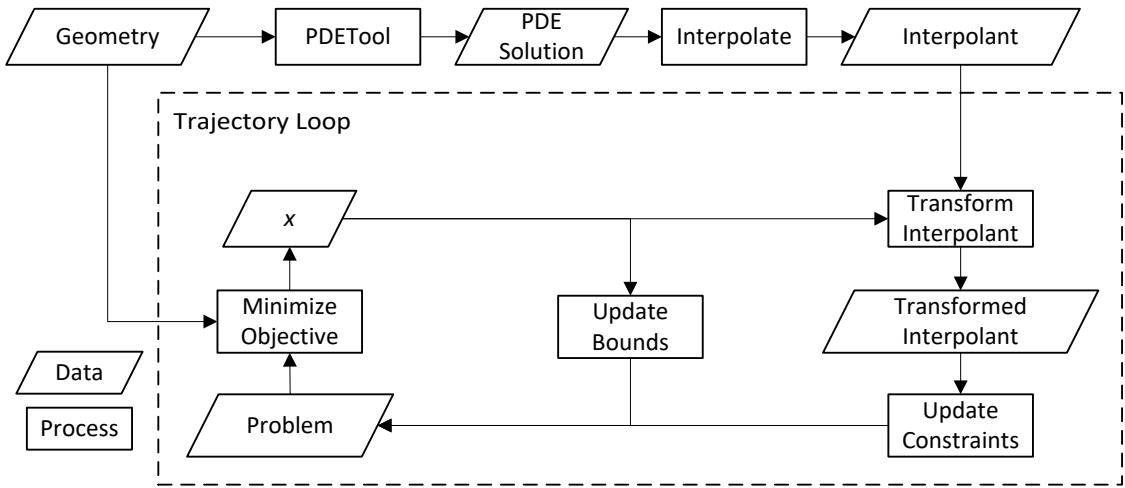

Figure 4.6: Overview of MATLAB prototype for tracking a moving phase-change interface
The Laplace equation is only solved once during initialization, and then this temperature field is transformed at each pseudo-time step to emulate a phase-change interface (PCI) evolving in time.

The entire code, including many tests and examples, is publicly shared at a GitHub repository (Zimmerman, dimice-trajectory-matlab, 2016). The repository includes regression tests that showcase a variety of features. Some of these are listed in Table 1. The code uses a configuration data structure that can be saved as a `.mat` file and edited to easily produce new test cases, and save their settings, without modifying the code. Among the many options, here we highlight that the user can choose…

- which interior point optimization method to use: MATLAB's `fmincon`, or external ipopt (COIN-OR, 2016).
- whether to use exact gradients or to approximate them with finite differences. Both PDE solver options also provide the gradients of the fields, which are used in some of these test cases to compute the exact gradients of the constraints.
- the Dirichlet boundary conditions to apply at every boundary, including separately the left and right sides of the nose for differential heating to induce a turn.
- an arbitrary force vector, rather than only allowing gravity in one direction.
- whether to integrate the centroid as the center of mass, or to arbitrarily set its coordinates.

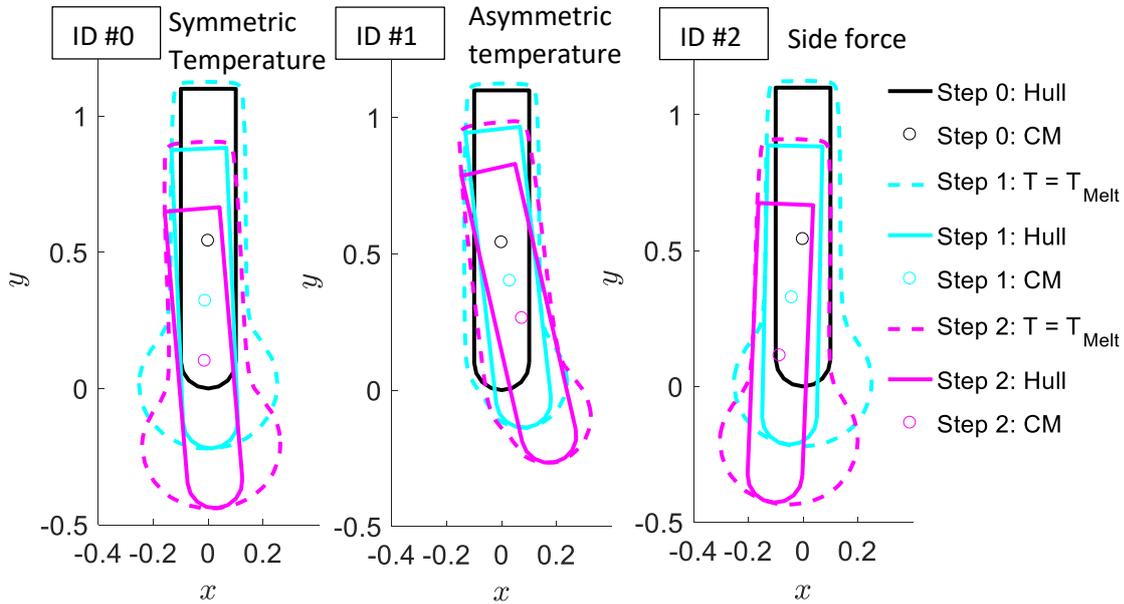

Figure 4.7: Example trajectories from the MATLAB prototype
The ID numbers at the top-left of each trajectory map to the ID column of Table 1.

| ID # | NLP | Gradients | Heating | Force Vector | Centroid |
|---|---|---|---|---|---|
| 0 | `fmincon` | Approx. | Symmetric | (-1, 0) | Integrated |
| 1 | `fmincon` | Approx. | Asymmetric | (-1, 0) | Integrated |
| 2 | `fmincon` | Approx. | Symmetric | (-0.9, -0.5) | Integrated |
| 3 | `fmincon` | Approx. | Asymmetric | (-1, 0) | Set arbitrarily |
| 4 | `fmincon` | Exact | Asymmetric | (-1, 0) | Integrated |
| 5 | `ipopt` | Exact | Symmetric | (-1, 0) | Integrated |
| 6 | `ipopt` | Exact | Asymmetric | (-1, 0) | Integrated |

Table 1: Subset of regression test matrix for MATLAB prototype of $\mathcal{H}$

# 5  $\mathcal{L}$: Unsteady convection-diffusion of a temperature field

To test the coupling of the rigid body dynamics (RBD) operator $\mathcal{H}$ and the ambient dynamics operator $\mathcal{L}$, we require some unsteady temperature field from which we may extract the phase-change interface (PCI) that drives the motion.

Accurately simulating the physical problem requires modeling the PC process; but we can obtain a qualitatively interesting unsteady temperature field by solving the heat equation. This is essentially assuming that the phase-change material (PCM) has zero latent heat of melting $h_m$. This will not predict the true physical behavior. But if we demonstrate the split operator approach with this simplified ambient model, which is easier to implement, then we can confidently proceed to coupling a more advanced model, e.g. the enthalpy-porosity model from section 2.1.2.

## 5.1  The strong form

Viewed from a fixed global reference frame, where the RB moves through the reference frame, the ambient dynamics operator (3.2) would be written

$$\frac{\partial}{\partial t}\boldsymbol{u}(\boldsymbol{x},t) = \mathcal{L}\big(t, \boldsymbol{u}(\boldsymbol{x},t), \boldsymbol{\xi}(t), \dot{\boldsymbol{\xi}} = [-\boldsymbol{v}, \boldsymbol{\omega}]^T\big)$$

Rather than viewing the rigid body (RB) as moving through the ambient field, we can instead fix our reference frame to the RB when solving $\mathcal{L}$. Hence we allow the temperature field to move relative to the RB. To this end, we keep $\boldsymbol{\xi}$ constant in time when solving the ambient dynamics $\mathcal{L}$. This is a fixed-grid approach. We interpret the RB velocity as the opposite of the convection velocity $\boldsymbol{v}$ of the temperature field. Note that if the RB is rotating (i.e. $\boldsymbol{\omega} \neq 0$), then the convection velocity $\boldsymbol{v}(\boldsymbol{x})$ must vary in space. This allows us to model the conservation of energy with the unsteady convection-diffusion problem

$$u_t(\boldsymbol{x},t) + \boldsymbol{v}(\boldsymbol{x}) \cdot \boldsymbol{\nabla} u(\boldsymbol{x},t) - \boldsymbol{\nabla} \cdot \big(\alpha(\boldsymbol{x})\boldsymbol{\nabla} u(\boldsymbol{x},t)\big) = s(\boldsymbol{x},t) \quad \forall\ \boldsymbol{x},t \in \boldsymbol{\Omega} \times (t_0, t_f) \qquad (5.1)$$

$$u(\boldsymbol{x},0) = u_0(\boldsymbol{x}) \quad \forall\ \boldsymbol{x} \in \boldsymbol{\Omega} \qquad (5.2)$$

$$u(\boldsymbol{x},t) = g(\boldsymbol{x},t) \quad \forall\ \boldsymbol{x},t \in \boldsymbol{\Gamma}_D \times (t_0, t_f) \qquad (5.3)$$

$$\alpha(\boldsymbol{x})(\hat{\boldsymbol{n}} \cdot \boldsymbol{\nabla})u(\boldsymbol{x},t) = h(\boldsymbol{x},t) \quad \forall\ \boldsymbol{x},t \in \boldsymbol{\Gamma}_N \times (t_0, t_f) \qquad (5.4)$$

where $\alpha$ is the diffusivity of the ambient material, $s$ is a source term (also known as a forcing function), $u_0$ are initial values, $g$ is a Dirichlet boundary function to be projected onto $\boldsymbol{\Gamma}_D$, and $h$ is a Neumann boundary function (representing the diffusive flux) to be projected onto $\boldsymbol{\Gamma}_N$. While we do not model any heat generation on the interior of the ambient domain $\boldsymbol{\Omega}$, we include the source term $s(\boldsymbol{x},t)$ for two reasons. First, this term is required to verify our implementation with the method of manufactured solutions (in sections 6.3.3 and 6.3.4). Second, including the source term in our implementation may allow for quick future implementation of some simple melting models, though such work will not be attempted in this thesis.

The temperature of the external environment will determine $g$, and also for a cold start will determine $u_0$. Typically, $h$ will be a control input which represents a heat flux. Controlling this boundary condition is how we wish to control the heat source's trajectory.

## 5.2 Discretization via the finite element method (FEM)

For many years and still to this day, finite volume methods have been most popular in the field of computational fluid dynamics. Alternative, the finite element method (FEM) has also existed in some form for about sixty years, when it was originally applied to the structural dynamics of flexible aircraft wings. More recently a rigorous mathematical description of FEM has been developed and is continuing to be research by many groups. FEM has proven to be an incredibly versatile method for discretizing partial-differential equations (PDE's), which has led to a surge of new ideas for tackling difficult problems. Perhaps most importantly, at least per the author's limited experience, today's quality of open source finite element code libraries, particularly for academic research, are on a higher tier than what is available for the finite volume method.

In short, industrial computational fluid dynamics (CFD) applications leverage advanced FV solvers that benefit from many decades of work, while the FEM CFD community is relatively new. The FEM community has the advantages of rigorous mathematical formulations with high generality, and a vibrant open source community for the development of academic research codes. The generality of FEM has led to many innovations that may prove useful for this work's research topic; but in this work, we will only employ FEM in its most basic form, commonly referred to as the Ritz-Galerkin method.

Applying FEM to fluid flow problems is detailed in the book by (Donea & Huerta, 2003). We use their formulation of the space discretization of the unsteady convection-diffusion problem using finite elements, and for time discretization using a finite differencing scheme, which is common for unsteady problems involving finite element spatial discretizations.

### 5.2.1 The weak form and spatial discretization

We now wish to discretize the unsteady convection-diffusion equation in space via the finite element method, which first requires derivation of the problem's weak form. In typical fashion, we accomplish this by multiplying the strong form by a weighting function $\phi$ and integrating the term including the Laplacian by parts, yielding

$$(\phi, u_t) + c(\boldsymbol{v}; \phi, u) + a(\phi, u) = (\phi, s) + (\phi, h)_{\Gamma_N} \tag{5.5}$$

where

$$a(\phi, u) = \int_\Omega \alpha \boldsymbol{\nabla}\phi \cdot \boldsymbol{\nabla} u \, d\Omega \tag{5.6}$$

$$c(\boldsymbol{v}; \phi, u) = \int_\Omega \phi(\boldsymbol{v} \cdot \boldsymbol{\nabla} u) d\Omega \tag{5.7}$$

$$(\phi, h)_{\Gamma_N} = \int_{\Gamma_N} \phi h \, d\Gamma \tag{5.8}$$

Note that a term that allows us to apply Neumann boundary conditions has arisen naturally in the weak form. On the contrary, special care will be needed to apply Dirichlet boundary conditions, as will be explained after we formulate the linear system.

Define the discrete solution as the linear combination

$$u^h(\boldsymbol{x}, t) = \sum_{A \in \eta \setminus \eta_D} \phi_A(\boldsymbol{x}) u_A(t) + \sum_{A \in \eta_D} \phi_A(\boldsymbol{x}) u_D(\boldsymbol{x}_A, t) \tag{5.9}$$

where $\eta$ is the set of global node indices and $\eta_D \subset \eta$ is the subset on the Dirichlet boundary $\Gamma_D$. We employ the well-established standard Galerkin method, whereby we choose the same functions $\phi$ for both the solution basis and the weak form's weighting function. This yields the semi-discrete system of ODE's

$$\boldsymbol{M}\dot{\boldsymbol{u}} + (\boldsymbol{C} + \boldsymbol{K})\boldsymbol{u} = \boldsymbol{f} \tag{5.10}$$

where $\boldsymbol{M}$ is the mass matrix, $\boldsymbol{C}$ is the convection matrix, and $\boldsymbol{K}$ is the diffusion (a.k.a. stiffness or Laplace) matrix, assembled element-wise with

$$M_{ab}^e = \int_{\Omega^e} \phi_a \phi_b \, d\Omega \tag{5.11}$$

$$C_{ab}^e = \int_{\Omega^e} \phi_a (\boldsymbol{v} \cdot \boldsymbol{\nabla}\phi_b) \, d\Omega \tag{5.12}$$

$$K_{ab}^e = \int_{\Omega^e} \alpha \boldsymbol{\nabla}\phi_a \cdot \boldsymbol{\nabla}\phi_b \, d\Omega \tag{5.13}$$

$$f_a^e = (\phi_a, s)_{\Omega^e} + (\phi_a, h)_{\Gamma^e \cap \Gamma_N} \tag{5.14}$$

where the right-hand-side vector $\boldsymbol{f}$ includes the source $s$ and the diffusive flux $h$.

### 5.2.2 Time discretization

Having used the finite element method (FEM) to discretize the unsteady convection diffusion equation in space, we now use a finite difference scheme to discretize in time. Space-time finite element formulations do exist; but they are a young and active area of research, while the method presented in this section is well established. We present the $\theta$-family of time discretization schemes, which is used by (Donea & Huerta, 2003) and also in the heat equation tutorial program on which our implementation will be based (Bangerth W. , Tutorial 26, 2016). For constant (in time) $\boldsymbol{v}$ and $\alpha$, the $\theta$-family time discretization of the strong form (5.1) is

$$\frac{\Delta u}{\Delta t} + \theta(\boldsymbol{v}\cdot\boldsymbol{\nabla} - \boldsymbol{\nabla}\cdot(\alpha\boldsymbol{\nabla}u))\Delta u = \theta s^{j+1} + (1-\theta)s^j - (\boldsymbol{v}\cdot\boldsymbol{\nabla} - \boldsymbol{\nabla}\cdot(\alpha\boldsymbol{\nabla}u))u^j \quad (5.15)$$

which combined with our weak form (5.5) is

$$\left(\phi, \frac{\Delta u}{\Delta t}\right) + \theta\left[c(\boldsymbol{v}; \phi, \Delta u) + a(\phi, \Delta u)\right] =$$
$$-[c(\boldsymbol{v}; \phi, u^j) + a(\phi, u^j)] + (\phi, \theta s^{j+1} + (1-\theta)s^j) + (\phi, \theta h^{j+1} + (1-\theta)h^j) \quad (5.16)$$

$\Delta u$ and $\Delta t$ can be chosen to yield different methods within the $\theta$-family; but for the purposes of this thesis, we take the simplest approach, with $\Delta u = u^{j+1} - u^j$ and $\Delta t = t^{j+1} - t^j$. Substituting the $\theta$-family scheme and $\Delta u$ into our discrete system of ODE's (5.10), and moving all known values to the right-hand-side, yields the discrete linear system

$$\left(\boldsymbol{M} + \Delta t^j \theta(\boldsymbol{C} + \boldsymbol{K})\right) \boldsymbol{U}^{j+1}$$
$$= \boldsymbol{M}\boldsymbol{U}^j - \Delta t^j (1-\theta)(\boldsymbol{C}+\boldsymbol{K})\boldsymbol{U}^j + \Delta t^j \left((1-\theta)\boldsymbol{f}^{j+1} + \theta\boldsymbol{f}^j\right) \quad (5.17)$$

Recall that $\boldsymbol{f}$ includes the contributions from both $\boldsymbol{s}$ and $\boldsymbol{h}$. This has second-order time accuracy when $\theta = 1/2$, which is referred to as the Crank-Nicholson method (Crank & Nicolson, 1996). $\theta = 0$ and $\theta = 1$ respectively yield the fully explicit forward and fully implicit backward Euler methods, which are only first order accurate. The scheme is A-stable when $\theta \geq 1/2$.

The linear system (5.17) is in a standard form that is solved by widely available solvers, including the solvers built in to the deal.II library, which we will use for our implementation in the next chapter. Since $\boldsymbol{K}$ is symmetric, the pure diffusion case of $\boldsymbol{C} = 0$ is especially easy to solve with the conjugate gradient (CG) algorithm. Asymmetric system matrices can be handled by an extension of CG called `BiCGStab` by the deal.II library, or more commonly by the well established GMRES (generalized minimum residual) method.

Note that (5.17) does not account for Dirichlet (strong) boundary conditions. We must now augment the linear system so that the solution is properly constrained. We apply the method described in (Bangerth W. , Boundary conditions, 2016). First decompose the discrete solution into the homogeneous part $U_0$ and an augmenting part $\tilde{G}$. $\tilde{G}$ is sometimes called a *lifting function*. The values of $\tilde{G}$ on the interior of $\boldsymbol{\Omega}$ are arbitrary, but the values on $\boldsymbol{\Gamma}_\mathrm{D}$ are constrained to the Dirichlet boundary conditions (5.3). In the practical implementation (Bangerth W. , Boundary conditions, 2016), a more complex $\tilde{G}$ is used; but for this discussion we simply choose

$$\tilde{G}(t, \boldsymbol{x}) = \begin{cases} g(t, \boldsymbol{x}) & \boldsymbol{x} \in \boldsymbol{\Gamma}_\mathrm{D} \\ 0 & \text{otherwise} \end{cases} \quad (5.18)$$

Now solve the linear system with homogeneous Dirichlet boundary conditions by subtracting $A\widetilde{G}$ from the right-hand side of (5.17), where $A = M + \Delta t^j \theta (C + K)$ is the system matrix. Finally, recover the solution to the non-homogeneous problem (5.1)-(5.4) with

$$U^{j+1} = U_0^{j+1} + \widetilde{G} \qquad (5.19)$$

### 5.2.3 The issue of discretizing asymmetric operators with FEM

A fundamental theorem for proving the stability of FEM, the best approximation theorem, assumes that the linear operator is symmetric. The diffusion operator (i.e. the Laplacian) is perfectly symmetric when diffusivity is constant in space. Unfortunately, the convection operator is asymmetric. Discretizing the convection operator with the standard Galerkin form of FEM induces an artificial loss of diffusion. The severity of this problem is characterized by the Peclet number, which is the ratio of the advective and diffusive rates of transport. The global Peclet number is based on a global characteristic length scale $L$, while the local element Peclet number is based on the local grid cell size $h$

$$Pe = \frac{vL}{\alpha}, \qquad Pe_h = \frac{vh}{2\alpha} \qquad (5.20)$$

The problem is said to be *convection dominated* when the Peclet number is high. The practical problem is that, for high $Pe_h$ and sharp boundary layers, large numerical oscillations appear in the solution. The community researching the application of FEM to fluid flow problems is well acquainted with this issue. Many stabilization methods, some of which are presented in (Donea & Huerta, 2003), have been developed to handle convection dominated flows. Some fall under the class of artificial diffusion methods, which exactly compensate for the artificial loss of diffusion, hence yielding a symmetric system matrix. For example, one of the most popular methods is Streamline Upwind/Petrov-Galerkin (SUPG) (Brooks & Hughes, 1982).

In the current work, we instead show that the asymmetry can be handled with local grid refinement, which effectively reduces the local element Peclet $Pe_h$. This is easily implemented for complex geometries in arbitrary dimensions with the finite element library deal.II (Bangerth, Davydov, & Heister, 2016). We demonstrate this in Figure 6.6. Generally, an ideal implementation of the convection-diffusion equation would include a robust stabilization method. The grid refinement approach may be prohibitively expensive for large 3D problems. That being said, it suffices for the work presented in this thesis. Furthermore, the boundary refinement approach is convenient for problems involving heat transfer, since an accurate temperature gradient is required near the boundary.

# 6 Peclet: An implementation of $\mathcal{L}$ in C++ with deal.II

Let us allocate an entire chapter to the implementation of the ambient dynamics operator $\mathcal{L}$, which is the unsteady convection diffusion problem (5.1)-(5.4), and its verification. For this thesis, the author independently developed a C++ code based on the deal.II finite element method (FEM) library (Bangerth, Davydov, & Heister, 2016) . We refer to the implementation by the name of the code, Peclet. The source code is shared in a public GitHub repository (Zimmerman, Peclet, 2016). The name Peclet alludes to the Peclet number, which is the ratio of convective to diffusive transport (5.20), and of course to the physicist Jean Claude Eugène Péclet.

This chapter first introduces the deal.II library and its tutorial program on which Peclet is based. The remaining sections highlight unique developments in Peclet that were necessary for this thesis. Many of the techniques should be generally applicable to other FEM codes. In fact, only a small fraction of the Peclet source code is specific to the convection-diffusion equation, as shown in Figure 6.2. Finally, we verify the correctness of our implementation. First we reproduce a result from (Donea & Huerta, 2003) which uses a unit source term and homogeneous Dirichlet boundary conditions. We then employ the method of manufactured solutions (MMS) to verify the spatial and temporal orders of accuracy for the general problem (5.1)-(5.4) in 1D and 2D, though we only verify simple geometries with constant diffusivity, which is a small subset of the theoretical capability of Peclet.

## 6.1 deal.II basics

deal.II is a widely used open source C++ library that is essentially a collection of low-level FEM tools which help the user quickly implement advanced FEM codes. Usability and generality have highest priority. That being said, the developers keep performance in mind, and many of the expensive methods have been optimized both for shared and distributed memory computing.

Other FEM libraries do exist; but deal.II was chosen for this thesis primarily because of its flexibility and the quality of its documentation. deal.II is powerful and flexible, which also means it requires a substantial amount of knowledge from the user. Thankfully, the developers are eager to teach, often through their responsive mailing list. Fifty-four (as of this writing) detailed tutorial programs explain the simple and advanced aspects of deal.II. The bulk of the work in this chapter borrows heavily from one of these tutorials (Bangerth W. , Tutorial 26, 2016).

Figure 6.1 shows the most important modules of deal.II. These modules are named and developed around the key aspects of most FEM programs. For the less experienced, the structure and documentation of deal.II may be a good FEM learning tool. The most effective way to learn about deal.II is to work through the first six tutorial programs. In the following sections, we will only briefly describe some of the most important features which were relevant to this thesis.

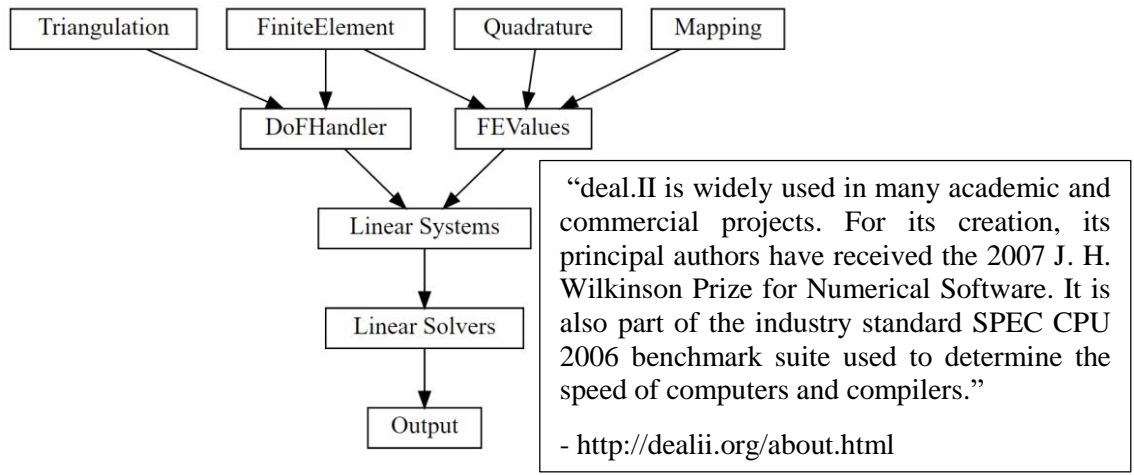

Figure 6.1: Primary modules of deal.II
Image retrieved from http://dealii.org/8.4.1/doxygen/deal.II/index.html on April 2016.

### 6.1.1 Matrix assembly and linear solvers

deal.II classifies the modules needed to assemble a linear system in a way that is analogous to the mathematical formulation. All geometric information and grid refinement information is stored within a `Triangulation`. Many elements are available in the `FiniteElement` module. `Quadrature` provides a library of quadrature rules. The `Mapping` module contains everything needed to map shape functions from the reference cell to the `Triangulation` cells.

As shown in Figure 6.1, evaluating shape functions for contribution to the system matrix requires information from the three classes of `FiniteElement`, `Quadrature`, and `Mapping`. Similarly, the `DoFHandler` (degrees of freedom handler) requires the `Triangulation` and the choice of `FiniteElement` to enumerate the degrees of freedom for the system matrix. What is still missing in this discussion is the equation for calculating element-wise contributions. To ensure flexibility of the library, the user writes the code which accumulates the system matrix.

Some widely used operators are already implemented into the `MatrixCreator` module, e.g. for the Laplace operator, `MatrixCreator::create_laplace_matrix`. Also for unsteady problems, deal.II provides `MatrixCreator::create_mass_matrix`. The community encourages users to share implementations of other operators which may be generally useful. Section 6.2.1 shows how we assemble the convection-diffusion matrix.

deal.II provides a suite of algorithms for the fast iterative solution of sparse linear systems of the standard form $Ax = b$. The available methods include many Krylov subspace methods such as GMRES and Conjugate Gradient (CG), among others. Furthermore, deal.II provides interfaces to PETSc and Trilinos libraries, perhaps the two most popular libraries for high performance scientific computing. In this work, the built-in CG method was used almost exclusively, and BiCGStab was used when high convection velocity makes the system matrix too asymmetric.

### 6.1.2 Mesh refinement, input/output, parallel computing, and compiler support

deal.II provides advanced mesh refinement tools, so users are encouraged to begin with the coarsest possible mesh which adequately represents their geometry. deal.II works only with unstructured grids, which simplifies local grid refinement (see Figure 6.3 for some examples). Hanging nodes are handled with an additional constraint matrix. While the refinement methods can be applied to complicated meshes in any number of dimensions, there are two fundamental limitations. First, deal.II assumes cells have two faces per dimension, i.e. quads in 2D or hexes in 3D. Second, grid refinement always takes the form of a tree of bisections, i.e. a binary tree in 1D, a quad-tree in 2D, or an oct-tree in 3D. The hidden limitation here is that the user has no control over the growth/refinement rate of cells. This has not been an issue for this thesis.

deal.II is well known for its adaptive grid refinement capabilities. Peclet supports adaptive refinement; but for this thesis, global refinement and the custom boundary layer refinement routine (see section 6.2.2) have been most useful. For future work where a realistic melting model is implemented, it will be important to adaptively refine near the phase-change interface (PCI).

deal.II supports many common data formats such as VTK and HDF5, though HDF5 support does not seem fully developed. Furthermore, there is a powerful `ParameterHandler` class which facilitates the design of user input files which can control all aspects of the program. For this work, only VTK output was used. This format is easily understood by open source visualization tools, e.g. ParaView (Kitware, 2016).

deal.II uses Thread Building Blocks (TBB) for shared memory parallel programming, and MPI for distributed memory parallel programming. Additionally there are some internal BLAS calls which use OpenMP for multi-threading. According to their website, deal.II has been shown to scale to more than 16,000 processors. For this thesis, the assembly of the convection-diffusion matrix was written as a small extension to the built-in Laplace matrix assembler, which uses TBB. Therefore, in theory, Peclet should achieve some speed-up when run with multithreading; but evaluating the performance and scalability of Peclet is beyond the scope of this thesis.

The deal.II developers only support the GCC compiler. That being said, there are at least two significant movements in the deal.II developer community that could eventually extend support to more compilers. At least one developer is actively working towards compiling with the Microsoft Visual C++ compiler. Secondly, the community is moving towards dropping support for standards older than C++11, which may simplify the task of supporting additional compilers.

## 6.2 Developments in the Peclet code

The development of Peclet began with a copy of deal.II's heat equation example (Bangerth W., Tutorial 26, 2016). The best way to learn about Peclet is to first work through that tutorial. That being said, there are many extensions and changes that together make Peclet. In this section we will attempt to explain some of the more unique or interesting aspects of the implementation. Some of the notable developments are that Peclet …

- 6.2.1: Builds the convection-diffusion matrix instead of the Laplace matrix.
- Supports Neumann boundary conditions.
- Uses the `ParameterHandler` class for parameter input file handling. Most aspects of the program are exposed as input parameters which do not require re-compiling any code.
- 6.2.3: Generalizes the handling of functions for all terms of the convection diffusion equation and auxiliary terms, including the velocity, diffusivity, initial values, source, and boundary conditions (both strong and natural), as well as for an exact solution function for verification. This is all accessible via the parameter input file.
- 8.3.3: Writes the final solution to disk as a `FEFieldFunction`, which contains all of the solution information in a format that can be interpolated/projected. This allows for restarting a simulation with the same grid or with a transformed grid.

The structure of Peclet source files is shown in Figure 6.2.

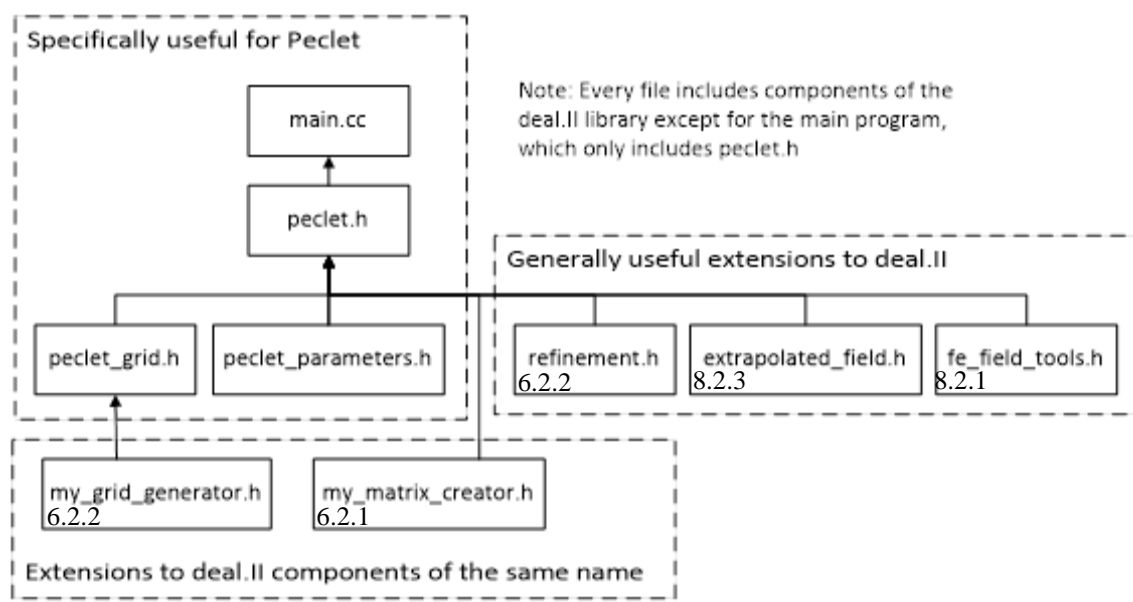

Figure 6.2: Overview of the convection-diffusion implementation, Peclet
Some components are described in this chapter, and are labeled with the corresponding section numbers.

One might notice that Peclet is entirely composed of header files. In C++, templates are restricted to header files. The heavily templated nature of deal.II complicates the use of the conventional design pattern in C/C++, where one declares classes and methods in a header, and implements them in .cc/.cpp files. For small projects, another design pattern exists for C++ code development, where one writes everything except for a small main program in header files. This can lead to prohibitively long compile times for large projects; but it is an effective strategy for small agile projects. deal.II itself is an extremely large code base; but user codes such as Peclet can be small enough to succeed with the header-only approach.

### 6.2.1 Assembling the convection-diffusion matrix

The deal.II step-26 tutorial (Bangerth W. , Tutorial 26, 2016) implements the heat equation, and uses a built-in method for assembling the Laplace matrix $K$. Comparing the discrete linear system (5.17) to the one derived in the tutorial, it is obvious that only a small modification to the implementation is needed. Namely, $K$ must be augmented by the convection matrix $C$.

One could leave the Laplace matrix $K$ assembly routine untouched, assemble $C$ independently, and add $K + C$. Keeping the matrices separate could be beneficial, especially considering that $K$ is symmetric and $C$ is asymmetric. An easier approach to implement, and perhaps also the most efficient approach in terms of number of floating point operations, is to assemble the combined matrix including the contributions both from $K$ (5.13) and from $C$ (5.12).

Hence Peclet replaces the Laplace matrix $K$ assembly routine with one for the convection-diffusion matrix $K + C$. The method is about one-hundred lines of code in the source file `my_matrix_creator.h` (Zimmerman, Peclet, 2016). Most of these lines are copied directly from the Laplace assembly routine. Some minor changes to the interface were needed, e.g. to accept the convection velocity as input. The only interesting internal change to the routine is the addition of one line of code which accumulates (5.12) for a given quadrature point.

### 6.2.2 Grid generation and refinement

Many grids are already included in the deal.II `GridGenerator` module. Among these, the hyper-shell provides a useful starting point for both geometries in this thesis. The hyper-shell domain exists between inner and outer spheres, forming a shell. Additionally deal.II provides a library of geometric manifolds that can be attached to a `Triangulation`. Peclet applies a spherical manifold to the hyper-shell domain, allowing the exact geometry to be used for grid refinement and quadrature. The coarse grid and refined grids from different refinement routines are shown in Figure 6.3. The adaptive refinement method is taken directly from deal.II's step 26 tutorial. Alternatively, the boundary method, shown more clearly in Figure 6.4, was developed for this thesis. The boundary refinement option is useful for resolving sharp boundary layers.

Given that boundary information exists in the `Triangulation` for the purpose of applying boundary conditions, writing a routine to refine the grid near a boundary is straightforward when using deal.II. The entire routine, `refine_mesh_near_boundaries`, is contained in the source file `peclet/source/pde_refinement.h` (Zimmerman, Peclet, 2016). Calling this routine only refines the cells on the specified boundary. After repeatedly calling this routine, the result is always such that the two inner-most layers have the same cell size, while the remaining layers double in size while moving outward from the boundary, as is clearly shown in Figure 6.4.

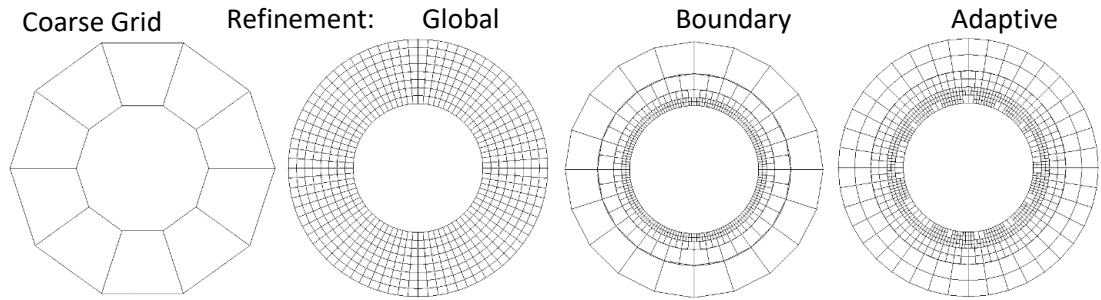

Figure 6.3: Hyper-shell grid and refinement strategies
Leftmost: Coarse grid, Second from left: Globally refined grid, Third from left: Boundary refined grid. Rightmost: Adaptively refined grid (with maximum cell count limit). Note now new points are generated on the exact geometric manifold.

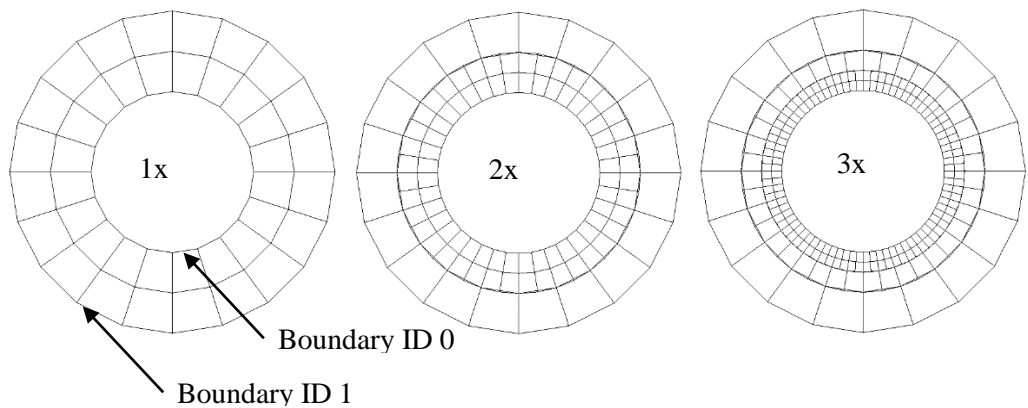

Figure 6.4: Boundary layer grid refinement
Here the `refine_mesh_near_boundaries` routine has been called with the specified boundary ID number zero.

To obtain a geometry more representative of a melting probe, we begin with a shell, transform four of the nodes, apply a spherical manifold to the domain at the nose, and apply a cylindrical manifold to the aft-body domain. The result is shown in Figure 6.5.

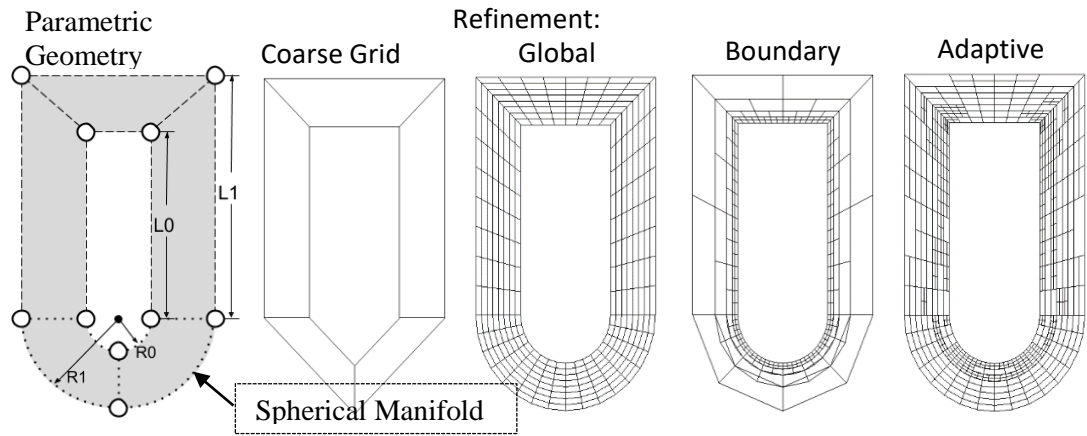

Figure 6.5: Sphere-cylinder geometry, coarse grid, and refinement strategies
Lengths are prefixed with *L* and radii are prefixed with *R*. The same four parameters uniquely define the geometry both in 2D and in 3D. The 3D geometry is axisymmetric.

### 6.2.3 Generalizing the handling of functions

The step-26 tutorial (Bangerth W. , Tutorial 26, 2016) only uses homogeneous Dirichlet boundary conditions (BC's). Peclet was extended to support general non-homogeneous Dirichlet and Neumann BC's. In this work, a powerful design pattern emerged. The most important concept is that, in C++, a pointer to some base class can point to any derived class. In deal.II, the base class for all functions is `Function<dim>`, with the pointer `Function<dim>*`. The code stores all BC functions as a vector of pointers to the deal.II Function<dim> base class, `std::vector<Function<dim>*>`. In execution, different derived classes of BC functions can be instantiated, and then pointers to these functions are pushed into this all-encompassing vector, which greatly simplifies the actual application of the BC's when building the system matrix and right-hand-side. The use of Function<dim>*> makes the code quite flexible and extensible. A very small amount of coding is required to support a new derived function class. For example, this allows for initializing a solution with a `FEFieldFunction` which is read from the hard drive. This allows one to restart a simulation with a transformed grid, the need for which is shown in chapter 8.

Most terms in the PDE are handled by the `ParsedFunction<dim>` class, including the velocity, source, and exact solution (for verification) functions. This lets the user write the function as a mathematical expression into the input file. For example, for the verification via the method of manufactured solution in sections 6.3.3 and 6.3.4, complicated functions are required for the Neumann boundary conditions and the source term. In initial versions of Peclet, implementing these required hundreds of lines of code. This was primarily overhead so that the functions would be derived from `Function<dim>` and could hence be used as arguments to the many methods in deal.II. In the current version of the code, the functions are entirely written in the input file, and the syntax is correct when the symbolic equations are copied and pasted directly from MATLAB outputs, where MATLAB has been used to derive symbolic functions.

## 6.3 Verification

Now we must verify the correctness of our implementation. First we reproduce a result from (Donea & Huerta, 2003), then we derive the exact 1D steady state solution with a non-homogeneous Neumann boundary condition, and finally we employ the method of manufactured solutions (MMS) to verify the spatial and temporal orders of accuracy in 1D and 2D.

### 6.3.1 Comparison to unsteady 1D model problem from Donea & Huerta

Since we have essentially implemented the equations written in (Donea & Huerta, 2003), it is useful to reproduce one of their results. They discuss the 1D unsteady convection-diffusion model problem with constant coefficients, unity source term, homogeneous Dirichlet boundary conditions, and initial values of zero, i.e.

$$u_t + v u_x - \alpha u_{xx} = 1 \tag{6.1}$$
$$u(0, t) = 0, \quad u(1, t) = 0 \tag{6.2}$$
$$u(x, 0) = 0 \tag{6.3}$$

In their book, (Donea & Huerta, 2003) use this problem to show the inadequacy of the standard Galerkin formulation for convection dominated flows. They show results for local element Peclet numbers $Pe_h = 1/2$ and $Pe_h = 5$, calculated with uniform grid spacing $h = 1/10$. Using Peclet, we reproduce these results in Figure 6.6. We set the velocity $v$ to unity, and set the diffusivity $\alpha$ according to $Pe_h$ (5.20). See the Appendix for the respective parameter input files to Peclet.

The only difference in the result, compared to (Donea & Huerta, 2003), is that the cell size is $h = 1/8$. This is because the bisection refinement method of deal.II cannot yield $h = 1/10$ when starting from a single cell of unit size.

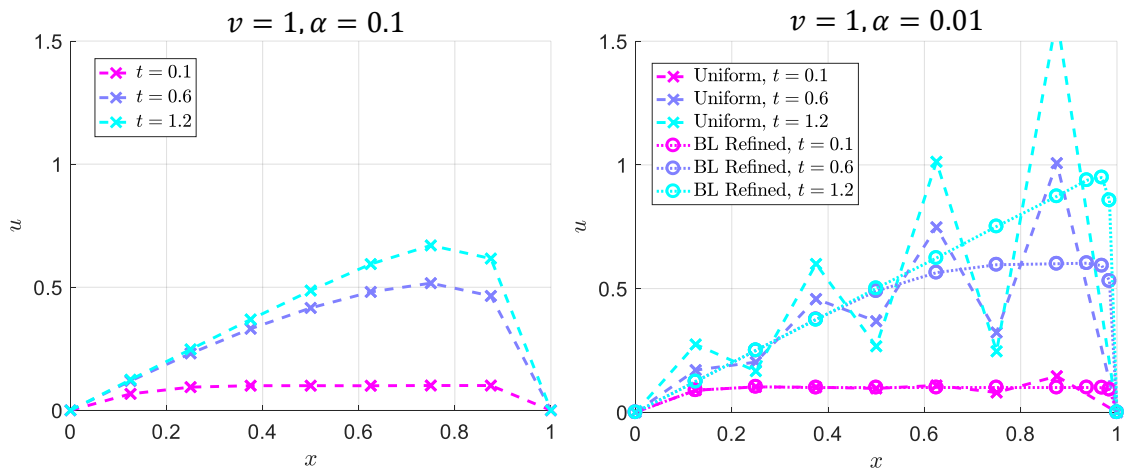

Figure 6.6: Reproduced unsteady convection-diffusion result from Donea & Huerta
In (Donea & Huerta, 2003), Figure 5.5, pertaining to solutions to their model problem 5.17. Left: $Pe = 0.5$ poses no numerical difficulties. Right: $Pe = 5$ and a sharp boundary layer together cause numerical oscillations that can be handled with local refinement. The original figure from (Donea & Huerta, 2003) only shows the globally refined result.

We also take this opportunity to verify our argument from section 5.2.3. Figure 6.6 includes a solution where boundary grid refinement was used to reduce the local element Peclet number $Pe_h$, and hence eliminate the numerical oscillations that arise when applying the standard Galerkin method to convection dominated problems with sharp boundary layers. As we admitted in section 5.2.3, an ideal convection-diffusion solver implementation would also include a robust suite of stabilization schemes; but the local grid refinement approach suffices for this thesis, where we have yet to solve large 3D problems.

### 6.3.2 1D exact steady solution

Now we wish to verify a problem with non-homogeneous Neumann and Dirichlet boundary conditions. No such example is available in (Donea & Huerta, 2003), so we design our own verification problem. First we use the method of exact solutions (MES) to verify the steady state solution of a 1D problem, which admits an easily obtainable exact solution. The 1D steady problem with zero source is

$$v u_x^\infty(x) - \alpha u_{xx}^\infty(x) = 0 \tag{6.4}$$

$$-\alpha u_x^\infty(0) = h \tag{6.5}$$

$$u^\infty(1) = g \tag{6.6}$$

which is a linear second-order ODE with exact solution

$$u_e^\infty(x) = g + \frac{h}{v}\left(\exp\left(\frac{v}{\alpha}\right) - \exp\left(\frac{vx}{\alpha}\right)\right) \tag{6.7}$$

with a pure diffusion limiting case of

$$\lim_{v \to 0} u_e^\infty(x) = g + \frac{h}{\alpha}(1 - x) \tag{6.8}$$

To obtain a physically interesting example, here we frame this example problem such that we are essentially solving the conservation of energy within the solid domain of the Stefan problem from Figure 2.1. For a given velocity, diffusivity, and external temperature $g$, we can derive a Neumann boundary condition $h$ which maintains a steady state temperature $u_e^\infty(0) = 0$ at the boundary. With this we derive $h$ from (6.7)

$$h = \frac{vg}{1 - \exp\left(\frac{v}{\alpha}\right)} \tag{6.9}$$

$$\lim_{v \to 0} h = -g\alpha \tag{6.10}$$

For this study, we will always use (6.9). The exact steady state defined by (6.4)-(6.9) is

$$u_e^\infty(x) = g\frac{\exp\left(\frac{vx}{\alpha}\right) - 1}{\exp\left(\frac{v}{\alpha}\right) - 1} \tag{6.11}$$

$$\lim_{v \to 0} u_e^\infty(x) = gx \tag{6.12}$$

The exact steady state solutions shown in Figure 6.7 are for a variety of convection velocities, diffusivities, Dirichlet boundary conditions, with the corresponding Neumann boundary condition (6.9). Peclet was run in fully implicit mode, $\theta = 1$, to produce approximately steady state solutions which are not shown here, because they do not visually differ from the exact solutions. The code produces the correct steady result; but in this thesis we will only present the rigorous calculation of empirical orders of convergence for the unsteady problem, since this is what Peclet actually solves.

These exact steady state solutions are useful for the design of our manufactured solutions in the following sections. They provide a way for us to anchor our manufactured solutions to reality.

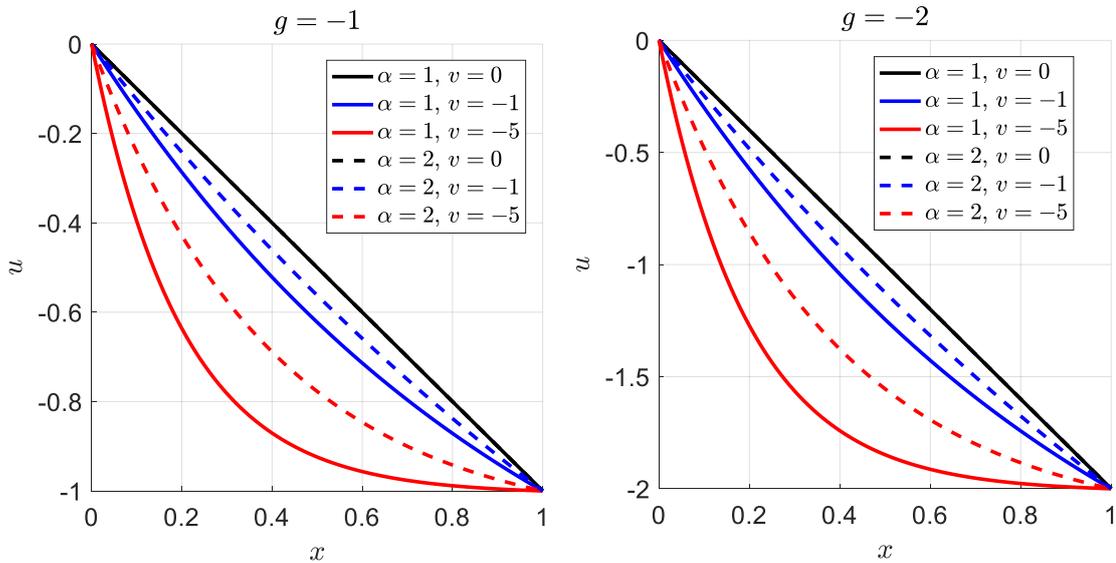

Figure 6.7: Exact 1D steady solutions of the convection-diffusion equation
A Dirichlet BC is set on the right, while the Neumann BC set on the left is exactly the value such that it yields a steady state temperature of $u = 0$ on the Neumann boundary, given by (6.9). Note how the pure diffusion solutions are a straight line. Increasing the convection velocity increases the curvature of the solution, which begins to develop a sharp boundary layer at the Neumann boundary. Increasing the velocity to larger values forces the boundary layer even closer to the edge, increasing its sharpness.

### 6.3.3 1D unsteady manufactured solution

The unsteady 1D problem is

$$u_t(x,t) + vu_x(x,t) - \alpha u_{xx}(x,t) = s(x,t) \quad \forall\, (x,t) \in [0,1] \times (0,1] \tag{6.13}$$

$$u(x,0) = u_0 \tag{6.14}$$

$$-\alpha u_x(0,t) = h(t) \tag{6.15}$$

$$u(1,t) = g \tag{6.16}$$

A closed form solution for the unsteady 1D problem (6.13)-(6.16) is not easily obtainable. Therefore, this work employs the method of manufactured solutions (MMS) to verify the spatial and temporal orders of convergence of the unsteady problem.

Generally, MMS does not require a solution design with any physical meaning. This property can be useful for manufacturing solutions that cover a wide range of code capabilities in a small number of coverage tests. For our present problem (6.13)-(6.16), it is easy enough to construct a somewhat physically meaningful solution. We do this by making a small modification to the exact solution of the steady solution $u^\infty$. We design an unsteady solution where the initial values $u^0$ are equal to the cold temperature $g$. We use an exponential term to transition from $u^0$ to $u^\infty$, i.e.

$$u^0 = g \tag{6.17}$$

$$u^\infty = g \frac{\exp\left(\frac{vx}{\alpha}\right) - 1}{\exp\left(\frac{v}{\alpha}\right) - 1} \tag{6.18}$$

$$u_M(x,t) = u^0 + (u^\infty - u^0)(1 - \exp(-\beta t^2)) \tag{6.19}$$

Altogether (6.17)-(6.19) yield

$$u_M(x,t) = g\left(1 + \left(\frac{\exp\left(\frac{vx}{\alpha}\right) - 1}{\exp\left(\frac{v}{\alpha}\right) - 1} - 1\right)(1 - \exp(-\beta t^2))\right) \tag{6.20}$$

$$\lim_{v \to 0} u_M(x,t) = -g((\exp(-\beta t^2) - 1)(x - 1) - 1) \tag{6.21}$$

where we choose a sufficient coefficient $\beta$ such that $u_M(t=1) \approx \lim_{t \to \infty} u_M$. The steady solution of (6.21) is equivalent to (6.11). Figure 6.8 shows the unsteady behavior.

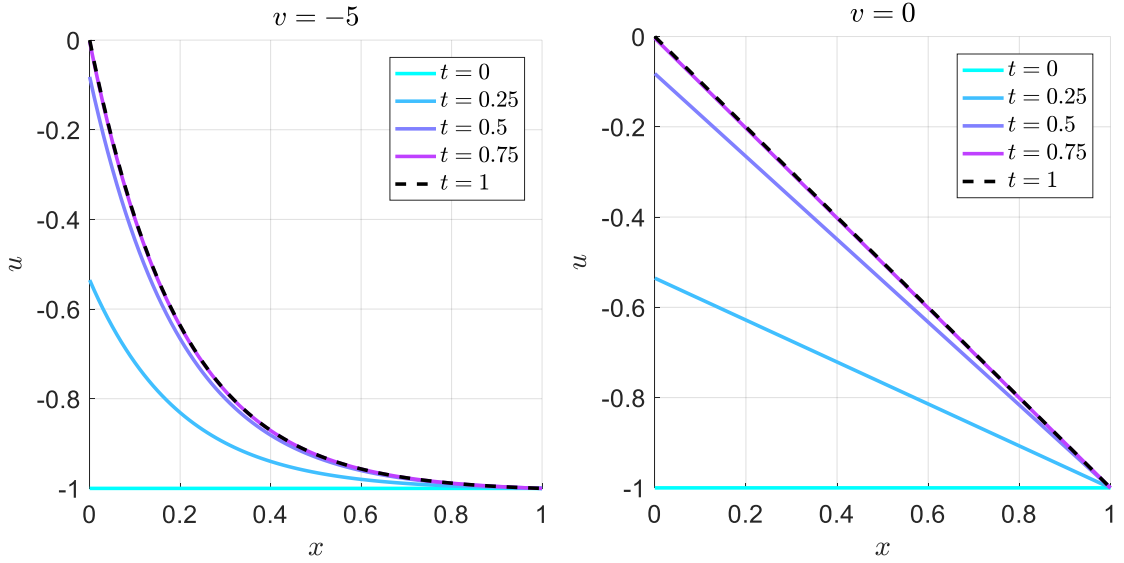

Figure 6.8: Manufactured 1D unsteady solutions
These are solutions to (6.20) and (6.21). Parameters not shown are $\alpha = 1$, $g = -1, \beta = 10$.

Via MMS we derive the Neumann boundary condition $h_M(t)$ and source $s_M(x,t)$ that together produce $u_M(x,t)$. Rather than doing all of this by hand, the author employed MATLAB's Symbolic Toolbox. The script and output are in Appendix A.

Derive $h_M$ from (6.15)

$$h_M(t) = -\alpha \frac{\partial}{\partial x} u_M(0,t) = vg \frac{\exp(-\beta t^2) - 1}{\exp\left(\frac{v}{\alpha}\right) - 1} \quad (6.22)$$

$$\lim_{v \to 0} h_M(t) = g\alpha(\exp(-\beta t^2) - 1) \quad (6.23)$$

Note that the steady ($t \to \infty$) limit of (6.22) is indeed equivalent to (6.9).

Derive $s_M$ from (6.13)

$$s_M = u_t + vu_x - \alpha \frac{\partial^2}{\partial x^2} u = 2\beta gt \exp(-\beta t^2) \left( \frac{\exp\left(\frac{vx}{\alpha}\right) - 1}{\exp\left(\frac{v}{\alpha}\right) - 1} - 1 \right) \quad (6.24)$$

$$\lim_{v \to 0} s_M(x,t) = 2\beta gt \exp(-\beta t^2)(x-1) \quad (6.25)$$

The steady state is consistent with the exact solution with zero source, so the manufactured source term $s_M$ vanishes as $t \to \infty$. Given sufficient $\beta$, $s_M(t)$ is nearly zero at $t = 1$. The Neumann boundary condition $h_M$ also approaches a steady state. Both of these propreties are shown in Figure 6.9.

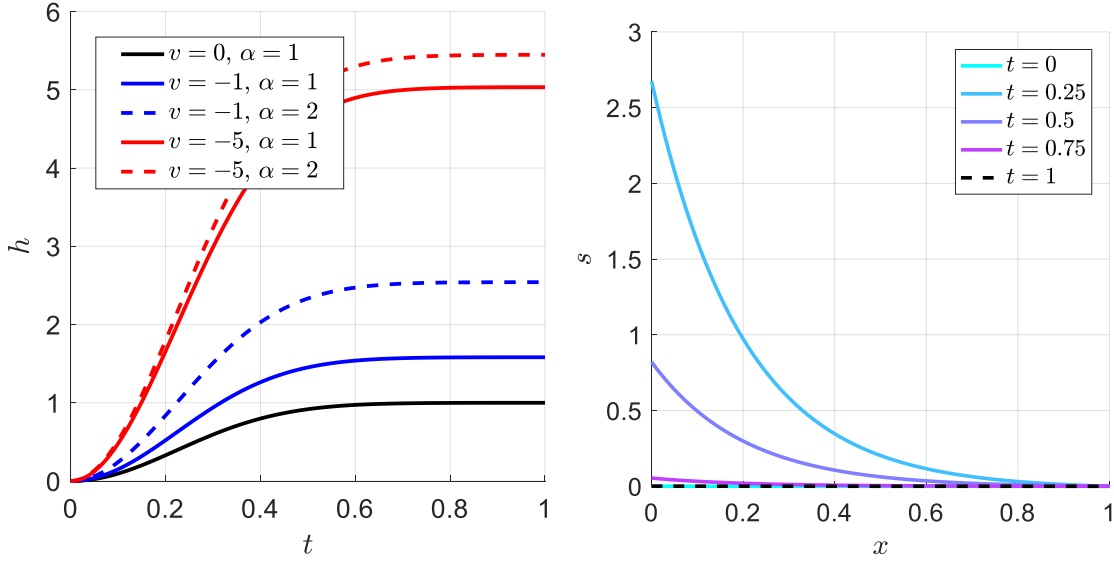

Figure 6.9: Derived auxiliary conditions for manufactured solution.
Left: Neumann boundary condition, $g = -1$, Right: Source term, $v = -5, \alpha = 1, g = -1$

Spatial and temporal convergence studies were run for a variety of $\alpha$, $v$, and $g$. Table 2 shows the results. See Appendix A for one of these parameter input files to Peclet, for which only the respective parameters must be changed to reproduce all results. The theoretical convergence orders are $p = 2$ and $q = 2$, so this verification was successful in all cases. The best match shown here are the three cases which match to four significant digits. It should be possible to further refine any of these cases to increase the precision of the empirical convergence order.

| v | $\alpha$ | g | p | q | $\tilde{v}$ | $\alpha$ | $\tilde{g}$ | p | q |
|---|---|---|---|---|---|---|---|---|---|
| -5 | 2 | -2 | 1.996 | 2.000 | -1 | 1 | -2 | 2.015 | 2.001 |
| -5 | 2 | -1 | 1.992 | 1.999 | -1 | 1 | -1 | 2.008 | 2.002 |
| -5 | 1 | -2 | 1.998 | 1.999 | 0 | 2 | -2 | N/A | 2.001 |
| -5 | 1 | -1 | 1.997 | 1.999 | 0 | 2 | -1 | N/A | 2.001 |
| -1 | 2 | -2 | 2.022 | 2.001 | 0 | 1 | -2 | N/A | 2.000 |
| -1 | 2 | -1 | 2.008 | 2.002 | 0 | 1 | -1 | N/A | 2.000 |

Table 2: MMS 1D spatial ($p$) and temporal ($q$) convergence table

### 6.3.4 2D unsteady manufactured solution

Next we verify a 2D case with a spatially variable convection velocity. Keeping with the theme of manufacturing physically relevant solutions, let us consider the case of a plate melting through a block of ice with a temperature differential which causes rotation about one of its endpoints. At the point of rotation, the convection velocity is zero. The velocity increases linearly with y. This problem is sketched in the left side of Figure 6.10.

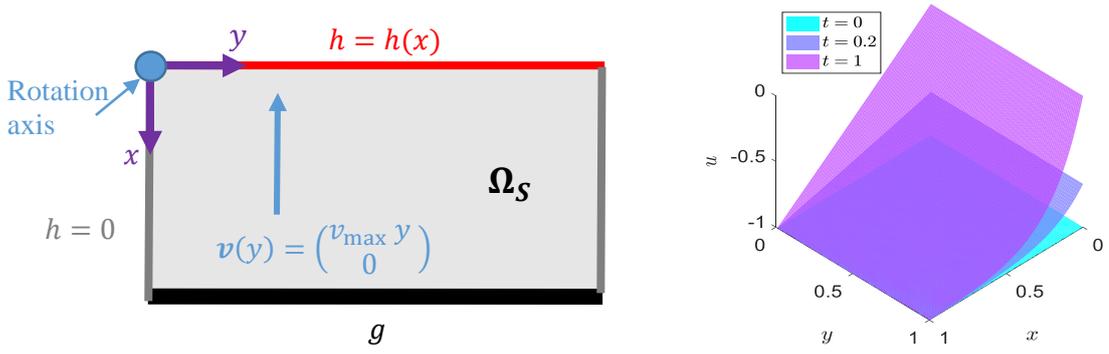

Figure 6.10: Sketch of 2D MMS problem, and exact unsteady solution.
Left: Sketch of the problem domain, boundaries, and parameters. $v$ is the convection velocity. $h$ are Neumann boundary conditions. $g$ is a Dirichlet boundary condition.
Right: Unsteady solution for $\alpha = 1, v_{max} = 5, g = -1$ given the manufactured source term.

Similar to the 1D manufactured solution, we will choose initial and steady states

$$u^0 = g \tag{6.26}$$

$$u^\infty(x, y) = g \frac{\exp\left(\frac{v_{max} xy}{\alpha}\right) - 1}{\exp\left(\frac{v_{max} y}{\alpha}\right) - 1} \tag{6.27}$$

and use the same exponential function of time to transition between them

$$u_M(x, t) = u^0 + (u^\infty - u^0)(1 - \exp(-\beta t^2))$$
$$= g\left(1 + \left(1 - \frac{\exp(v_{max} xy/\alpha) - 1}{\exp(v_{max} y/\alpha) - 1}\right)(\exp(-\beta t^2) - 1)\right) \tag{6.28}$$

The Neumann boundary conditions and source term are again derived per the strong form. There are many more terms than for the 1D problem, so the equations are in Appendix A. Convergence studies were again run for a variety of $\alpha, v_{max}$, and $g$, with the results shown in Table 3.

| $\alpha$ | $v_{max}$ | $g$ | $p$ | $q$ | $\alpha$ | $v_{max}$ | $g$ | $p$ | $q$ |
|---|---|---|---|---|---|---|---|---|---|
| 2 | -5 | -2 | 2.036 | 2.058 | 1 | -5 | -2 | 1.999 | 2.058 |
| 2 | -5 | -1 | 2.036 | 2.058 | 1 | -5 | -1 | 1.999 | 2.059 |
| 2 | -1 | -2 | 2.013 | 2.042 | 1 | -1 | -2 | 2.007 | 2.029 |
| 2 | -1 | -1 | 2.012 | 2.042 | 1 | -1 | -1 | 2.007 | 2.029 |

Table 3: 2D MMS convergence table.
The columns are diffusivity, max velocity, Dirichlet BC, spatial convergence order, and temporal convergence order.

# 7 A realistic phase-change model: Moving melt film

So far we have neglected the actual phase-change process within the ambient dynamics $\mathcal{L}$, which physically requires some non-zero latent heat $h_m$. Ultimately the successful prediction of melting probe trajectories will require a model for $\mathcal{L}$ which resolves both the contact melting/sublimation on the contact surface $\mathbf{\Gamma}_c$, shown in Figure 3.1, and the refreezing throughout the entire ambient domain $\mathbf{\Omega}$. In future work, we hope to apply a realistic melting model throughout $\mathbf{\Omega}$, perhaps being the enthalpy-porosity model which we briefly presented in 2.1.2. For the purposes of this thesis, we take a smaller step by only modeling an aspect of the phase-change near $\mathbf{\Gamma}_c$.

This chapter models the phase-change near $\mathbf{\Gamma}_c$ by considering the Stefan problem, previously formulated in section 2.1.1. We embed the Stefan problem micro-scale melt film boundary which is fixed to the rigid body (RB) and therefore moving in the global reference frame.

This is the first known attempt to use the convection-diffusion equation to model the unsteady evolution of the temperature field in the domain around a boundary characterized by the Stefan problem. Ideally with a consistently chosen velocity $v$ of the phase-change interface (PCI), melt film thickness $\delta$, and wall temperature $T_w$, the temperature near $\mathbf{\Gamma}_c$ would remain steadily at the melting temperature $T_m$. We should expect to see deviations in the 2D domain. Also we can study the sensitivity of the temperature profile near $\mathbf{\Gamma}_c$ in respect to all parameters.

### 7.1.1 Modeling the domain near a melt film boundary

To model the domain near $\mathbf{\Gamma}_c$, consider a case where the rigid body (RB) moves with a constant velocity $v$. Figure 7.1 shows a simplified rectangular domain that results if we zoom in to the contact surface from Figure 3.1 and ignore curvature. Most of the boundaries $\mathbf{\Gamma}$ in Figure 7.1 map directly to boundaries in Figure 3.1. There are two additional boundaries. Cutting out this section near the contact surface introduces a new boundary $\mathbf{\Gamma}_s$ which slices through the original solid domain. On $\mathbf{\Gamma}_s$ we apply a homogeneous Neumann (i.e. adiabatic) boundary condition, enforcing zero lateral heat conduction out of the domain. Note that the grid is fixed in this reference frame, so the RB's velocity appears as the opposite of the convection velocity $v$.

The domain described by Figure 7.1 is entirely solid on the interior. A PC occurs on $\mathbf{\Gamma}_c$ and $\mathbf{\Gamma}_m$. On $\mathbf{\Gamma}_m$ we set the temperature directly and ignore the physical details of the PC; but on $\mathbf{\Gamma}_c$ we apply a heat flux boundary condition based on the Stefan condition from (2.1), which physically models the latent heat of melting $h_m$. Rearranging the Stefan condition equation yields the heat flux to be applied to the melt boundary for a given velocity and probe temperature

$$q^+(\boldsymbol{x}) = q^-(\boldsymbol{x}) - \rho_S h_m v = -k_L \frac{T_m - T_w(\boldsymbol{x})}{\delta(\boldsymbol{x})} - \rho_S h_m v \qquad (7.1)$$

Note that this is a heat flux, while we are simulating the convection and diffusion of a temperature field. Consider the state equation, given by the first law of thermodynamics

$$\rho c_p \frac{\partial T}{\partial t} - \nabla \cdot (k \nabla T) = \dot{q}_V \tag{7.2}$$

where $\dot{q}_V$ is the volumetric heat flux. Normalizing by $\rho c_p$ yields the heat equation

$$\frac{\partial T}{\partial t} - \nabla \cdot (\alpha \nabla T) = \frac{\dot{q}_V}{\rho c_p} \tag{7.3}$$

With this knowledge we obtain a function for the Neumann boundary values

$$h(\mathbf{x}) = \frac{q^+(\mathbf{x})}{\rho_S c_{p,S}} \tag{7.4}$$

We must assume that we have some function to compute the film thickness $\delta(\mathbf{x}) \approx \mathcal{O}(1)$ µm. Accurately predicting the melt film thickness requires solving the micro-scale problem. Note that (2.1) and therefore (7.1) is a one-dimensional model which we apply in the normal direction at every point $\mathbf{x}$ on $\Gamma_c$. We use the thermal properties of ice near its melting temperature, which are

$$k_S = 2.14 \frac{\text{W}}{\text{m K}}, \quad \rho_S = 917 \frac{\text{kg}}{\text{m}^3}, \quad c_{p,S} = 2110 \frac{\text{J}}{\text{kg K}}, \quad h_m = 3.34 \cdot 10^6 \frac{\text{J}}{\text{kg}}$$

The thermal diffusivity is then

$$\alpha_S = \frac{k_S}{\rho_S c_{p,S}} = 1.11 \cdot 10^{-6} \frac{\text{m}^2}{\text{s}}$$

And finally we need the thermal conductivity of water near its freezing temperature

$$k_L = 0.5611 \frac{\text{W}}{\text{m K}}$$

Indeed, for a given $v$, $\delta$, $T_w$, and therefore $h$, the bubble of a steady melt interface forms. Figure 7.2 shows how the size of the "bubble" inside of the melt interface decreases as we apply a heat flux approaching the one specified by the exact Stefan condition, which is labeled $h^*$. With the exact $h^*$, the temperature near the wall only nearly reaches the melting temperature, and so there is no $T = T_m$ isoline. This is exactly the expected behavior. This means that the convection-diffusion equation is indeed useful for modeling the temperature distribution within the solid domain near the heating surface of a close-contact melting problem with a moving heat source.

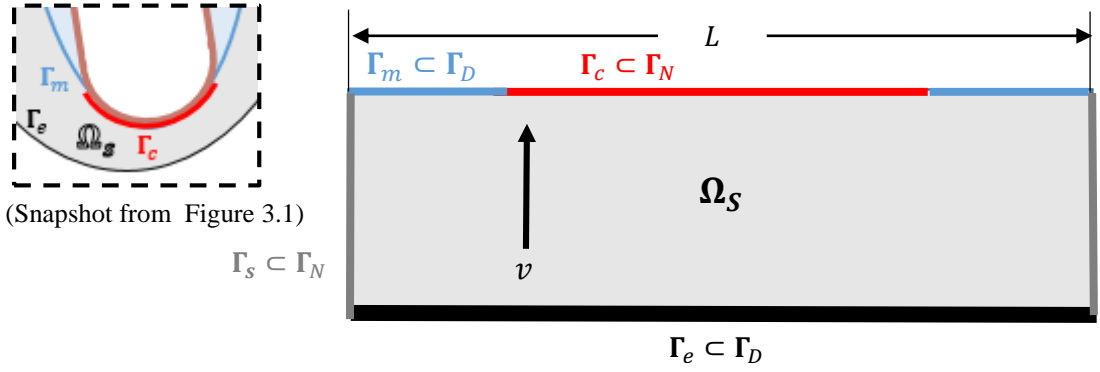

Figure 7.1: Domain near close-contact melt film
$L$ is a reference length. $v$ is the convection velocity. Boundary conditions are Dirichlet on $\Gamma_D$ and Neumann on $\Gamma_N$.

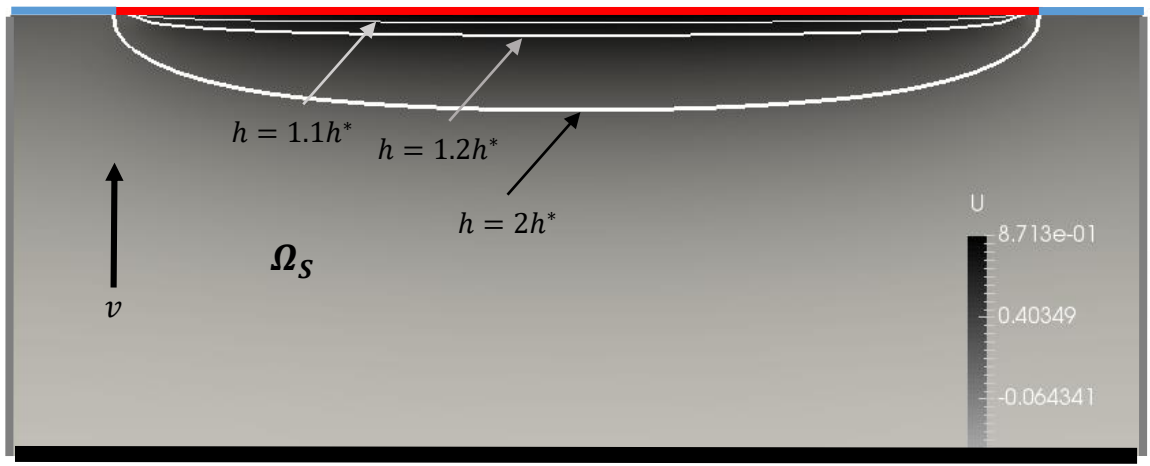

Figure 7.2: Steady-state melt temperature contour for the Stefan condition based 2D problem
This is Peclet output visualized with ParaView. White isolines mark the phase-change interface (PCI). The boundaries are color-coded according to Figure 7.1. $h$ is the Neumann BC prescribed on the contact surface. $h^*$ is the exact Neumann BC corresponding to the Stefan condition.

### 7.1.2 Sensitivity of the melt film boundary

To analyze the sensitivity of the Stefan-condition based Neumann boundary to the parameters, consider a 1D slice down the center of the domain from Figure 7.2. Here we initialize with the known steady state solution for $v^*$, but evolve the unsteady convection-diffusion equation with a perturbed velocity of $v$. Small perturbations on $v$ significantly change the steady state temperature at the boundary. The same occurs for small perturbations in the wall temperature $T_w$ or melt film thickness $\delta$. The heat flux is very sensitive to the melt film thickness. Physically the melt film thickness is dependent on the velocity and the wall temperature, which would have a regulating effect if we were resolving the dynamics of the melt film itself.

In section 8.4.2 we will demonstrate a coupled problem where we perturb the heat flux and simulate the dynamic response of the rigid body's velocity. It would be interesting to instead control the wall temperature embedded in the Stefan condition. The sensitivity of the boundary value to small changes in the parameters makes this impractical without a function to update the melt film thickness, which regulates the heat flux and therefore the velocity. This makes intuitive sense, since with the methods developed in this thesis, there is no means of resolving features on the scale of the melt film thickness.

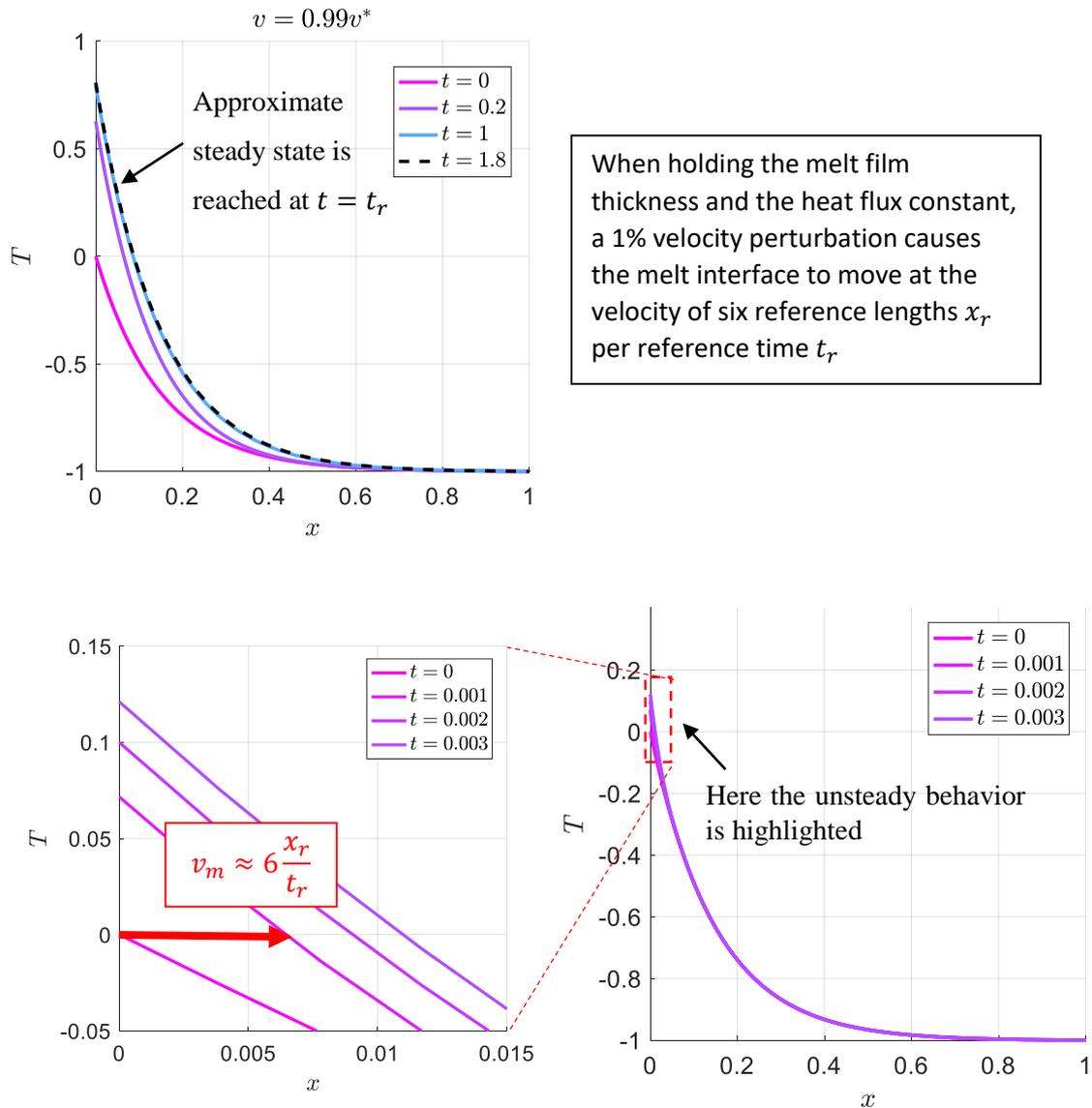

Figure 7.3: Sensitivity of the Stefan condition based Neumann boundary in 1D
The top-left visualizes large time steps reaching the steady state solution. The bottom-right visualizes small time steps toward the beginning of the simulation, highlighting the unsteady behavior. $T$ is the temperature. $v$ is the convection velocity. $v^*$ is the steady velocity corresponding to the Stefan condition. $v_m$ is the velocity of the phase-change interface (PCI).

# 8 Coupling $\mathcal{H}$ and $\mathcal{L}$: The complete forward problem

In the previous chapters, we have already presented mathematical models for the split operators $\mathcal{H}$ and $\mathcal{L}$, which are respectively the rigid body dynamics (RBD) operator and the ambient dynamics operator. We further have already implemented these models and tested them independently. In chapter 3, we briefly introduced the abstract mathematical formulation of the coupled problem. For quick reference, we repeat here the coupled split operators

$$\dot{\xi}(t) = \mathcal{H}\big(t, \xi(t), u(t,x)\big), \qquad u_t(t,x) = \mathcal{L}(t, u(t,x), \xi(t), \dot{\xi}(t)) \tag{8.1}$$

## 8.1 Temporal coupling

In chapter 4, we formulated $\mathcal{H}$ as a minimization problem which requires the solution of a constrained nonlinear program (4.9)-(4.12) at every time step. In chapter 5, we formulated $\mathcal{L}$ as an initial boundary value problem that requires the solution of a linear system (5.17) at every time step. In chapter 3, we introduced how these two problems are coupled via the shared rigid body (RB) state $\xi$ and ambient state $u$. As discussed in chapter 3, both $\xi(t)$ and $u(t)$ depend on time $t$. We couple $\mathcal{H}$ and $\mathcal{L}$ by requiring that they share an equivalent $\xi^i = \xi(t_i)$ and $u^i = u(t_i)$ at every discrete time $t_i$. In the current chapter, we detail the coupling procedure for (8.1). We will write the algorithm for the iterative time evolution procedure, followed by a detailed description of the software implementation, finally followed by results.

Different discrete time step sizes are appropriate for each operator. In general, the trajectory integration $\mathcal{H}$ allows for larger time steps. We view this as a multi-index into discrete time, as visualized in Figure 8.1. We discretize each $\mathcal{H}$ time interval $t_i$ onto a smaller scale for $\mathcal{L}$. When discussing $\mathcal{H}$, we may use a single index for time interval $t_i$ and discrete time $t_i$. When discussing $\mathcal{L}$, which includes the smaller time scale, we must use the multi-indexed time interval $t_{i,j}$ and discrete time $t_{i,j}$.

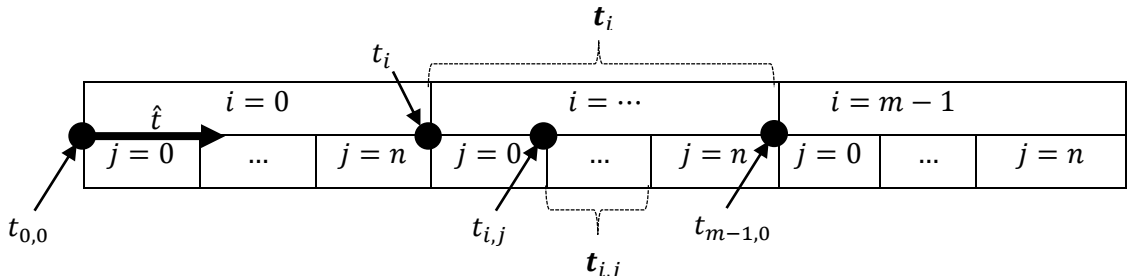

Figure 8.1: Multi-indexed time discretization for the split operators
The rigid body dynamics $\mathcal{H}$ is split into $m$ time intervals. Within each $i$th interval, the ambient dynamics $\mathcal{L}$ are further split into $n$ time sub-intervals. Note that as time increases from left to right along the axis $\hat{t}$, each time interval is closed on the left and open on the right, i.e. $t = \{t \in \mathbb{R}: t_i \leq t < t_{i+1}\}$.

In the following equations, the superscripted $\xi^i$ denotes the rigid body state evaluated at time $t_i$, and the superscripted $u^{i,j}$ denotes the evaluation of the ambient state at time $t_{i,j}$, i.e.

$$\xi^i = \xi(t_i) \tag{8.2}$$

$$u^{i,j}(x) = u(x, t_{i,j}) \tag{8.3}$$

Now we label the solutions for a discrete time step of $\mathcal{H}$ or $\mathcal{L}$ as the discrete operators $\hbar$ or $\ell$, which respectively advance $\xi$ or $u$ a discrete time step, i.e.

$$\xi^{i+1} = \hbar(u^i, \xi^i), \qquad u^{i,j+1} = \ell(t_{i,j+1} - t_{i,j}, u^{i,j}, \xi^i, \dot{\xi}^i) \tag{8.4}$$

Evolving the coupled problem in time, and hence obtaining the time evolution both of $\xi(t)$ and $u(t)$, requires an iterative procedure explained by Algorithm 1. To help explain the examples in this chapter, we also show here the time-dependent heat flux input $h_i$. There is one additional complication shown in the algorithm: We superpose the velocity $\dot{\xi}$ when solving $\mathcal{H}$ and $\mathcal{L}$, so we must track an additional state in order to view the rigid body from a global reference frame. We name this auxiliary state $\xi_V$.

1. $\xi^0 := \xi$, $\dot{\xi}^0 := \dot{\xi}$     Set initial values for the rigid body (RB) state and its derivative.
2. $\xi_V^0 := \xi^0$     At the initial RB state, the global auxiliary RB state is equivalent.
3. for $i \in [0,1,2, \ldots, m)$:     Loop through the discrete rigid body dynamics (RBD) time steps.
    a. for $j \in [0,1,2, \ldots, n)$:     Loop through ambient dynamics time steps.
        i. $u^{i,j+1} := \ell(t_{i,j}, u^{i,j}, \xi^i, \dot{\xi}^i; h_i)$     Solve an ambient dynamics step.
    b. $u^{i+1,0} := u^{i,n}$     Set the correct index per Figure 8.1.
    c. $\xi^{i+1} := \hbar(u^{i+1,0}, \xi^i)$     Solve a RBD time step.
    d. $\dot{\xi}^{i+1} := \dot{\xi}^i + (\xi^{i+1} - \xi^i)/t_i$     Update the state time derivative.
    e. $\xi_V^{i+1} := \xi_V^i + (\xi^{i+1} - \xi^i) + t_i \dot{\xi}^i$     Update the global auxiliary state.

Algorithm 1: Iterating the coupled problem through time

Line 1.a.i. of Algorithm 1 shows that the previous ambient state is used to initialize $\ell$, so that the ambient state is evolving in time through all steps of the trajectory. The phase-change interface (PCI) is a feature of $u(t)$ and hence is also evolving in time. Recall that the convection velocity profile $v(x)$ within $\mathcal{L}$ is derived from the state time derivative $\dot{\xi}$. For a sufficiently high heat flux $h$ and sufficiently low velocity $v$, the PCI moves away from the rigid body. This behavior is shown in Figure 7.2. In such cases, solving $\hbar$ yields an update to $\xi$. Dividing the difference of the updated state $\xi^{i+1}$ and the previous state $\xi^i$ by the length of the time interval $t_i$ yields an update for $\dot{\xi}$. For constant $h$, iterating $i$ converges $\dot{\xi}$ toward a steady state.

Here a fundamental limitation of the superposed velocity concept can be seen. When the velocity is too high or the heat flux is too low, no PCI will form, and hence $\mathcal{H}$ cannot update the position, and Algorithm 1 cannot update the velocity. This means that the heat flux input $h$ must be non-decreasing, i.e. the rigid body cannot slow down.

In this work, Algorithm 1 was implemented in Python and C++. In addition to the algorithm, the implementation requires pre-processing, post-processing, and an interface between the Python driver and the C++ implementation of $\mathcal{L}$. Recall from chapters 5 and 6 that $\mathcal{L}$ is a partial-differential equation (PDE), which is assembled as a linear system and solved by the finite element library deal.II (Bangerth, Davydov, & Heister, 2016). It is not trivial to use an existing solution to initialize the PDE solver in Line 1.a.i. of Algorithm 1. We explain these difficulties in section 8.3.

## 8.2 The overall trajectory simulator software system

For this work, Algorithm 1 was implemented in Python and C++. The primary components of the implementation, and how they interact, are shown in Figure 8.2.

Figure 8.2 shows the overall flow of the trajectory simulator, how it interfaces with the PDE solver, and how the PDE solver re-initializes based on input from the simulator. The entire source code is shared at a public GitHub repository (Zimmerman, dimice-python-cpp, 2016). Note that this repository includes the PDE solver as a submodule. While we presented the most recent and complete version of the PDE solver, Peclet, in chapter 6, the submodule points to an older version that is named dimice-pde-dealii.

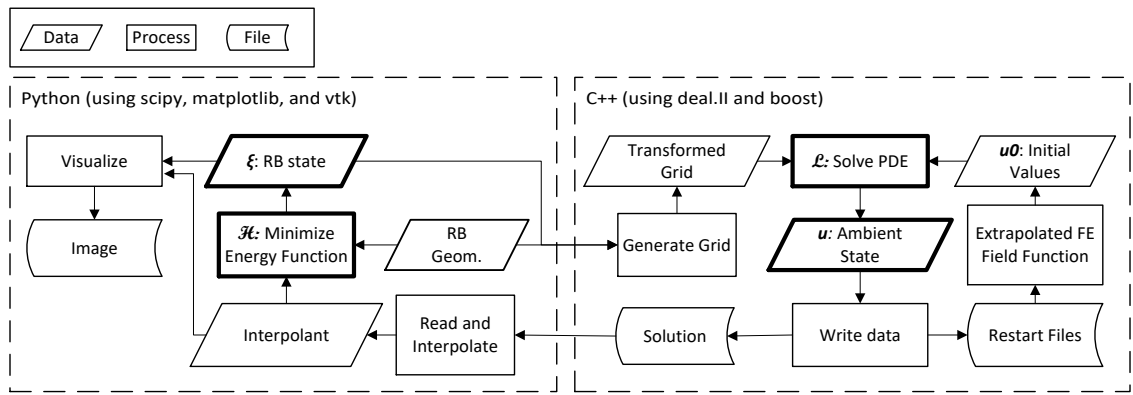

Figure 8.2: Software system overview for coupled $\mathcal{H}$ and $\mathcal{L}$.
$\mathcal{H}$ determines the rigid body dynamics (RBD), while $\mathcal{L}$ determines the ambient dynamics. The components outlined in **bold** map directly to steps in Algorithm 1. The majority of the components are for pre-processing, post-processing, and implementation details.

All time steps output by the PDE solver are archived so that they can be explored in ParaView (Kitware, 2016). The final solution for each trajectory time step is read by VTK and interpolated with SciPy (The Scipy community, 2016). matplotlib (Hunter, Dale, Firing, & Droettboom, 2016) is used to visualize the state of the rigid body (RB) and the phase-change interface (PCI). Additionally, a contour is drawn where the ambient temperature is 80% of the farfield temperature, to get a sense for the time evolution of the ambient away from the PCI.

The Python program is decomposed into a group of modules shown in Figure 4.4. The object-oriented design of this program is flexible, making it simple to test new ideas. The main class is `Trajectory`, which has instances of the `Body` and `PDE` classes as members. With this framework, the user imports the `trajectory` module, instantiates a `Trajectory` object in their own script, and modifies the parameters that are members of their instance before finally running the trajectory with the `Trajectory.run_steps` or `Trajectory.run_step` method. The entire user script for generating Figure 8.8 is provided in Appendix A. After iterating the method `Trajectory.run` for half of the total time steps, the Neumann boundary condition was increased, and the method `Trajectory.run` was iterated again for the remaining time steps. The state is always saved within the `Trajectory` instance, making it simple to generate trajectory simulations with a sequence of different inputs.

As detailed in section 4.3.2, SciPy's SLSQP routine within `scipy.minimize` (The Scipy community, 2016) was used to solve the minimization problem. SciPy is a widely used library for scientific computing; and it is convenient to try this as a first approach. In future work when extending to 3D and more complex ambient dynamics models, it will likely become necessary to use interior point optimization instead. A third-party Python interface, `pyipopt`, does exist for `ipopt`, but it does not appear to be actively supported.

## 8.3 Implementation of the interface to $\mathcal{L}$ in deal.II

Three capabilities were needed which extend the PDE solver:

- Transform a grid according to a state vector
- Write and read an entire finite element solution to and from the hard drive
- Initialize a solution based on interpolation and extrapolation of an old solution

Figure 8.3 visualizes why interpolation and extrapolation are necessary. Reading and writing to and from the disk is necessary so that the external driver, the trajectory simulator, can run the PDE solver from the command line, and restart each execution based on the solution from the previous step. The details of implementing each of these three features in deal.II presents some difficulties, especially with reading solutions from the hard drive. In the next sections, we will discuss some of the more interesting points.

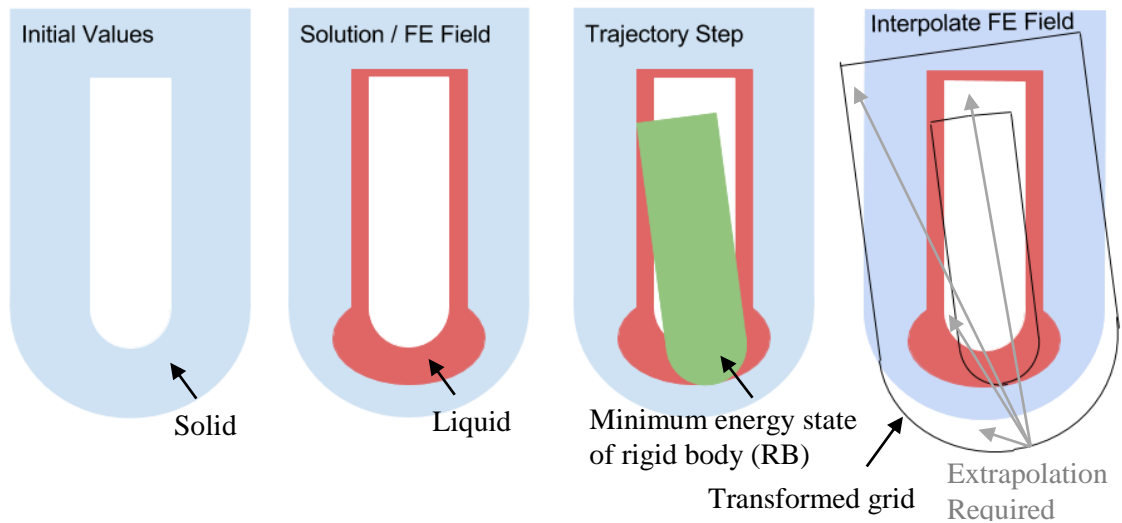

Figure 8.3: Sketch of grid transformation and initial values interpolation/extrapolation
The liquid domain is colored red and the solid domain is colored blue. The minimum energy state of the rigid body (RB) for the given ambient is colored green. The outline of the transformed grid corresponding to the new RB state is shown on the right-hand sketch, which illustrates regions of the PDE solution which must be interpolated or extrapolated.

### 8.3.1 Grid transformation

The rigid body (RB) transformation can be applied either absolutely from the original RB position, or relatively at each sequential RB position. For RB transformations either should suffice; but the absolute approach is simpler. The next question is whether to solve the ambient dynamics numerically on the reference grid or on the transformed grid. Either approach will require multiplying the entire grid by the transformation matrix. During pre-processing either the current grid must be transformed to conform with the reference grid, or vice versa.

A careful read of the deal.II documentation leaves us with only one option that will work for parallel computations. In the detailed description of `GridTools::Transform`, the deal.II developers say that transforming a refined parallel (distributed) `Triangulation` will result in invalid ghost cells, which breaks many methods, e.g. the Kelly error estimator which is needed for adaptive grid refinement. Therefore, we will only transform the coarse grid, and not solutions which are always on refined grids.

### 8.3.2 Interpolation and extrapolation

Initializing the solution values based on a previous solution is a difficult task, because the grid has moved. Essentially the previous solution must be interpolated onto a transformation of the previous grid. There are at least three ways one could attempt to do this, in decreasing order of the fidelity of the projection:

1. Create a field function that uses all of the finite element discretization and solution information

2. Use VTK methods to sample VTK solution files
3. Treat the nodal data as a scattered set and use a scattered data interpolation algorithm

The first option is ideal. Thankfully deal.II has a built-in function that is almost what we need, except for a major drawback. Given a finite element discretization and solution, one can create a `FEFieldFunction`, which represents the finite element solution that can be continuously sampled within the domain. The key drawback here is that the sample must be *within* the domain, which is not good enough for us, since we wish to move the grid. Already for the simple geometry in Figure 8.3, we see four regions where data must be sampled from outside of the domain of the existing finite element solution.

In this work, `FEFieldFunction` was extended with exception handling that performs nearest-neighbor extrapolation. For the outer domain, nearest-neighbor is fine, since in our ambient dynamics model $\mathcal{L}$ we specify a constant environment temperature $T_e$ which is normally used as a Dirichlet boundary condition $g$. On the other hand, near the heat source surface, nearest-neighbor extrapolation is inaccurate. Fortunately, extrapolation only occurs aftward of the movement, and we are particularly interested in the forward region driving the movement. Figure 8.4 shows how, even for a reasonably coarse grid, the majority of the melt interface does not move during the interpolation procedure. There is an obvious error highlighted in red, far aft of the body, where the grid is coarsest. This is acceptable for our problem.

Regarding the detailed implementation, a new class was designed which contains a `FEFieldFunction` member, along with a routine to extrapolate to the nearest boundary after catching the proper exception. The entire class only required 84 lines of code, including the crude custom routine for finding the nearest boundary vertex. The entire source code is at `peclet/source/extrapolated_field.h` (Zimmerman, Peclet, 2016). A slightly abbreviated version is in Appendix B.

In theory, the second option, interpolating the field as a set of scattered data points, should be simple and worth trying. Unfortunately, libraries that support the interpolation of scattered data are not widely available. MATLAB's `ScatteredInterpolant`, a relatively new feature, supports n-dimensional natural neighbor interpolation. The natural neighbor method is the only known option for smoothly interpolating unstructured finite element solutions, where we use the strict definition of "interpolate", meaning that the interpolant must exactly match the data where data is given. The best available option in Python is SciPy's `LinearNDInterpolator`, which, as the name implies, only currently supports linear (*i.e.* non-smooth, only $C^0$ continuous) interpolation. No options have been found in C/C++ for general scattered data interpolation.

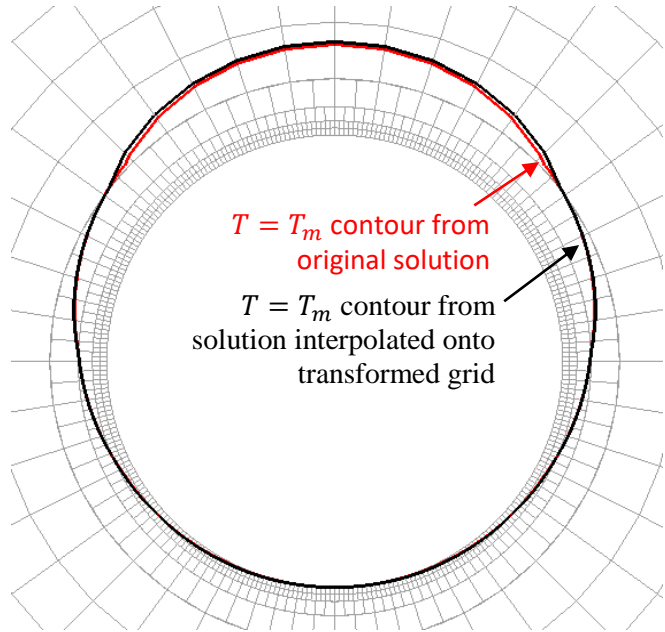

Figure 8.4: Error from interpolating onto a transformed grid
$T_m$ is the melting temperature of the phase-change material (PCM). The error is most evident in the regions with a coarse mesh.

### 8.3.3 Restarting onto a transformed grid with a finite element field function

Developing the capability to write and read the entire finite element solution, rather than a simple projection of it for a visualization tool, to and from the file system was substantially more difficult than expected. The technique is not clearly defined in the deal.II documentation, and the problem becomes much more difficult when using parallel distributed triangulations.

The idea employed in this work is to serialize, using the `boost` library (Dawes & Abrahams, 2016), all data needed to construct the `FEFieldFunction` that samples the solution. When restarting the PDE solver, the data is deserialized, the `FEFieldFunction` is constructed, and this Function is passed as an argument to the initial values function. In Peclet, the methods to write and read this data are encapsulated in the source file `peclet/source/fe_field_tools.h` (Zimmerman, Peclet, 2016).

### 8.3.4 Sampling the finite element solution outside of the solver

As shown in Figure 8.2, we simulate the coupled trajectory with a Python driver. Since the minimization problem $\mathcal{H}$ is solved within Python, we need to share a representation of the finite element solution outside of the PDE solver, which is a compiled C++ program. This will allow us to sample the temperature field for evaluating the inequality constraints (4.5). The finite element solution resides in a complex data structure that is not easily shared externally. A typical solution output file, e.g. one formatted for a visualization tool, omits information about the basis used to construct the solution, and also omits information about geometric manifold, i.e. most visualization tools assume linear interpolation between all nodes.

There are three primary ways we could sample the solution output. In order of decreasing complexity:

A. Program an interface that uses the internal data structure. This would require 1. using a library such as `pybind` to create Python bindings for the C++ code, and 2. compiling a shared library. Neither of these tasks are trivial.
B. Interface with the visualization library to benefit from the discretization structure. This would require 1. constructing an object that represents the geometry we wish to sample onto (the hull of the body), 2. reading the VTK solution, and 3. applying a `VTKProbeFilter` with the VTK solution as the data source, and the geometry as a sampling input.
C. Construct an interpolator based on a scattered interpretation of the output data. This would require 1. reading the solution as a set of unstructured data points, 2. constructing an interpolator using either linear or natural neighbor interpolation

Option A was briefly attempted, but not yet with any success. Option B was attempted with more effort, but working directly with VTK has proven difficult. Instead scripting ParaView (Kitware, 2016) with Python might be a good approach to try in the future. Option C has been successful and is what we use in this chapter. The scattered data representation has already been shown to work in the MATLAB example from section 4.3.3, and a scattered data interpolation class exists in SciPy.

SciPy provides the `LinearNDInterpolator` class to interpolate scattered data. MATLAB's `ScatteredInterpolant` class is more advanced, e.g. it supports a natural neighbor interpolation which is once differentiable, but SciPy's routine was sufficient for this work.

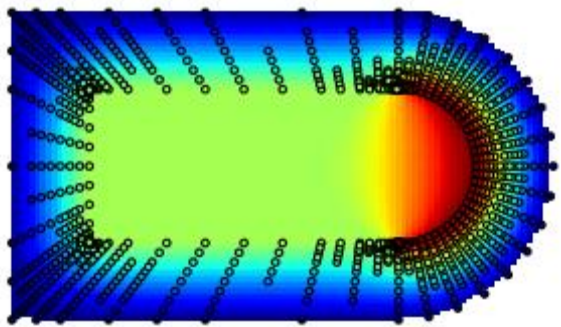

Figure 8.5: Scattered data interpolation with SciPy LinearNDInterpolator.
The circles mark nodal values from the VTK solution output file. The domain is shaded with bilinear interpolation. Since the data is considered scattered, values are interpolated inside of the rigid body (RB) surface boundary as well.

## 8.4 Example results for the coupled trajectory

The coupled problem is implemented in such a way that trajectory simulations can be run for a variety of geometries and boundary conditions, with or without coupling the rigid body's velocity as the convection velocity. First in this section we show three examples where only the position has been coupled for a 2D sphere-cylinder geometry, and then we show an example where the velocity has been coupled for a circular geometry.

### 8.4.1 Coupled position

Turning trajectories for different parametric geometries are shown in Figure 8.6. Here the temperature of the heat source surface was prescribed, rather than a heat flux. Asymmetric temperatures were applied to the nose of a 2D sphere-cylinder.

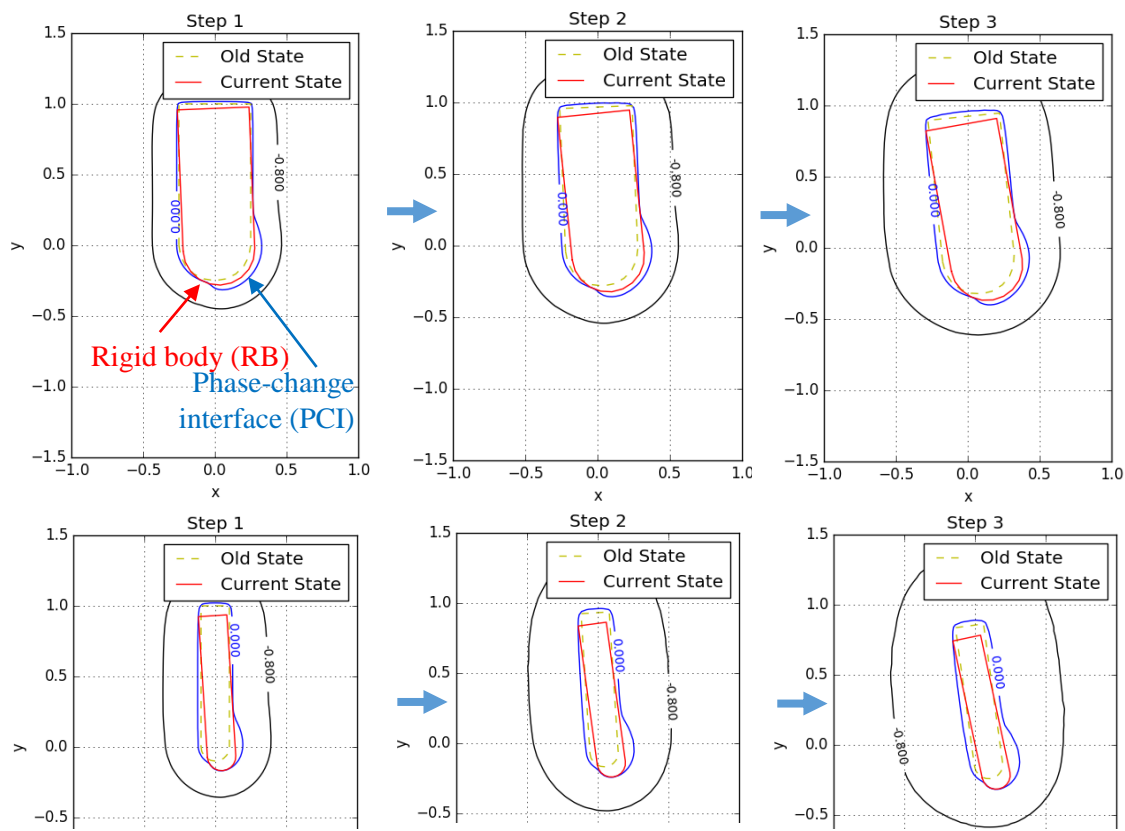

Figure 8.6: Example trajectories with coupled position
Top: Large ratio of aft-body length to nose radius. Bottom: Small ratio of aft-body length to nose radius. Time is stepping from left to right. The rigid body (RB) surface at the current time step is drawn in red, while the RB from the previous time step is drawn with a dashed yellow line. A blue isoline marks the phase-change interface (PCI) in each plot, while a black isoline marks where the temperature is 80% of the environment temperature, to illustrate the time evolution of the entire ambient. Warm Dirichlet BC's have been applied to the body, except for a hot BC on one side of the nose to cause a turn.

To generate Figure 8.7, rather than setting the body's temperature as a Dirichlet BC, the heat flux out of the body was set as a Neumann BC. A larger value was set at the nose, so the melt forms near the nose before the aft-body. In our current implementation of $\mathcal{H}$, the optimization problem must be initialized with a feasible state. The constraints $g$ require that the temperature is strictly greater than the melting temperature, so the state of the RB is invalid until melt has formed around the entire RB, effectively restricting the RB's movement. As soon as the optimizer begins with a feasible state, it is free to minimize the problem, which for our examples in this work is simply minimizing gravitational potential (as detailed in section 4.2). This leads to a very large amount of movement in a single time step of the coupled problem.

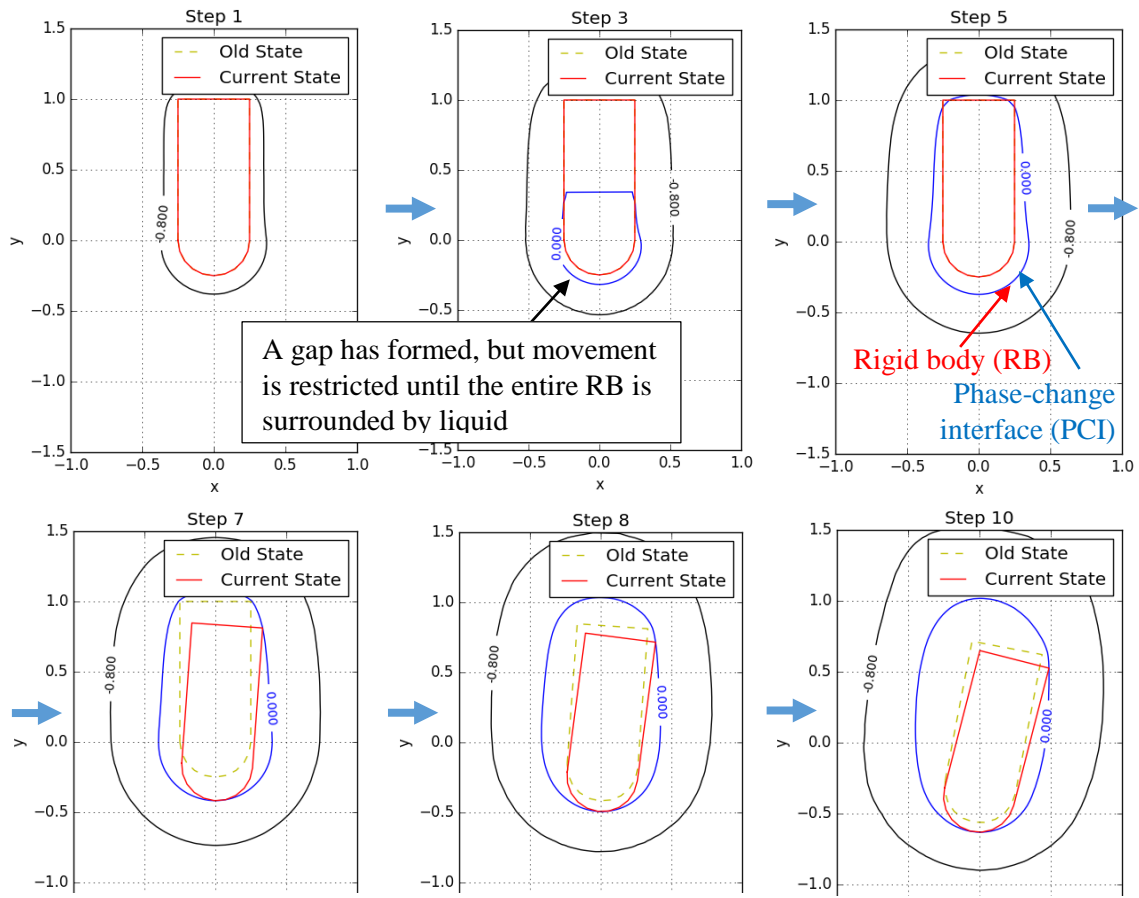

Figure 8.7: Example of restricted movement when the rigid body is not surrounded by liquid
A larger heat flux was prescribed on the nose than the aft-body, and so the melt forms around the nose before forming around the rest of the body. Time is stepping from left to right. The rigid body (RB) surface at the current time step is drawn in red, while the RB from the previous time step is drawn with a dashed yellow line. A blue isoline marks the phase-change interface (PCI) in each plot, while a black isoline marks where the temperature is 80% of the environment temperature, to illustrate the time evolution of the entire ambient. The body is restricted from movement until it is entirely surrounded by melt, and is first permitted to move in step seven.

### 8.4.2 Coupled velocity and position

To explore the behavior when coupling the convection velocity to the velocity of the rigid body (RB), we consider a circular geometry with a spatially constant Neumann BC $h(t)$ applied on the heat source surface $\Gamma_h$, and with a scalar convection velocity $v$ pointing upward. For some initial velocity $v_0$ and a constant heat flux in time $h$ that is sufficiently high to form a melt, an iteration of $\mathcal{H}$ in Algorithm 1 moves the RB and increases the velocity $v$. The next PDE solution will in turn have reduced temperatures near $\Gamma_h$, which then reduces the size of the position and velocity update in the next iteration of $\mathcal{H}$. This eventually reaches a steady state, where no gap forms between $\Gamma_h$ and the PCI, and hence an iteration of $\mathcal{H}$ does not yield any RB state update.

To produce the result in Figure 8.8, after an approximately steady state velocity was reached, the heat flux $h$ was increased by 20%. As we would expect, this causes a dynamic response in the velocity, which eventually reaches a new steady state.

The right side of Figure 8.8 verifies the expected linear relationship between the heat flux and the steady state velocity. This qualitative result is promising. For the method of this thesis to quantitatively predict such a trajectory, the ambient dynamics model $\mathcal{L}$ must be either augmented or replaced to account for the latent heat of melting $h_m$. With a model accounting for $h_m$, we would expect the same steady state velocities as shown here, but different unsteady behavior.

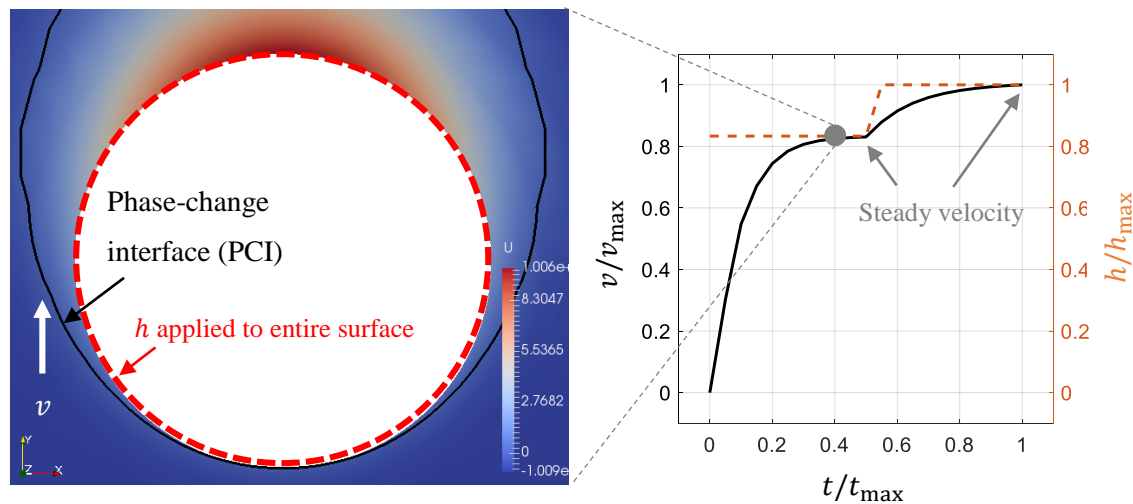

Figure 8.8: Example of coupled velocity with circular geometry
Left: Annotated PDE solution at time $t = 0.25\, t_f$, viewed in ParaView.
Right: Dynamic response of velocity given a step function (in time) for the heat flux input. $v$ is the velocity and $h$ is the heat flux. The Python script to generate this result is in Appendix A.

# 9 Summary, conclusions and proposals for future work

In meeting its primary goals, this thesis has

- formulated a mathematical model for the general problem of rigid body heat source trajectories through phase-change materials. We split the rigid body dynamics and the ambient dynamic into two operators and the coupling of these operators.
- formulated and implemented concrete examples of each of the two operators.
- verified the correctness, and the spatial and temporal orders of accuracy, of our finite element method (FEM) implementation for the ambient dynamics.
- implemented the coupled problem with Python and C++.
- demonstrated the coupled problem by showing the dynamic velocity response to a rapid change in the heat flux, which causes the velocity to approach a new steady state. The result verifies the expected linear relationship between the applied heat flux and the steady state velocity.

Further, this thesis has shown that

- interior-point optimization and sequential least-squares quadratic programming (SLSQP) are both suitable for solving the presented minimization problem when sampling a FEM solution for constraints. Interior-point optimization will likely be superior for larger problems.
- the convection-diffusion equation can model the temperature distribution within the solid domain near the close-contact melting interface of a heat source moving through a phase-change material. This is accomplished by representing the micro-scale melt film attached to the heating surface as a Neumann boundary, with a value based on the heat flux specified by the Stefan condition.
- the Stefan condition based Neumann boundary condition is super-sensitive to changes in the wall temperature, the micro-scale melt film thickness, and the velocity of the melting interface. This suggests that this phase-change model might not be useful for predicting the unsteady physical behavior. This motivates the future work of implementing a realistic melting model within the ambient domain.
- convection dominated problems with sharp boundary layers can be solved with the standard Galerkin formulation of the finite element method. Generally, this can be handled with stabilization schemes that exactly compensate for the artificial loss of diffusion when discretizing the convection operator. We demonstrated the option to instead locally refine the grid at the boundary layer, effectively reducing the local element Peclet number. We admit that this could be prohibitively expensive for large 3D problems.

Moving forward, we should first and foremost quantify the errors introduced by the operator splitting. Error sources include the truncation error from time discretization, and interpolation/extrapolation errors when transferring solutions to transformed grids.

The next critical step is to implement a realistic melting model. This will allow validation against experimental trajectories. Also important, this will yield new geometric properties in the melt interface, and hence qualitatively change the constraints of the minimization problem which models the rigid body dynamics. We must verify that interior point optimization can robustly handle these constraints.

Before attempting any larger problems, another important step is to optimize the routine for interpolating and extrapolating the ambient field onto transformed girds. The critical issue is efficient and accurate extrapolation.

Altogether the approach of this work shows promise and warrants further investigation. A key decision point will be whether to continue with the transformed grid and interpolation approach for the ambient dynamics, or to invent a new approach which does not require interpolation/extrapolation. If such a new approach is found for the ambient dynamics, then the rigid body dynamics formulation and the temporal coupling from this thesis should be equally applicable.

# Appendix

## A. Peclet code verification

Peclet parameter input file to reproduce Donea & Huerta Figure 5.17 with $Pe(h = 0.1) = 0.5$

```
subsection meta
    set dim = 1
end

subsection output
    set write_solution_table = true
end

subsection pde

  subsection parsed_velocity_function
      set Function expression = 1.
  end

  subsection parsed_diffusivity_function
      set Function expression = 0.1
  end

  subsection parsed_source_function
      set Function expression = 1.
  end

end

subsection initial_values
    subsection parsed_function
        set Function expression = 0.
    end
end

subsection boundary_conditions
    set implementation_types = strong, strong
    subsection parsed_function
        set Function expression = 0.
    end
end

subsection refinement
    set initial_global_cycles = 3
end

subsection time
    set end_time = 1.2
    set step_size = 0.01
end
```

Set to 0.01 to run for $Pe = 5$

Edit this section to reproduce boundary refined case

```
subsection refinement
  set boundaries_to_refine = 1
  set initial_boundary_cycles = 3
  set initial_global_cycles = 3
end
```

Symbolic derivation of 1D MMS auxiliary conditions in MATLAB

```
syms x t g v alpha beta real
syms ue(x) ue_v0(x) um(x,t) um_v0(x)
syms sm(x,t) sm_v0(x,t) hm(t) hm_v0(t)
ue = g*(exp(v*x/alpha) - 1)/(exp(v/alpha) - 1);
ue_v0 = limit(ue, v, 0);
um = g*(1 + ((exp(v*x/alpha) - 1)/(exp(v/alpha) - 1) - 1)*...
    (1 - exp(-beta*t^2)));
um_v0 = limit(um, v, 0);
umx = diff(um, x);
hm = subs(-alpha*umx, x, 0);
hm_v0 = limit(hm, v, 0);
umt = diff(um, t);
umxx = diff(umx, x);
sm = umt + v*umx - alpha*umxx;
sm_v0 = limit(sm, v, 0);
```

Symbolic derivation of 2D MMS auxiliary conditions in MATLAB

```
syms x y t g alpha beta vmax real
syms vx(y) u0 uinfty(x,y) u(x,y,t)
syms h_y0(x,t) h_x0(y,t) h_x1(y,t), s(x,y,t)
vx = vmax*y;
u0 = g;
uinfty = g*(exp(vx*x/alpha) - 1)/(exp(vx/alpha) - 1);
u = u0 + (uinfty - u0)*(1 - exp(-beta*t^2));
u = simplify(u);
u_y0 = limit(u, y, 0);
h_x0 = -alpha*diff(u, x);
h_x0 = subs(h_x0, x, 0);
h_y0 = -alpha*diff(u, y);
h_y0 = limit(h_y0, y, 0);
h_y1 = alpha*diff(u, y);
h_y1 = subs(h_y1, y, 1);
s = diff(u, t) + vx*diff(u, x) - ...
    alpha*(diff(diff(u, x), x) + diff(diff(u, y), y));
s_y0 = limit(s, y, 0);
```

Example Peclet parameter input file for 1D MMS

```
subsection meta
    set dim = 1
end

subsection geometry
    set grid_name = hyper_cube
    set sizes = 0., 1.
End

subsection verification
    set enabled = true
    set exact_solution_function_name = parsed
    subsection parsed_exact_solution_function
        set Function constants = alpha=2, v=-5, g=-2, beta=10
        set Function expression = -g*(((exp((v*x)/alpha) - 1)/
            (exp(v/alpha) - 1) - 1)*(exp(-beta*t^2) - 1) - 1)
    end
end

subsection parsed_velocity_function
    set Function constants = v=-5
    set Function expression = v
end

subsection parsed_diffusivity_function
    set Function constants = alpha=2
    set Function expression = alpha
end

subsection parsed_source_function
    set Function constants = alpha=2, v=-5, g=-2, beta=10
    set Function expression = 2*beta*g*t*exp(-beta*t^2)*((exp(
1)/(exp(v/alpha) - 1) - 1)
end

subsection initial_values
    subsection parsed_function
        set Function expression = -2*1.000000001
    end
end

subsection boundary_conditions
    set implementation_types = natural, strong
    set function_names = parsed, constant
    set function_double_arguments = -2.
    subsection parsed_function
        set Function constants = alpha=2, v=-5, g=-2, beta=10
        set Function expression=(g*v*(exp(-beta*t^2) - 1))/
            (exp(v/alpha) - 1)
    end
end
subsection refinement
  set initial_global_cycles = 6
end
subsection time
  set global_refinement_levels = 6
end
```

# Example Peclet parameter input file for 2D MMS

```
subsection meta
    set dim = 2
end
subsection geometry
    set grid_name = hyper_rectangle
    set sizes = 0., 0., 1., 1.
end
subsection verification
    set enabled = true
    subsection parsed_exact_solution_function
        set Function constants = epsilon=1.e-14, alpha=2, vmax=-5, g=-2, beta=10
        set Function expression = if(y < epsilon,g + (g - g*x)*(exp(-beta*t^2) - 1),
            g + (g - (g*(exp((vmax*x*y)/alpha) - 1))/(exp((vmax*y)/alpha) - 1))*
            (exp(-beta*t^2) - 1))
    end
end
subsection parsed_velocity_function
    set Function constants = vmax=-5
    set Function expression = vmax*y; 0
end
subsection parsed_diffusivity_function
    set Function constants = alpha=2
    set Function expression = alpha
end
subsection parsed_source_function
    set Function constants = epsilon=1.e-14, alpha=2, vmax=-5, g=-2, beta=10
    set Function expression = if(y<epsilon,(g*exp(-beta*t^2)*(x - 1)*(2*vmax^2*x^2 - vma
2*vmax^2*x^2*exp(beta*t^2) + 12*alpha*beta*t + vmax^2*x*exp(beta*t^2)))/(6*alpha),- alph
beta*t^2) - 1)*((g*vmax^2*exp((vmax*y)/alpha)*(exp((vmax*x*y)/alpha) -
1))/(alpha^2*(exp((vmax*y)/alpha) - 1)^2) -
(2*g*vmax^2*exp((2*vmax*y)/alpha)*(exp((vmax*x*y)/alpha) - 1))/(alpha^2*(exp((vmax*y)/al
1)^3) - (g*vmax^2*x^2*exp((vmax*x*y)/alpha))/(alpha^2*(exp((vmax*y)/alpha) - 1)) +
(2*g*vmax^2*x*exp((vmax*y)/alpha)*exp((vmax*x*y)/alpha))/(alpha^2*(exp((vmax*y)/alpha)
(g*vmax^2*y^2*exp((vmax*x*y)/alpha)*(exp(-beta*t^2) - 1))/(alpha^2*(exp((vmax*y)/alpha)
2*beta*t*exp(-beta*t^2)*(g - (g*(exp((vmax*x*y)/alpha) - 1))/(exp((vmax*y)/alpha) - 1))
(g*vmax^2*y^2*exp((vmax*x*y)/alpha)*(exp(-beta*t^2) - 1))/(alpha*(exp((vmax*y)/alpha) -
end
subsection initial_values
    subsection parsed_function
        set Function expression = -2*1.000000001
    end
end
subsection boundary_conditions
    set implementation_types = natural, strong, natural, natural
    set function_names = parsed, constant, parsed, parsed
    set function_double_arguments = -2
    subsection parsed_function
        set Function constants = epsilon=1.e-14, alpha=2, vmax=-5, g=-2, beta=10
        set Function expression = if(x<epsilon,(g*vmax*y*(exp(-beta*t^2) - 1))/
            (exp((vmax*y)/alpha) - 1),if(y<epsilon,-alpha*(exp(-beta*t^2) -
            1)*((g*vmax*x)/(2*alpha) - (g*vmax*x^2)/(2*alpha)),alpha*(exp(-beta*t^2)
            - 1)*((g*vmax*exp(vmax/alpha)*(exp((vmax*x)/alpha) - 1))/
            (alpha*(exp(vmax/alpha) - 1)^2) - (g*vmax*x*exp((vmax*x)/alpha))/
            (alpha*(exp(vmax/alpha) - 1)))))
    end
end
subsection refinement
    set initial_global_cycles = 4
end
subsection time
    set global_refinement_levels = 4
end
```

## A. Example Python scripts for running the trajectory simulator

circle_single_step.py: Script for circle geometry in Figure 4.5

```python
#!/usr/bin/python
import trajectory
import pandas

t = trajectory.Trajectory()
r = 1.
t.body.sizes[0] = r
t.max_change = [0., r/2., 0.]

t.pde.geometry.grid_name = 'hyper_shell'
t.pde.geometry.sizes = [r, 2*r]
t.pde.bc.implementation_types = ['natural', 'strong']
t.pde.bc.function_names = ['constant', 'constant']
t.pde.bc.function_double_arguments = [2., -1.]
t.pde.iv.function_name = 'constant'
t.pde.iv.function_double_arguments = -1.
t.pde.refinement.initial_global_cycles = 5
t.pde.time.step_size = 0.2

t.run_step()
```

sphere-cylinder_single_step.py: Script for sphere-cylinder geometry in Figure 4.5

```python
#!/usr/bin/python
import trajectory

t = trajectory.Trajectory()

t.body.geometry_name = 'sphere-cylinder'
r = 0.1
L = 1
t.body.sizes = [r, L]

t.pde.geometry.grid_name = 'hemisphere_cylinder_shell'
t.pde.geometry.sizes = [r, r + 3*r, L, L + 3*r]
t.pde.bc.implementation_types = ['strong', 'strong', 'strong',
      'strong', 'strong', 'strong', 'strong', 'strong',
      'strong', 'strong']
t.pde.bc.function_names = ['constant', 'constant', 'constant',
      'constant', 'constant', 'constant', 'constant', 'constant',
      'constant', 'constant']
t.pde.bc.function_double_arguments = [-1, -1, -1,
      -1, -1, 1, 0.1, 0.1,
      0.1, 0.2]
t.pde.iv.function_name = 'constant'
t.pde.iv.function_double_arguments = -1.
t.pde.refinement.initial_global_cycles = 5
t.pde.time.semi_implicit_theta = 1.
t.pde.time.step_size = 0.2

t.run_step()
```

Python script to simulate response of velocity to a step in the heat flux

```
#!/usr/bin/python
import trajectory
import pandas

def make_time_history_row(traj):
    return pandas.DataFrame(
        {'step': traj.step, 'time': traj.time,
         'velocity': traj.state_dot.get_position()[1],
         'depth': traj.state.get_position()[1],
         'pde_depth': traj.pde.state.get_position()[1],
         'heat_flux': traj.pde.bc.function_double_arguments[0]},
        index=[traj.step])

t = trajectory.Trajectory()
r = 1.
t.body.sizes[0] = r
t.max_change = [0., r/2., 0.]

t.pde.geometry.dim = 2
t.pde.geometry.grid_name = 'hyper_shell'
t.pde.geometry.sizes = [r, 2*r]
t.pde.bc.implementation_types = ['natural', 'strong']
t.pde.bc.function_names = ['constant', 'constant']
t.pde.bc.function_double_arguments = [2., -1.]
t.pde.iv.function_name = 'constant'
t.pde.iv.function_double_arguments = -1.
t.pde.refinement.boundaries_to_refine = 0
t.pde.refinement.initial_boundary_cycles = 3
t.pde.refinement.initial_global_cycles = 2
t.pde.time.semi_implicit_theta = 1.
t.pde.time.step_size = 0.2

time_history =  make_time_history_row(t)

for step in range(0,10):
    t.run_step()
    time_history = time_history.append(make_time_history_row(t))

t.pde.bc.function_double_arguments[0] =
1.2*t.pde.bc.function_double_arguments[0]

for step in range(10, 20):
    t.run_step()
    time_history = time_history.append(make_time_history_row(t))

print(time_history)
```

## B. Code excerpts

`ExtrapolatedField`, derived from deal.II's `FEFieldFunction`

```
#include <deal.II/base/function.h>
#include <deal.II/grid/grid_tools.h>
#include <deal.II/numerics/fe_field_function.h>
using namespace dealii;

template<int dim>
class ExtrapolatedField : public Function<dim>
{
public:
    ExtrapolatedField(const DoFHandler<dim> &dof_handler,
            const Vector<double> &f)
      : Function<dim>(),
        field(dof_handler, f),
        dof_handler_sp(&dof_handler, "ExtrapolatedField")
    {}
    virtual double value(const Point<dim>  &point,
                         const unsigned int component = 0) const;
private:
    Functions::FEFieldFunction<dim> field;
    SmartPointer<const DoFHandler<dim>,ExtrapolatedField<dim>> dof_handler_sp;
    Point<dim> get_nearest_boundary_vertex(const Point<dim> &point) const;
};

template<int dim>
double ExtrapolatedField<dim>::value(const Point<dim> &point,
                                     const unsigned int component) const
{
    Assert(component == 0, ExcInternalError());
    double val;
    try
    {
        val = field.value(point, component);
    }
    catch (GridTools::ExcPointNotFound<dim>)
    {
        val = field.value(get_nearest_boundary_vertex(point));
    }
    return val;
}

template <int dim>
Point<dim> ExtrapolatedField<dim>::get_nearest_boundary_vertex
    (const Point<dim> &point) const
{
    double arbitrarily_large_number = 1.e32;
    double nearest_distance = arbitrarily_large_number;
    Point<dim> nearest_vertex;
    for (auto cell : dof_handler_sp->active_cell_iterators())
    {
    if (!cell->at_boundary())
        {
            continue;
        }
        for (unsigned int f=0; f<GeometryInfo<dim>::faces_per_cell; ++f)
        {
            if (!cell->face(f)->at_boundary())
            {
                continue;
            }
            for (unsigned int v=0; v < GeometryInfo<dim>::vertices_per_face; ++v)
            {
                Point<dim> vertex = cell->face(f)->vertex(v);
                double distance = (point - vertex).norm_square();
                if (distance < nearest_distance)
                {
                    nearest_vertex = vertex;
                    nearest_distance = distance;
                }
            }
        }
    }
```